\documentclass[11pt,twoside]{report}
\pdfoutput=1
\usepackage{suthesis-2e}
\usepackage{amsmath}
\usepackage{txfonts}
\usepackage{textcomp}
\usepackage{gensymb}
\usepackage[numbers,sort&compress]{natbib}
\usepackage{units}
\usepackage[pdftex]{graphicx}
\usepackage[pdftex]{color}
\usepackage[amsmath]{benc}
\usepackage{amssymb}
\usepackage{fancyhdr}

\DeclareMathOperator{\erf}{erf}

\renewcommand{\comment}[1]{{\color{blue} [#1]}}

\newcommand{\delt}{\del_\perp}

\begin{document}

\title{Photonic Crystal Laser-Driven Accelerator Structures}
\author{Benjamin M. Cowan}
\submitdate{April 2007}
\dept{Physics}

\titletop{
\thispagestyle{fancy}
\renewcommand{\headrulewidth}{0pt}
\rhead{%
SLAC--R--877\\
AARD--482%
}
\cfoot{}
}

\titlebottom{%
\begin{center}
\emph{Stanford Linear Accelerator Center, 2575 Sand Hill Road, Menlo Park, CA 94025}\\
\hrulefill\\
Work supported by Department of Energy contracts DE-AC02-76SF00515 (SLAC) and
DE-FG03-97ER41043-II (LEAP).
\end{center}%
}

\principaladvisor{Robert Siemann}
\secondreader{Robert Byer \\ (Applied Physics)}
\thirdreader{Todd Smith}

\beforepreface
\prefacesection{Abstract}
Laser-driven acceleration holds great promise for significantly
improving accelerating gradient.  However, scaling the conventional
process of structure-based acceleration in vacuum down to optical
wavelengths requires a substantially different kind of structure.  We
require an optical waveguide that (1) is constructed out of dielectric
materials, (2) has transverse size on the order of a wavelength, and
(3) supports a mode with speed-of-light phase velocity in vacuum.
Photonic crystals---structures whose electromagnetic properties are
spatially periodic---can meet these requirements.

We discuss simulated photonic crystal accelerator structures and
describe their properties.  We begin with a class of two-dimensional
structures which serves to illustrate the design considerations and
trade-offs involved.  We then present a three-dimensional structure,
and describe its performance in terms of accelerating gradient and
efficiency.  We discuss particle beam dynamics in this structure,
demonstrating a method for keeping a beam confined to the waveguide.

We also discuss material and fabrication considerations.  Since
accelerating gradient is limited by optical damage to the structure,
the damage threshold of the dielectric is a critical parameter.  We
experimentally measure the damage threshold of silicon for picosecond
pulses in the infrared, and determine that our structure is capable of
sustaining an accelerating gradient of \unit[300]{MV/m} at
\unit[1550]{nm}.  Finally, we discuss possibilities for manufacturing
these structures using common microfabrication techniques.

\prefacesection{Acknowledgments}
I have been fortunate during my graduate career to have had the
benefit of interacting with so many talented and inspirational
people.  None of this would have been possible without their help,
guidance, and encouragement.

First and foremost, I would like to thank my advisor, Bob Siemann.  He
placed his trust in me and allowed me the freedom to find my own way.
At the same time, he was always available to offer his insights and
advice, and steered me out of many an intellectual quagmire.  His
leadership has helped me mature as a scientist and as a person.

Eric Colby has been a mentor to me from the beginning and has garnered
my respect and admiration from day one.  His intensity and brilliant
creativity have been an inspiration to me.

\begin{sloppypar}
My colleagues on the LEAP experiment, Tomas Plettner, Chris Barnes,
Chris Sears, Jim Spencer, Bob Byer, and Todd Smith, shared with me the
excitement of turning an idea into reality, and the extra excitement
of turning an idea into reality at 3 a.m.  For that I will always be
grateful.
\end{sloppypar}

I have enjoyed working with my colleagues on E-163 structure
development, namely Bob Noble, Rasmus Ischebeck, Chris McGuinness, and
Melissa Lincoln.  I have also enjoyed the company and lunchtime
companionship of all those in AARD who I was not fortunate enough to
work with directly.  In addition, I wish to thank Stephanie Santo for
all of her help.

There are many others who deserve recognition for making this work
possible.  Alf Wachsmann has diligently maintained the MPI cluster
here at SLAC which has been critical to many of my simulations.  Sami
Tantawi, Cho Ng, and Shanhui Fan have been generous with helpful
advice, and I have had several thoroughly enlightening conversations
with Mary Tang at SNF regarding microfabrication.  I wish to extend
special thanks to George Marcus, Keith Cohn, Dmitrii Simanovskii, and
Daniel Palanker.  They graciously allowed me to use valuable time on
their OPA system and many times helped me to tame that particular
beast.

The support of friends and family has been indispensable throughout my
time in graduate school.  I extend my thanks to all those in the
Stanford Running Club who have gone running with me over the years,
from jaunts around campus to marathons far away.  To Robert and the
University Singers, I always looked forward to Tuesday nights for fun
and relaxation.  Sara, Martin, Jim, and Deirdre have been wonderful
friends for many years.

Being near my family has been one of the great benefits of being back
in the Bay Area.  I wish to thank my parents for their unwavering
support.  This thesis is dedicated to my grandmother, who is the best
grandmother anyone could wish for.

Finally, to Jason: Thank you for being by my side, and making me smile
every single day.

\afterpreface

\chapter{Introduction}
\label{ch:Introduction}
Humankind's understanding of the fundamental physical properties of
our universe has advanced greatly during the last century.  Particle
accelerators have been critical to these advances, and continual
improvements in accelerators have allowed continued expansion of our
scientific understanding.  From early explorations of nuclear
structure, through the discovery of quarks, to the current standard
model of particle physics, improvements in particle beam energy and
luminosity from accelerators have paved the way for new discoveries.

\begin{sloppypar}
However, significant improvements in fundamental accelerator
technology are necessary if accelerator-based particle physics is to
continue its pace of discovery.  New regimes of particle interactions
are probed using particle beams of higher energy.  In the past, it has
been possible to increase particle energy by simply increasing the
physical size of the accelerator.  In a linear accelerator, beam
energy is proportional to the length of the accelerator.  In an
electron storage ring, energy is limited by synchrotron radiation.
With synchrotron radiation power scaling as $p^4/r^2$, where $p$ is
the particle momentum and $r$ is the ring radius, larger accelerators
are necessary to reduce the synchrotron radiation loss.  While
synchrotron radiation is not significant in a proton storage ring due
to the greater proton mass, the energy is limited in that case by
dipole magnet strength.  The required average magnetic field scales as
$p/r$, so again the energy can be increased by making the accelerator
larger.  Now, with the largest storage ring having a circumference of
\unit[27]{km}, and proposed designs for the next-generation linear
collider reaching \unit[30]{km}, we are reaching the limit of
accelerator size within reasonable practical constraints.  Instead,
beam energy must be improved by increasing the
\emph{gradient}---the particle energy gain per unit length---in linear
accelerators.
\end{sloppypar}

As accelerator technology has been developed for particle physics, it
has found uses in other areas of research.  The same synchrotron
radiation which limits the beam energy in storage rings can be used to
probe matter on the molecular level.  Thus accelerators are used as
versatile x ray sources for a wide range of applications, such as
biochemistry, condensed matter, and materials science.  In addition,
particle beams are used in medicine for radiation treatment and
isotope production.  Advances in fundamental accelerator technology
may allow accelerators for these purposes to be built with greater
capability while being more compact and less expensive.

This dissertation presents an acceleration technique with the
potential to significantly improve the capabilities of particle
accelerators.  There are many bridges to cross between current
accelerator technology and the goal of a practical, operating
accelerator delivering a significant improvement in gradient.  It will
likely be years before that goal is realized.  However, this document
addresses some of the major technical challenges standing in the way
of that goal, and indicates a path forward toward reaching it.

\section{Structure-based laser-driven acceleration in vacuum}
\label{sec:LaserAcceleration}

The key motivation behind the concepts presented here is to use lasers
instead of microwave klystrons as the power sources for an accelerator
because of the much larger intensities available from lasers.  To
compare intensities, let us take $P/\lambda^2$ as the intensity of a
source, where $P$ is the power and $\lambda$ is the wavelength, since
the mode area is limited by diffraction.  A typical source for a
conventional RF accelerator has $P\sim\unit[100]{MW}$ and
$\lambda\sim\unit[10]{cm}$.  In contrast, ultrafast lasers exist which
can produce power well in excess of \unit[1]{TW} \cite{Zhou:Terawatt}.
For $\lambda\sim\unit[1]{\micro m}$, such a laser exceeds the klystron
in intensity by 14 orders of magnitude, corresponding to a factor of
$10^7$ in gradient.

The question then becomes one of how to utilize properly the
extraordinary power available from lasers to accelerate a charged
particle beam.  We require two characteristics of an accelerating
laser field: First, we must have an electric field in the direction of
propagation of the particle beam.  Second, the mode must have phase
velocity in that direction equal to the speed of light in vacuum, for
phase matching with the particle bunch.  In a conventional RF
accelerator, this is accomplished using a disk-loaded metallic
waveguide.  Because of the linearity of the Maxwell equations, the
same structure with the same material properties, simply scaled down
to optical wavelengths, would in principle serve our purposes.
However, in practice this is not realistic.  First of all, because of
the significant loss of metals at optical frequencies, we wish instead
to use dielectric materials for a structure.  Second, manufacturing a
circular disk-loaded waveguide on such small scales poses a
significant challenge.  For these reasons, we must consider structures
which differ significantly from those used in conventional
accelerators.

The Laser Electron Acceleration Program (LEAP), an experimental
program to address the question posed above, is currently underway,
and the first stage in that program was recently completed.  In those
experiments, a free-space mode was used to modulate an unbunched
electron beam \cite{Plettner:LEAPResults2004}.  A single TEM\tsub{00}
Gaussian beam was propagated at a small angle with respect to the
electron bunch, giving the $\vect{E}$-field a small longitudinal
component.  A general theorem of acceleration (often dubbed the
``Lawson-Woodward theorem'' but stated clearly by Palmer
\cite{Palmer:AccelerationTheorems}) prohibits net energy
exchange between a free-space mode and a charged particle.  Since the
theorem only applies in the absence of materials, a single boundary
was used to terminate the fields and allow net acceleration.  The
experiment demonstrated the expected linear scaling of energy
modulation with laser electric field as well as the expected
polarization dependence.  The next stage of the experimental program,
to be performed at the E163 facility at SLAC, seeks to demonstrate net
energy gain by first optically bunching the electron beam using an
IFEL \cite{Barnes:LEAPPhase2}.  Beyond that, however, the experimental
program is less clear.  The desire is to demonstrate a scalable
acceleration mechanism.

Let us consider the possibility of using free-space modes to
accelerate a particle beam.  For instance, we might try to design a
structure to refocus and reset the phase of a propagating mode
periodically in order to overcome the diffraction and phase mismatch
associated with a free-space mode.  We take the particle beam to
propagate in the $z$-direction, and the laser field to co-propagate
with it.  We can write a free-space field as a superposition of
Gauss-Hermite modes.  The simplest such field with a longitudinal
electric field component on axis is the lowest-order odd mode; let us
therefore consider the $x$-polarized TEM\tsub{10} mode.  The amplitude
of the transverse component at the beam waist is given by
\cite{Siegman:Lasers}
\[ E_x = E_0\frac{x}{w_0}e^{-(x^2 + y^2)/w_0^2}, \]
where $w_0$ is the beam waist parameter.
We can find the longitudinal component amplitude using the paraxial
approximation, which assumes that the amplitude envelope of the fields
varies slowly, on a scale much longer than a wavelength.
Specifically, if $\psi$ is a field component, we can let $\psi =
ue^{-ik_0z}$, where $u$ is the envelope function and $k_0 =
2\pi/\lambda$.  Then we assume
\[ \abs{\frac{\ptl u}{\ptl z}}\ll \abs{k_0u}, \]
so that we have the approximation
\[ \frac{\ptl\psi}{\ptl z}\approx -ik_0\psi. \]
The Maxwell equation $\del\cdot\vect{E} = 0$ then becomes
\[ 0 = \frac{\ptl E_x}{\ptl x} + \frac{\ptl E_z}{\ptl z}
 \approx \frac{\ptl E_x}{\ptl x} - ik_0E_z, \]
so
\[ E_z \approx -\frac{i}{k_0w_0}E_0\paren{1 - \frac{2x^2}{w_0^2}}
e^{-(x^2 + y^2)/w_0^2}. \]
The ratio of the magnitude of the longitudinal component of the
$\vect{E}$-field on axis, which we denote $E\tsub{acc}$, and the
maximum transverse field magnitude is then given by
\[ \frac{E\tsub{acc}}{|E_x|\tsub{max}} =
\frac{e^{1/2}}{\sqrt{2}\pi}\frac{\lambda}{w_0}. \]

We must place a boundary within one Rayleigh length of the focus due
to the Guoy phase shift inherent in free-space modes.  The Rayleigh
length, defined as $z_0 = \pi w_0^2/\lambda$, sets the length scale
for both diffraction and dephasing of the beam.  The laser field will
acquire a Guoy phase shift with respect to a particle bunch equal to
$\phi(z) = 2\tan^{-1}(z/z_0)$ \cite{Siegman:Lasers}.  When the laser
field is more than $\pi/2$ out of phase, it will decelerate rather
than accelerate the particles.  We therefore require that
\[ \abs{\tan^{-1}\paren{\frac{z}{z_0}}} \le \frac{\pi}{4}, \]
or $z < z_0$.  The boundary must therefore be placed within one
Rayleigh length of the focus, and at that short distance, the fields
are only reduced from their value at the focus by a factor of 2.  Thus
$|E_x|$ is limited by the damage threshold of the structure material,
so we must then make $w_0$ as small as possible to achieve the largest
possible accelerating gradient; namely, we must have $w_0\sim\lambda$.
However, in that case, $z_0
\sim\lambda$.  Therefore we must interrupt the beam approximately
every wavelength of propagation for refocusing and rephasing.  We are
now right back where we started: We are not considering a free-space
mode at all, but rather a waveguide which confines the mode in a
diameter on the order of a wavelength, and has speed-of-light phase
velocity.  This can be accomplished with dielectric materials using
photonic crystals, which we describe in the next section.

\section{Photonic crystals}

\subsection{Introduction and motivation}

A \emph{photonic crystal}, most broadly speaking, is a structure whose
electromagnetic parameters are spatially periodic, where the length of
each period is on the same size scale as the operating wavelength of
the structure \cite{Joannopoulos:Molding}.  A multilayer dielectric
reflector is an example of a one-dimensional photonic crystal, since
its permittivity varies periodically with depth.  As we shall see,
two- and three-dimensional photonic crystals can be considered
generalizations of the multilayer structure, since they have a similar
function: to reflect, and therefore confine, light.

Interesting and useful devices can be formed by introducing a
\emph{defect} into a photonic crystal lattice.  Here, the term
``defect'' does not refer to a manufacturing error or material
impurity, but rather describes a deliberate geometric feature embedded
within the structure in place of the lattice in that region.  Since
photonic crystals can function as reflectors, light can be confined to
such defects.  Therefore, a bounded defect can serve as a resonant
cavity, and an extended linear defect can be a waveguide.

While photonic crystals provide a mechanism for guiding fields, there
are other methods available.  In fact, because of the periodic
geometry, photonic crystal waveguides present more complex problems
for simulation and manufacturing than do simpler methods.  However, we
find that conventional methods of guidance are generally not suitable
for vacuum laser-driven acceleration.  First, as remarked above, we
are limited to low-loss dielectric materials as opposed to metals,
which are ubiquitous in microwave accelerators.

Second, the requirement of phase-matching to a speed-of-light particle
beam excludes conventional optical fibers, which guide by total
internal reflection (TIR).  In an axially uniform structure, within a
region of uniform material, each field component $\psi$ obeys the
equation
\[ (\delt^2 + \gamma^2)\psi = 0, \]
where $\delt$ denotes the gradient in the transverse dimensions, and
$\gamma^2 = \mu\epsilon\omega^2 - k_z^2$.  Here $\mu$ and $\epsilon$
are the permeability and permittivity of the material, respectively,
$\omega$ is the angular frequency of the field, and $k_z$ is the
longitudinal wavenumber.  For a properly phase-matched field, we must
have $k_z = \omega/c$, where $c = 1/\sqrt{\mu_0\epsilon_0}$ is the
speed of light in vacuum.  Then in a material with index of refraction
$n$, we have
\[ \gamma^2 = (\mu\epsilon - \mu_0\epsilon_0)\omega^2
= (n^2 - 1)\frac{\omega^2}{c^2}. \]
However, for guiding to occur, the waveguide must be surrounded by a
cladding region, in which the fields are evanescent, that is, decaying
exponentially away from the core.  Fields are evanescent only in
regions with $\gamma^2 < 0$, but for any conventional dielectric
material, $n\ge 1$, so $\gamma^2\ge 0$.  Thus TIR guiding is
unsuitable for particle acceleration.  The unsuitability of these simple,
widely-used mechanisms for guiding electromagnetic fields strongly
motivates the investigation of photonic crystal waveguides.

\subsection{Maxwell equations in periodic media}

We can understand the behavior of photonic crystals by solving the
Maxwell equations with periodic boundary conditions.  We start by
writing the Maxwell equations in the frequency domain, with
$e^{i\omega t}$ time dependence assumed:
\[ \del\cross\vect{E} = -i\omega\mu_0\vect{H},\quad
\del\cross\vect{H} = i\omega\epsilon_0\epsilon_r\vect{E}. \]
Here we are considering the materials to be nonmagnetic at optical
frequencies, so we use the free-space permeability $\mu_0$.  Also, we
separate the permittivity into two parts, the free-space permittivity
$\epsilon_0$, and the relative permittivity $\epsilon_r$, which is a
dimensionless parameter equal to $n^2$.  We also consider free charges
and currents to be absent from our system.

Combining these first-order equations into a single second-order
equation
\begin{equation}
\del\cross\paren{\frac{1}{\epsilon_r}\del\cross\vect{H}}
= \frac{\omega^2}{c^2}\vect{H},
\label{eq:MaxwellEigenproblem}
\end{equation}
we can see that we have an eigenvalue problem:  We define the
operator $\Theta$ on differentiable vector fields by
\[ \Theta\vect{H}
= \del\cross\paren{\frac{1}{\epsilon_r}\del\cross\vect{H}}. \]
Then $\Theta$ is a linear operator, since both the curl operator and
multiplication by a scalar field are linear, so we have a linear
eigenvalue problem with $\Theta$ as the operator and $(\omega/c)^2$ as
the eigenvalue.  Furthermore, we can use
the standard Euclidean inner product $\ip{,}$ on complex vector
fields:
\[ \ip{\vect{F}, \vect{G}} =
\int_{\mathbb{R}^3}\vect{F}(\vect{x})^*\cdot\vect{G}(\vect{x})\,d^3\vect{x}.
\]
Under this inner product, $\Theta$ is then a Hermitian operator, that
is, $\ip{\vect{F}, \Theta\vect{G}} = \ip{\Theta\vect{F}, \vect{G}}$.
Hermiticity implies that all the eigenvalues are real, and that the
eigenvectors are orthogonal (or can be chosen to be, in the case of
degenerate eigenvalues).  Not only that, but $\Theta$ is also positive
semidefinite, meaning that all its eigenvalues are nonnegative.  This
is expected for our case of lossless dielectric materials, since it
means that all the frequencies are real.

The power of photonic crystals derives from their periodic nature.
Mathematically, the periodicity requires that an eigenvector
$\vect{H}$ of the eigenvalue problem in
Eq.~(\ref{eq:MaxwellEigenproblem}) take the form $\vect{H} =
\vect{u}e^{-i\vect{k}\cdot\vect{x}}$, where the vector field $\vect{u}$
is periodic in the photonic crystal lattice and $\vect{k}$ is any
vector.  Here $\vect{k}$ is termed the \emph{Bloch wavevector}, and
the phase shift among unit cells is given by the
$e^{-i\vect{k}\cdot\vect{x}}$ term.  Writing
Eq.~(\ref{eq:MaxwellEigenproblem}) in terms of $\vect{u}$, we have
\begin{equation}
(\del - i\vect{k})\cross
\brckt{\frac{1}{\epsilon_r}(\del - i\vect{k})\cross\vect{u}}
= \frac{\omega^2}{c^2}\vect{u}.
\label{eq:BlochEigenproblem}
\end{equation}
This is again an eigenproblem, this time for a Hermitian operator
$\Theta_{\vect{k}}$ which depends on the wavevector and is given by
\begin{equation}
\Theta_{\vect{k}}\vect{u} = (\del - i\vect{k})\cross
\brckt{\frac{1}{\epsilon_r}(\del - i\vect{k})\cross\vect{u}}.
\label{eq:MaxwellOperator}
\end{equation}
Now since $\vect{u}$ is periodic, the spectrum of each
$\Theta_{\vect{k}}$ is discrete.  We can therefore separate the
frequencies into bands, i.e. the set of lowest frequencies
$\{\omega_1(\vect{k})\}$ for each $\vect{k}$ forms the first band, and so
on.  In addition, the wavevectors are only distinct modulo the
reciprocal lattice; in other words if $\vect{k}_1$ and $\vect{k}_2$
are separated by a reciprocal lattice vector, then
$\Theta_{\vect{k}_1}$ and $\Theta_{\vect{k}_2}$ have the same
spectrum.  We need only consider wavevectors in a single unit cell
(called the \emph{Brillouin zone}) of the reciprocal lattice.
Since a unit cell is compact, each frequency band has a minimum and a
maximum.  It is possible that the maximum of one band is less than the
minimum of the next higher band.  In that case, the range of
frequencies between the bands is known as a \emph{band gap}, and no
frequencies within that range can propagate through the lattice.  This
is the origin of the photonic band gap phenomenon which provides a
confinement mechanism for accelerating waveguides.

\section{Overview of photonic acceleration}
\label{sec:Overview}

The physics of charged particle acceleration is highly complex,
involving many phenomena which can all interact with one another.  An
accelerator requires many components, all working in concert, in order
to properly function.  However, there are certain key processes which
form the basis of accelerator operation and determine the performance
parameters of the machine.

Central to particle acceleration is the accelerating waveguide, and
much of this dissertation concerns the design of that waveguide and
its parameters.  As discussed in Sec.~\ref{sec:LaserAcceleration},
the transverse size of the waveguide must be the order of a
wavelength, which in our case of optical acceleration is on the scale
of $\unit[1]{\micro m}$.  A relativistic particle beam propagates
through the waveguide, and the drive laser field copropagates with it
with a phase velocity equal to the speed of light in vacuum.  In order
to be accelerated in the oscillating laser field, the particle beam
must form short optical bunches which have only a small phase extent
within a laser oscillation.  For instance, for a laser wavelength of
$\unit[1]{\micro m}$, a particle bunch with a phase extent of
$\Delta\phi = \unit[30]{mrad}$ has a duration of just
$\lambda\Delta\phi/2\pi c = \unit[16]{attosec}$.  One can use a single
optical bunch at a time, or a train of bunches spaced at the optical
wavelength or an integer multiple thereof.

Unlike conventional metallic disk-loaded waveguide structures, the
photonic waveguides we consider here are transmission-mode structures.
The distinction between the two lies in how we conceptualize the
fields in the structures.  We can think of a disk-loaded waveguide as
a series of weakly-coupled resonant cavities, each of which can store
energy.  The group velocity is only a small fraction of the speed of
light, and the structure takes many periods to fill.  In a photonic
waveguide, on the other hand, the mode group velocities are
substantial fractions of the speed of light, so electromagnetic energy
will propagate down the guide along with the particles.  We wish to
use ultrafast pulses, on the order of \unit[1]{ps} or shorter in
duration, because with shorter pulse duration, materials can generally
sustain larger fields before the onset of damage
\cite{Stuart:BreakdownPRB1996}.  Because the group velocities are only
fractions of the speed of light, however, a relativistic particle beam
will outrun a picosecond pulse in a very short distance, typically
$\unit[100]{\micro m}$--\unit[1]{mm}.  This sets the length of an
individual accelerator segment; we must couple a new laser pulse into
the waveguide after each segment.  Because the coupling must be so
frequent, the couplers must be both compact and efficient; we discuss
coupling in Sec.~\ref{sec:Coupling}.  In addition, the segments and
all the necessary coupling and power distribution optics would likely
be lithographically fabricated in the same process on the same
substrate (a silicon wafer, for instance), due to the small size of
each element.

Some segments would contain structures to focus, rather than
accelerate, the particle beam in order to keep it confined.  Because
the transverse dimensions of the waveguide are on the order of a
wavelength we must have an RMS spot size $\sigma\lesssim\lambda/6$,
which is only \unit[170]{nm} for a wavelength of \unit[1]{\micro m}.
In order to achieve such small spot sizes while keeping the emittance
requirements as loose as possible, we must endeavor to achieve very
strong focusing forces while still keeping the particle beam stable.
This is described in detail in Sec.~\ref{sec:BeamDynamics}.

The parameters of the laser pulses do not themselves determine the net
acceleration of the particle beam.  This is because a particle bunch
will generate its own fields in the waveguide, which can then react
back on that bunch or subsequent bunches.  These fields are opposite
in phase to the accelerating field, so they destructively interfere
with the incident laser pulse.  This makes sense from the point of
view of conservation of energy, which indicates that as the particle
beam gains energy, the laser field must lose energy.  Because the
beam-driven wakefield has amplitude proportional to the bunch charge,
wakefields limit the practical amount of charge one can accelerate:
If $q$ is the bunch charge, then the energy gain scales as $q$, but
the energy loss due to wakefields scales as $q^2$.  Increasing the
charge too much ultimately decreases the energy gain, and thus the
efficiency, of the structure.  We discuss structure efficiency in
detail in Sec.~\ref{sec:SymmMode}.

With these considerations, here then is how we envision a laser-driven
photonic particle accelerator:  The accelerator would consist of a
sequence of monolithic pieces several centimeters in length.  The
pieces would be several millimeters in width and several hundred
microns (the thickness of a typical silicon wafer) in height; though
the active structure would be only a few wavelengths tall.  They would
be mounted on separate mechanical stages for proper alignment and
enclosed in a vacuum pipe.  This differs from the case of conventional
RF structures, in which the accelerating waveguide also serves as the
vacuum chamber.  Each piece would be fed by one, or very few,
single-mode optical fibers, and the piece would contain all the
necessary optics to distribute the power to the various accelerating
and focusing segments.

At each accelerating segment, the laser power would be coupled into
the accelerating waveguide.  A train of optically bunched particles,
no longer than the laser pulse, would enter the waveguide near the
tail of the laser pulse.  Due to the group velocity mismatch, the
particle bunches would move forward with respect to the laser pulse,
and by the end of the segment would be near the head of the pulse.  As
the bunches did so they would deplete the laser field.  To improve the
overall efficiency of the accelerator, any remaining laser power at
the end of the segment would be recycled for use with the next bunch
train, a technique described in detail in \cite{Na:Efficiency}.  A
structure similar to the distribution optics could be used in reverse
to recombine the output pulses and then couple the power out to an
optical fiber.  That power would then be fed, along with the addition
of some incident power from a laser source, back into the input fiber.
Because the repetition rate of both the laser pulses and particle
bunch trains would be on the order of \unit[1]{GHz} in order to
achieve high average beam current, the delay line for the recycling
would be of manageable length.

While the above discussion paints a picture of what a future photonic
accelerator might look like, this dissertation largely concentrates on
the design of individual accelerator segments.  Many of the design
considerations for scaling an accelerator to its full length have yet
to be considered.

\section{Accelerator parameters}
\label{sec:Parameters}

There are several qualities which characterize the performance of a
particle accelerator that are determined by parameters of the
accelerating mode in a structure.  One of the most important, as
discussed in the introduction, is the accelerating gradient.  The
sustainable gradient in an accelerator structure is limited by
breakdown of the material from which the structure is constructed.
While we discuss the phenomenon of optical damage in detail in
Sec.~\ref{sec:damage}, here we relate the sustainable gradient of a
structure to the breakdown threshold of the material.  To this end we
define the \emph{damage impedance} of an accelerating mode by
\[ Z_d = \frac{E\tsub{acc}^2}{2u\tsub{max}c}, \]
where $E\tsub{acc}$ is the accelerating gradient on axis and
$u\tsub{max}$ is the maximum electromagnetic energy density anywhere
in the material.  Then, if $u\tsub{th}$ is the energy density in the
material at damage threshold, we can write the sustainable
accelerating gradient as
\begin{equation}
E\tsub{acc} = \sqrt{2Z_du\tsub{th}c}.
\label{eq:MaxGradient}
\end{equation}
While fluence is the laser pulse parameter most often quoted in the
optical damage literature, we choose to relate the accelerating
gradient to energy density for several reasons.  First, with this
description, Eq.~(\ref{eq:MaxGradient}) describes the accelerating
gradient in terms of two logically distinct factors.  The damage
impedance as defined above depends only on the mode field pattern and
does not involve the laser pulse width or wavelength, while
$u\tsub{th}$ depends only on the material, wavelength, and pulse
width, and is independent of structure geometry. Second, the energy
density reflects the likely importance of multiphoton ionization
in the damage process, as discussed in Sec.~\ref{sec:damage}.

Efficiency is another important performance parameter of an
accelerator.  With designs for future high-energy colliders specifying
beam power in excess of \unit[10]{MW}, high efficiency is a critical
requirement of an accelerator structure.  Laser-driven accelerator
efficiency was examined in \cite{Siemann:Efficiency}, and the
idea of incorporating the accelerator structure into a laser cavity
was explored.  That analysis was continued in
\cite{Na:Efficiency}, where the case of an accelerator
structure embedded in a passive, externally pumped optical resonator
was described.

The efficiency of the structure depends upon several parameters.
First, the characteristic impedance of the mode, which
describes the relationship between input laser power and accelerating
gradient \cite{Pierce:TWT}, is
\[Z_c = E\tsub{acc}^2\lambda^2/P, \] where $P$ is the laser power.
Second, the group velocity $v_g = \beta_gc$ of the mode affects the
efficiency, as modes with group velocity closer to the speed of light
couple better to a relativistic particle beam.  This is quantified by
the loss factor, which is given by \cite{Bane:GuideImpedance}
\[ k = \frac{1}{4}\frac{c\beta_g}{1 - \beta_g}\frac{Z_c}{\lambda^2};
\]
an optical bunch with charge $q$ will radiate fields in the
accelerating mode with decelerating gradient equal to $kq$.  Finally,
the \v{C}erenkov impedance $Z_H$ parametrizes the energy loss due to
wide-band \v{C}erenkov radiation.  Following
\cite{Siemann:Efficiency}, we can estimate $Z_H$ from its value
for a bulk dielectric with a circular hole of radius $R$, which is
\[ Z_H = \frac{Z_0}{2\pi(R/\lambda)^2}, \]
where $Z_0 = \unit[376.73]{\ohm}$ is the impedance of free space.  For
arbitrary accelerating structures, we can let $R$ be a length
characterizing the radius of the vacuum waveguide, even if the guide
is not circular.

\chapter{Two-dimensional accelerator structures}
\label{ch:2D}
In this chapter we consider geometries which are two-dimensional: we
take them to be infinite in the vertical ($y$) direction, transverse
to the electron beam, which we take to propagate in the $z$-direction
(see Fig.~\ref{fig:lattice} or \ref{fig:waveguide} for a description
of the coordinate system).  As such, these structures would not be
suitable for actual acceleration unless a method of vertical
confinement were found, and then their properties would be altered
from those computed here.  Rather, they provide a means to build
intuition for photonic crystal structures.  By considering
two-dimensional structures our computation time is greatly reduced,
allowing us to quickly explore multiple sets of geometric parameters.
The computational technique is discussed in
Sec.~\ref{sec:IterativeEigensolver}.

\section{Lattice geometry}

\begin{figure}
\begin{center}
\includegraphics{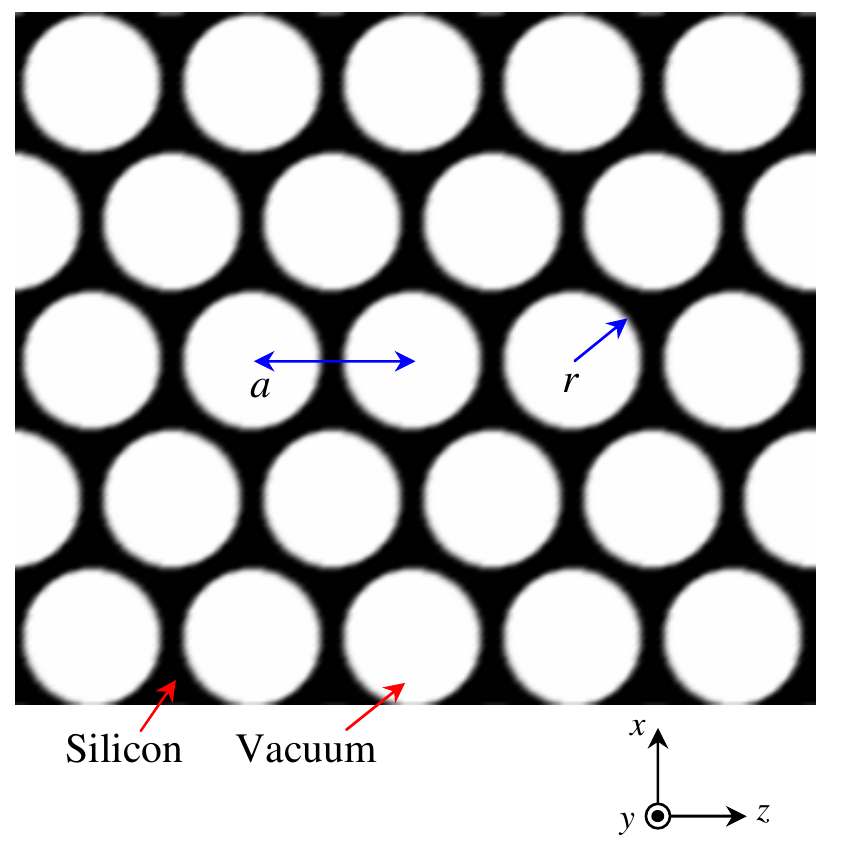}
\caption{The geometry of an optimized 2D photonic crystal lattice
consisting of vacuum holes in silicon.  Here $r = 0.427a$.  The
coordinate system used throughout this chapter is shown, and we take the
particle beam to propagate in the $z$-direction.}
\label{fig:lattice}
\end{center}
\end{figure}

Our underlying photonic crystal lattice is a triangular array of
vacuum holes in a silicon substrate, shown in Fig.~\ref{fig:lattice}.
We consider laser acceleration using a wavelength of \unit[1550]{nm},
in the telecommunications band where many promising sources exist
\cite{IMRAWeb}.  At this wavelength silicon has a normalized
permittivity of $\epsilon_r = \epsilon/\epsilon_0 = 12.1$
\cite{Edwards:HOCSilicon}.  As shown in the figure, the holes have
radius $r$, and the lattice constant is $a$.

Because of the vertical symmetry of the structure we can classify each
mode as either TE or TM, where the ``transverse'' designation is with
respect to the $y$-direction, not the $z$-direction in which the beam
propagates.  Thus the accelerating modes are TE since these have the
$\vect{E}$-field in the plane of the structure.

\begin{figure}
\begin{center}
\includegraphics{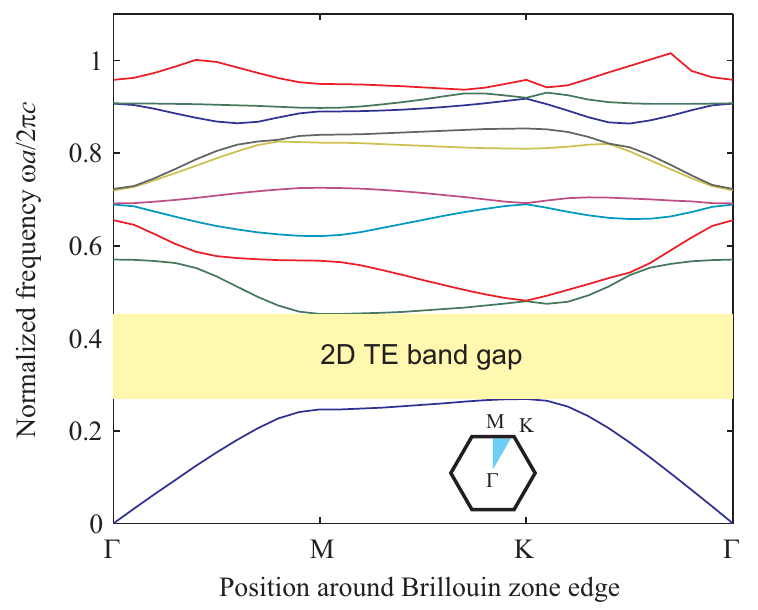}
\caption{The TE bandstructure of the photonic crystal lattice shown in
Fig~\ref{fig:lattice}, with $r = 0.427a$.  The symbols $\Gamma$,
$\mathrm{M}$, and $\mathrm{K}$ refer to the points at the corners of
the irreducible Brillouin zone, as shown in the inset.}
\label{fig:bandstructure}
\end{center}
\end{figure}

Like electronic states in a solid, electromagnetic modes in the
lattice fall in discrete bands, as described in
Ch.~\ref{ch:Introduction}.  The TE bandstructure of the lattice is
shown in Fig.~\ref{fig:bandstructure}.
As described in \cite{Joannopoulos:Molding}, the irreducible Brillouin
zone of this lattice forms a triangle in reciprocal space.  This
triangle has corners at $\vect{k} = 0$ (the $\Gamma$ point), $\vect{k}
= (2\pi/a)(\xhat/2)$ (the $\mathrm{M}$ point) and $\vect{k} =
(2\pi/a)(\xhat/2 + \zhat/2\sqrt{3})$ (the $\mathrm{K}$ point).  The
figure shows the frequencies for values of $\vect{k}$ around the edge
of the irreducible Brillouin zone, for $\Gamma$ to $\mathrm{M}$ to
$\mathrm{K}$ and back to $\Gamma$.  We see from this figure that the
lattice exhibits a band gap---in this case, a range of frequencies in
which no TE mode exists.  In fact, this lattice was chosen
specifically to exhibit such a gap based on energy considerations;
also, a high-index material was chosen in order to obtain a wide band
gap \cite{Joannopoulos:Molding}. Then, $r/a$ was optimized to give the
widest band gap, which occurs at $r/a = 0.427$.  With light at
frequencies in the band gap forbidden to propagate in the lattice,
modes at such frequencies can only exist in a defect, should one be
provided.  Thus we are able to create waveguides in all-dielectric
photonic crystal structures, and customize them to support
accelerating modes.  In addition, such a waveguide only confines light
at frequencies in the bandgap, so higher-order guided wakefield modes
are suppressed, potentially eliminating a major source of beam
break-up instability \cite{Li:PBGWakefields,Kroll:PBG90GHz}.

\section{Accelerating waveguide geometry}

\begin{figure}
\begin{center}
\resizebox{!}{3.5in}{\includegraphics{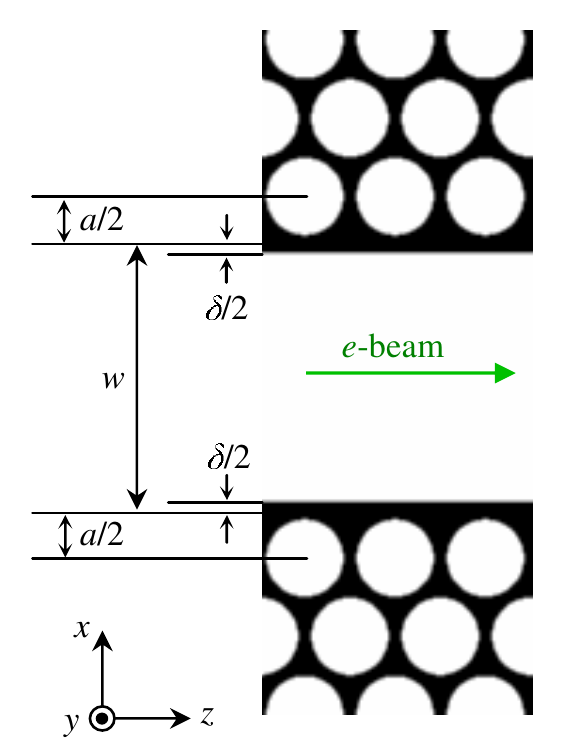}}
\caption{The waveguide design, illustrating the geometric
  parameters.  Here $r = 0.427a$, $w = 3.0a$, and $\delta = 0.25a$.}
\label{fig:waveguide}
\end{center}
\end{figure}

Our example accelerator structure consists of the lattice shown in
Fig.~\ref{fig:lattice} with a vacuum waveguide of width $w$
introduced.  The waveguide geometry is shown in
Fig.~\ref{fig:waveguide}.
Specifically, $w$ is defined so that the distance between the centers
of the holes adjacent to the waveguide is $w+a$.  In addition, we can
line the edge of the waveguide with dielectric, a concept similar to
the dielectric waveguide accelerator (DWA) described in
\cite{Keinigs:CWA}.  We define $\delta$ as the total width of
silicon in the guide, so there is $\delta/2$ of dielectric ``padding''
on each edge.

\begin{figure}
\begin{center}
\includegraphics{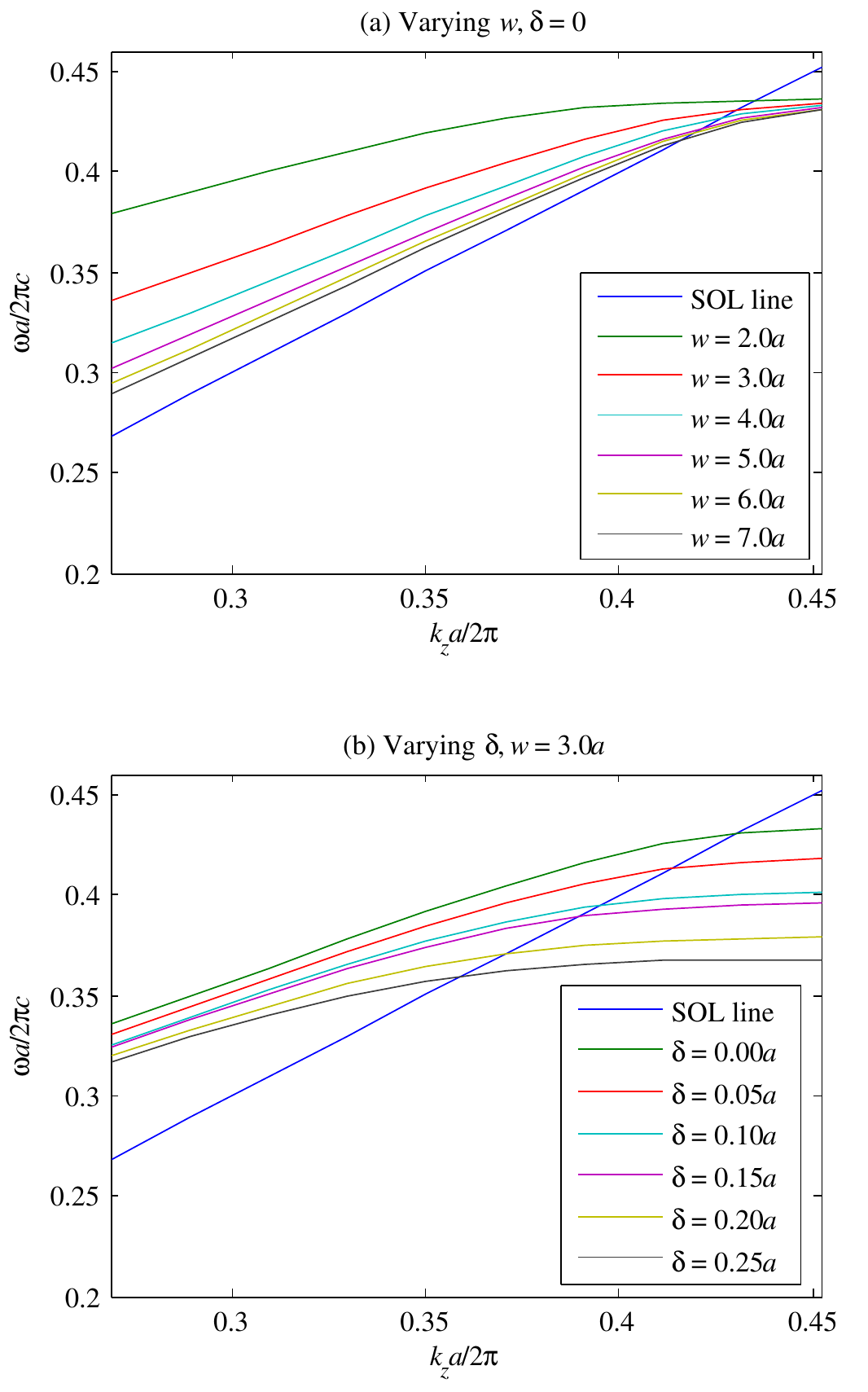}
\caption{Dispersion curves for a set of waveguide geometries, plotted
  with the speed-of-light (SOL) line shown.  The range of wavenumbers
  consists of all $k_z$ with $ck_z$ a frequency in the bandgap.  (a)
  We fix $\delta = 0$ and vary $w$; (b) we fix $w = 3.0a$ and vary
  $\delta$.}
\label{fig:Dispersion2D}
\end{center}
\end{figure}

\begin{figure}
\begin{center}
\includegraphics{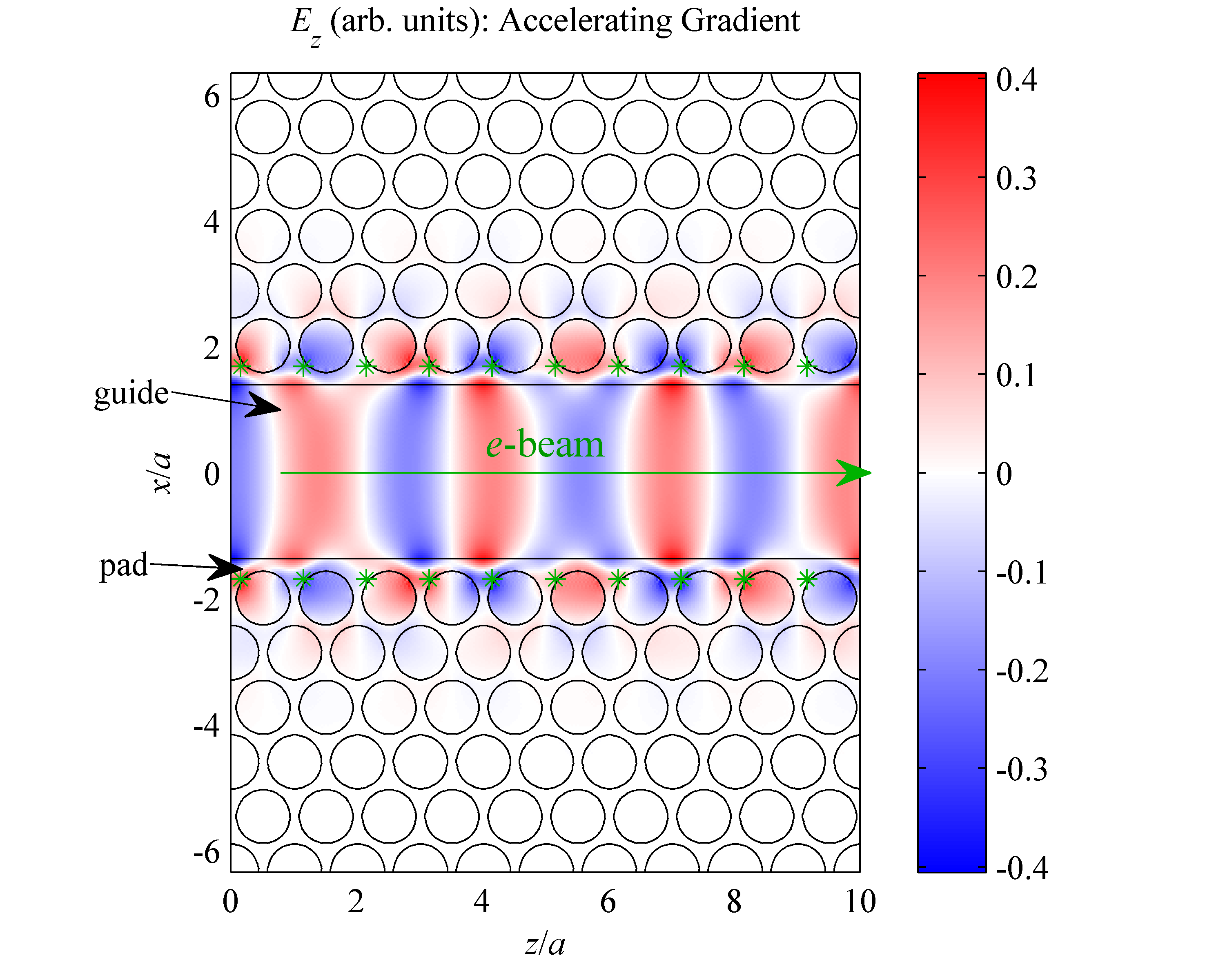}
\caption{An accelerator structure geometry with a speed-of-light
  waveguide mode.  The circles delineate the vacuum holes, and the
  blue/red coloring denotes the accelerating $E_z$ component.  Here $w
  = 3.0a$ and $\delta = 0.25a$, and for this mode $\lambda = 2.78a$.
  The asterisks show the damage sites, where the energy density within
  the dielectric is maximized.}
\label{fig:Mode2D}
\end{center}
\end{figure}

If we normalize the length scales of our structures to the lattice
constant $a$, we are left with three parameters: the guide width $w$,
the pad width $\delta$, and the wavelength.  Fixing the lattice
constant makes the wavelength, or equivalently the longitudinal
wavenumber $k_z$ or frequency $\omega$, a free parameter.  For a
general selection of $w$ and $\delta$, there will be a $k_z$ for which
the waveguide mode is synchronous, i.e. $\omega=ck_z$.  (Certainly if
there is too much dielectric material within the waveguide, the phase
velocity will be limited below $c$; consider for instance the case of
a guide entirely filled with silicon.)  This is demonstrated in
Fig.~\ref{fig:Dispersion2D}(a), which shows the dispersion diagrams
for a set of waveguide widths, with $\delta=0$.
Since each curve crosses the speed-of-light line, each geometry
supports a synchronous mode.  Adding dielectric material to the edges
of the guide brings the speed-of-light frequency down into the
interior of the bandgap, as shown in Fig.~\ref{fig:Dispersion2D}(b).
We show such a mode with its longitudinal electric field in
Fig.~\ref{fig:Mode2D}, for $w = 3.0a$ and $\delta = 0.25a$.
Observe that the mode is mostly confined to within $a$ of the edge of
the guide; the wide band gap makes such good confinement possible.

\section{Accelerating mode parameters}

\begin{figure}
\begin{center}
\includegraphics{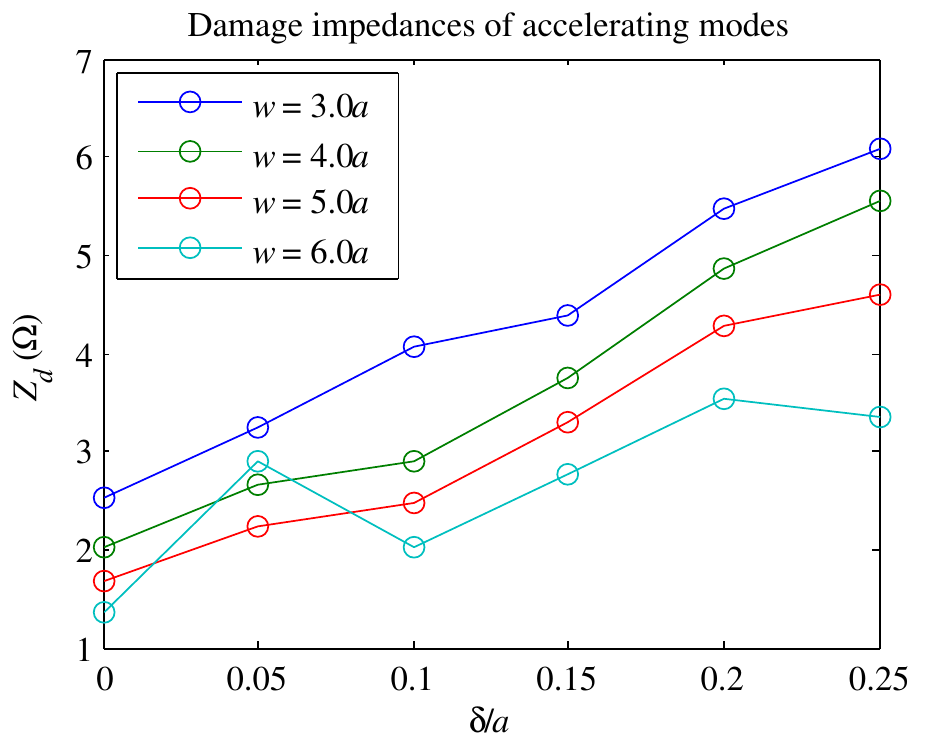}
\caption{Damage impedances of the various structure geometries.}
\label{fig:DamageImpedance2D}
\end{center}
\end{figure}

\begin{figure}
\begin{center}
\includegraphics{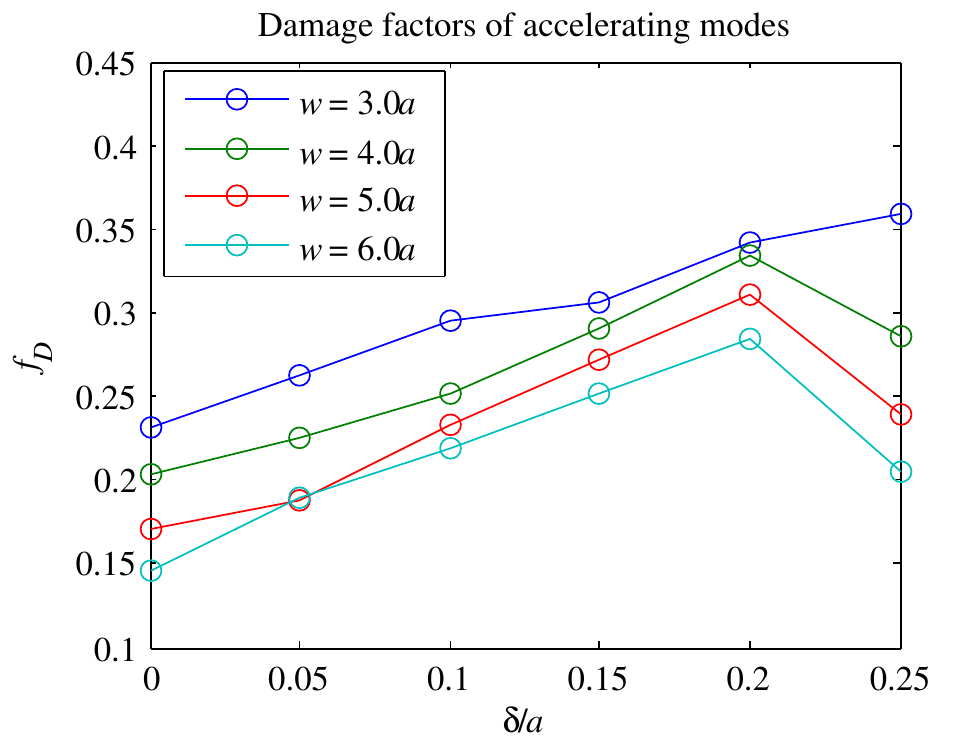}
\caption{Damage factors of various structure geometries.}
\label{fig:DamageFactor2D}
\end{center}
\end{figure}

We computed accelerating modes in our two-dimensional waveguides for a
variety of geometries, and we can then characterize the performance of
an these modes according to the parameters described in
Sec.~\ref{sec:Parameters}.  The sustainable gradient is determined by
the damage threshold of the material and the damage impedance $Z_d$ of
the mode.  We plot the damage impedances in
Fig.~\ref{fig:DamageImpedance2D}.
When we originally performed the computations on these two-dimensional
structures, we characterized their sustainable gradients in
terms of different parameter.  Instead of the damage impedance, we
used the \emph{damage factor}, defined as
\[ f_D =
\frac{E\tsub{acc}}{\abs{\vect{E}}\tsup{material}\tsub{max}},
\]
where $\abs{\vect{E}}\tsup{material}\tsub{max}$ is the maximum
electric field magnitude anywhere in the dielectric material.  We plot
the damage factors in Fig.~\ref{fig:DamageFactor2D}.
For both parameters we see that the sustainable gradient increases as
the guide is narrowed.  This is the behavior we would expect from the
discussion in Sec.~\ref{sec:LaserAcceleration}, in which we saw that
the on-axis accelerating field increases relative to the transverse
components as the mode size decreases.  The question of which
parameter yields the most accurate figure for sustainable gradient is
still an open question, and we discuss this issue further in
Sec.~\ref{sec:damage}.  However, for the rest of this dissertation, we
report the damage impedance exclusively for our computed modes.

\begin{figure}
\begin{center}
\includegraphics{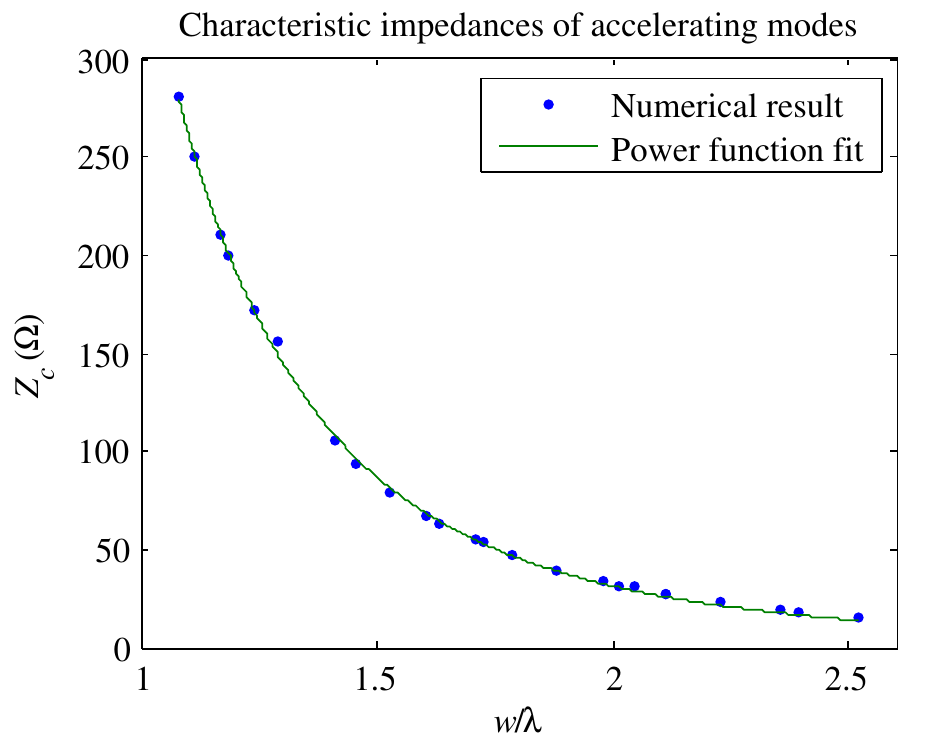}
\caption{Characteristic impedances of 2D photonic crystal waveguide
structures.  By normalizing the impedance to that of a structure one
wavelength high, we obtain a value of $Z_c$ in \ohm.  The
data shown here include multiple values of $\delta$ ranging from $0$
to $0.25a$.}
\label{fig:Impedance2D}
\end{center}
\end{figure}

\begin{figure}
\begin{center}
\includegraphics{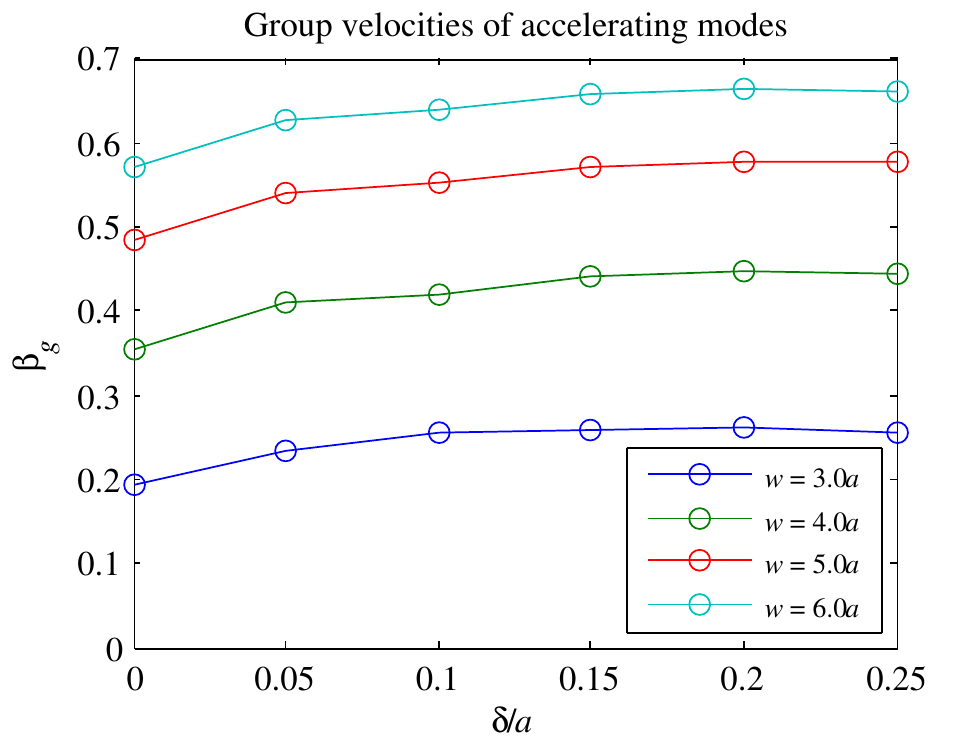}
\caption{Group velocities of accelerating modes for a variety of $w$
  and $\delta$ values.}
\label{fig:GroupVelocity2D}
\end{center}
\end{figure}

As discussed in Sec.~\ref{sec:Parameters}, the efficiency of the
accelerator is determined by both the characteristic impedance and the
group velocity.  We plot these in Figs.~\ref{fig:Impedance2D} and
\ref{fig:GroupVelocity2D} respectively.
Since our 2D structures only confine modes in one transverse
dimension, for these structures we normalize the impedance to that of
a structure one wavelength high, so
\[ Z_c = \frac{E\tsub{acc}^2\lambda}{P_h}, \]
where $P_h$ is the laser power per unit height.  We can derive $P_h$
from the computation by summing the $z$-component of the Poynting
vector over one transverse slice of the mode.  Plotting the
characteristic impedance versus $w/\lambda$, the width of the guide in
wavelengths, we observe a power law relation $Z_c\propto
(w/\lambda)^{-3.55}$.  Thus we see again that as the guide narrows,
the longitudinal electric field increases relative to other
components.  On the other hand, we see from
Fig.~\ref{fig:GroupVelocity2D} that the group velocity decreases for
narrower guides, which reduces the loss factor and thus the
efficiency.

We explore these issues in greater detail in the next chapter, where
we present an accelerator structure that confines a mode in both
transverse directions.

\chapter{The woodpile structure}
\label{ch:Woodpile}
In the previous chapter we presented a class of two-dimensional
structures.  As was pointed out in that chapter, such structures are
impractical over macroscopic distances because they only confine the
accelerating mode in one transverse dimension.  This naturally raises
the question of whether the technique of photonic crystal confinement
might be applied toward a three-dimensional structure.  As we shall
see in this chapter, not only is this possible, but the conceptual
development of the structure is quite similar:  (1) Find a photonic
crystal lattice with a bandgap; (2) introduce a defect; and (3) find
the point along the dispersion curve of the accelerating mode with
speed-of-light phase velocity, and (4) adjust the parameters of the
waveguide if necessary.

Here we present the design and simulation of such a three-dimensional
planar structure which fully confines the accelerating mode.  We
further examine the structure by exploring symmetry considerations,
mode coupling, and particle beam dynamics.  We begin, as in the
two-dimensional case, with the underlying PBG lattice.

\section{The woodpile lattice}
\label{sec:WoodpileLattice}

\begin{figure}
\begin{center}
\includegraphics{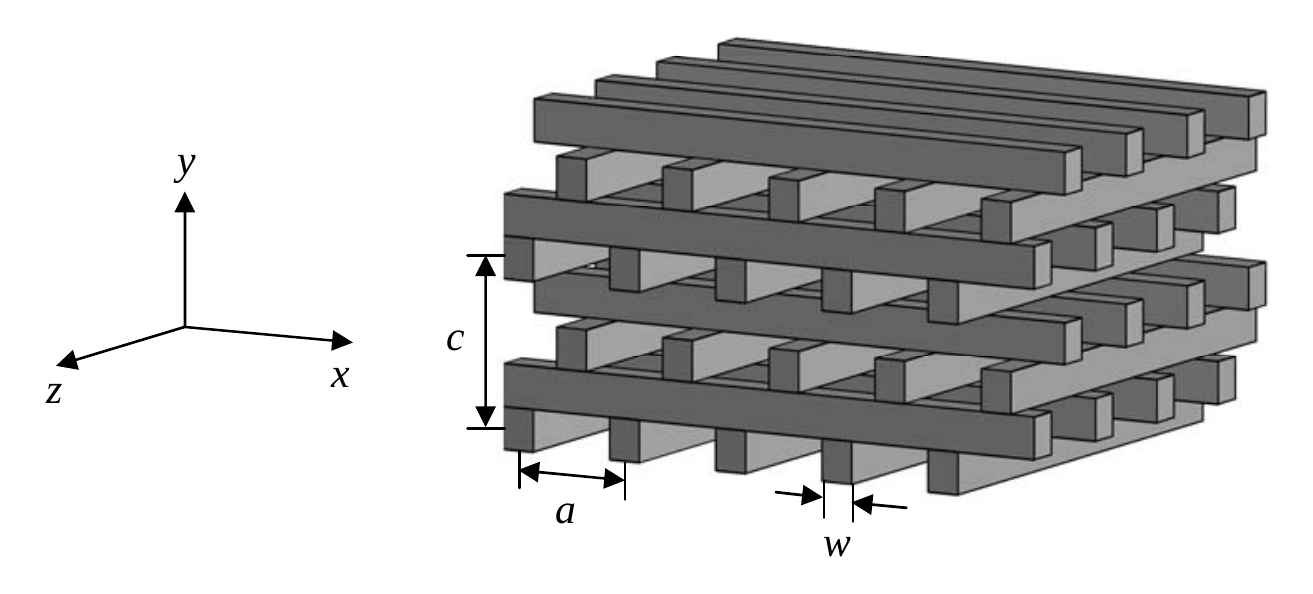}
\caption{A diagram of 8 layers (2 vertical periods) of the woodpile lattice.}
\label{fig:woodpileLattice}
\end{center}
\end{figure}

The so-called ``woodpile'' geometry is a well-established
three-dimensional photonic crystal lattice designed to provide
a complete photonic bandgap in a structure with a straightforward
fabrication process \cite{Ho:Woodpile}.  The lattice consists of
layers of dielectric rods in vacuum, with the rods in each layer rotated
90\degree\ relative to the layer below and offset half a lattice
period from the layer two below, as shown in
Fig.~\ref{fig:woodpileLattice}.
Figure~\ref{fig:woodpileLattice} also shows the coordinate system and
geometric parameters of the lattice, with $a$ being the horizontal
period (in the $z$ and $x$ directions) and $c$ being the vertical
period (in the $y$ direction).

At first glance, this structure appears to have a simple tetragonal
lattice structure, with $(a\zhat,\linebreak[0] a\xhat,\linebreak[0]
c\yhat)$ as the lattice basis.  However, the lattice in fact has a
more complex crystal structure that is more amenable to the presence
of a photonic bandgap.  Early work on photonic crystals suggested that
crystals with a Brillouin zone as close as possible to spherical in
shape are most likely to exhibit a complete photonic bandgap, and in
fact one of the first experimentally observed PBG's was found in a
face-centered cubic (FCC) lattice, which has a Brillouin zone closer
to spherical in shape than any other common crystal
\cite{Yablonovitch:BandStructure}.  The woodpile lattice in fact has
FCC crystal structure, as we now show.

In addition to symmetry under translation by the lattice basis vectors
given above, the woodpile lattice is also symmetric under translation
by half a period in each direction.  While this symmetry immediately
implies a body-centered tetragonal structure, the lattice also has
face-centered tetragonal structure under a different orthogonal
basis.  Consider the basis $(a\unitvec{Z}, a\unitvec{X},
c\unitvec{Y})$, where
\[
\unitvec{Z} = \zhat - \xhat,\quad
\unitvec{X} = \zhat + \xhat,\quad
\unitvec{Y} = \yhat.
\]
This is an orthogonal basis, and the woodpile lattice is certainly
symmetric under translations by the lattice vectors.  The lattice is
also symmetric under translation by the vectors
\begin{align*}
\frac{1}{2}(a\unitvec{Z} + a\unitvec{X}) &= a\zhat, \\
\frac{1}{2}(a\unitvec{X} + c\unitvec{Y})
&= \frac{a}{2}\zhat + \frac{a}{2}\xhat + \frac{c}{2}\yhat, \\
\frac{1}{2}(c\unitvec{Y} + a\unitvec{Z})
&= \frac{a}{2}\zhat - \frac{a}{2}\xhat + \frac{c}{2}\yhat.
\end{align*}
Thus the woodpile lattice has face-centered tetragonal structure.  In
particular, if $c/a = \sqrt{2}$, then the orthogonal basis given above
is cubic, so the lattice has FCC structure.  Thus we choose $c =
a\sqrt{2}$ for our geometry.

\begin{figure}
\begin{center}
\includegraphics{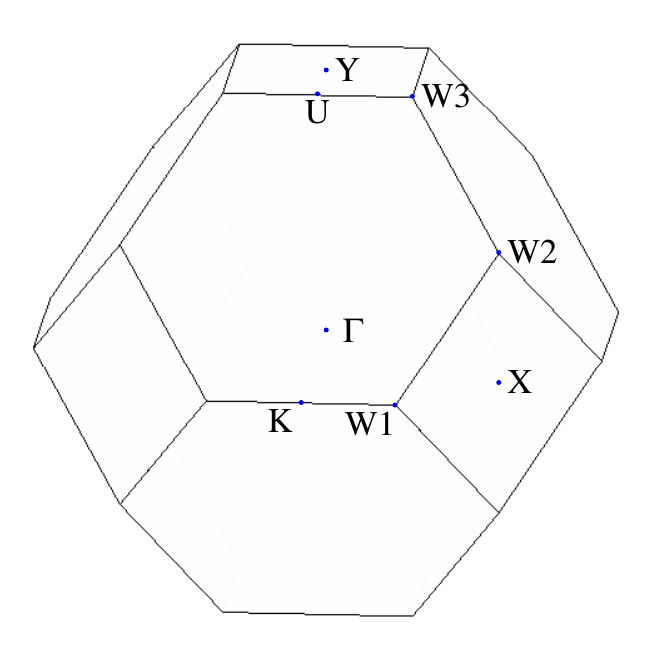}
\caption{The Brillouin zone of the FCC lattice, with symmetry points
  of the irreducible Brillouin zone for the woodpile lattice shown.}
\label{fig:WoodpileBZ}
\end{center}
\end{figure}

The Brillouin zone of the FCC lattice is shown in
Fig.~\ref{fig:WoodpileBZ}.
The woodpile lattice is symmetric under rotations by 90\degree\ in the
$zx$-plane, reflections in $x$, and reflections in $y$ followed by
half a period of horizontal translation.  Because of these symmetries,
the irreducible Brillouin zone of the woodpile lattice is reduced
significantly from the FCC Brillouin zone.
Fig.~\ref{fig:WoodpileBZ} indicates the symmetry points which bound
the irreducible Brillouin zone.  For reference, the indicated symmetry
points along the Cartesian axes are given by
\[
\mathrm{X} = \frac{\pi}{2a}\unitvec{X},\quad
\mathrm{Y} = \frac{\pi}{2a}(\sqrt{2}\unitvec{Y}).
\]
The other points shown are then given by
\begin{align*}
\Gamma &= 0, \\
\mathrm{K} &= \frac{\pi}{2a}
\cdot\frac{3}{4}(\unitvec{Z} + \unitvec{X}), \\
\mathrm{W1} &= \frac{\pi}{2a}
\paren{\frac{1}{2}\unitvec{Z} + \unitvec{X}}, \\
\mathrm{W2} &= \frac{\pi}{2a}
\paren{\unitvec{X} + \frac{\sqrt{2}}{2}\unitvec{Y}}, \\
\mathrm{W3} &= \frac{\pi}{2a}
\paren{\sqrt{2}\unitvec{Y} + \frac{1}{2}\unitvec{X}}, \\
\mathrm{U} &= \frac{\pi}{2a}
\paren{\sqrt{2}\unitvec{Y} + \frac{1}{2\sqrt{2}}\unitvec{X} +
  \frac{1}{2\sqrt{2}}\unitvec{Z}}.
\end{align*}
This notation is taken from \cite{Ibach:SolidState}, with several
points added because the lattice is not symmetric under all
interchanges of its basis vectors.

\begin{figure}
\begin{center}
\includegraphics{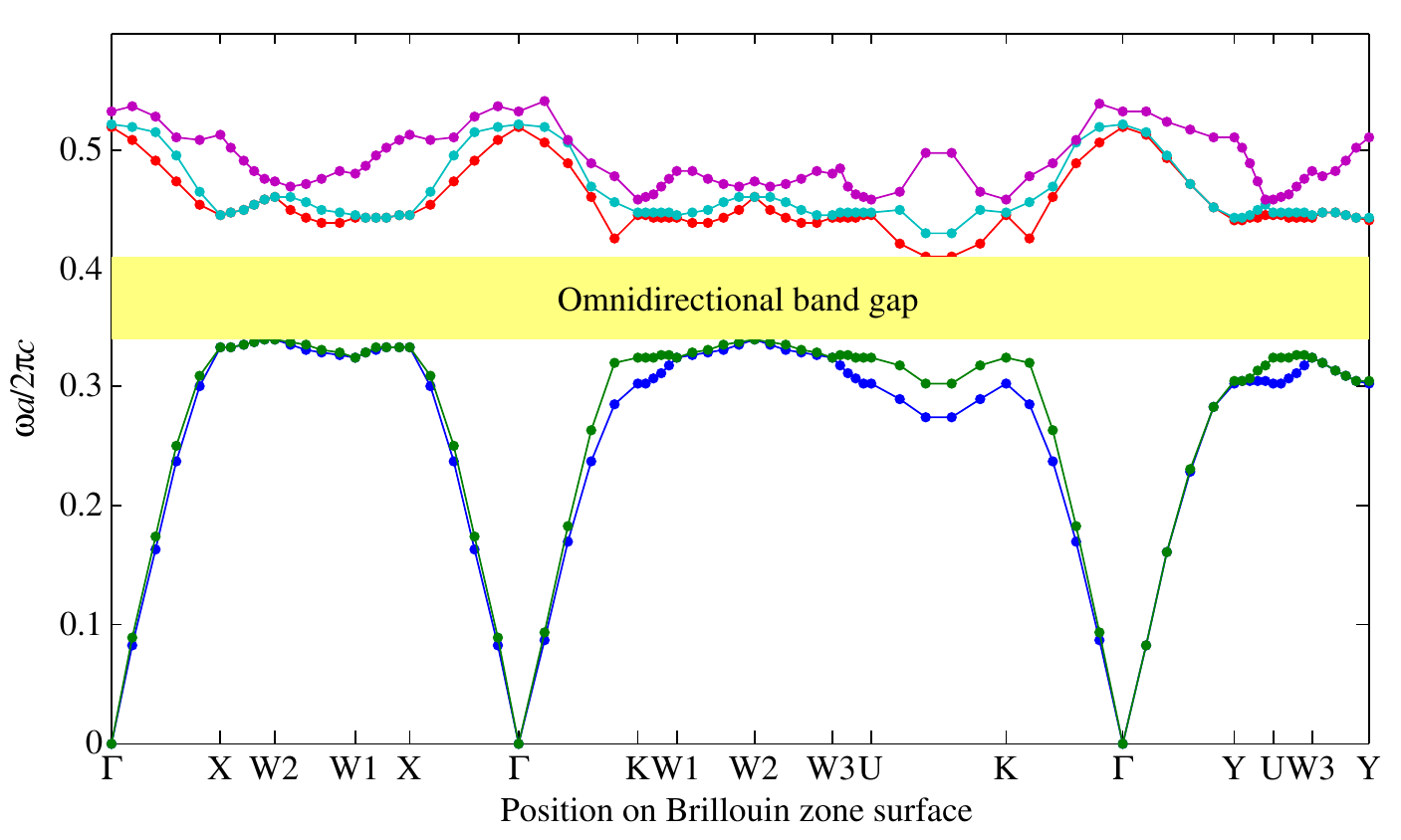}
\caption{The bandstructure of the woodpile lattice.}
\label{fig:WoodpileBandstructure}
\end{center}
\end{figure}

Based on the results in \cite{Ho:Woodpile}, we take $w/a = 0.28$ for
our geometry.  As in the previous chapter, we take the relative
permittivity of the dielectric material to be $\epsilon_r = 12.1$.
With these parameters, the bandstructure of this woodpile lattice is
shown in Fig.~\ref{fig:WoodpileBandstructure}.
We see from this figure that the lattice exhibits an omnidirectional
bandgap---a range of frequencies in which no mode, of any wavevector
or polarization, exists.  The ratio of the width of the gap to its
center frequency is 18.7\%.  The method of computation is described in
Sec.~\ref{sec:IterativeEigensolver}.

\section{Mode in asymmetric lattice}
\label{sec:AsymMode}

\begin{figure}
\begin{center}
\resizebox{.8\textwidth}{!}{\includegraphics{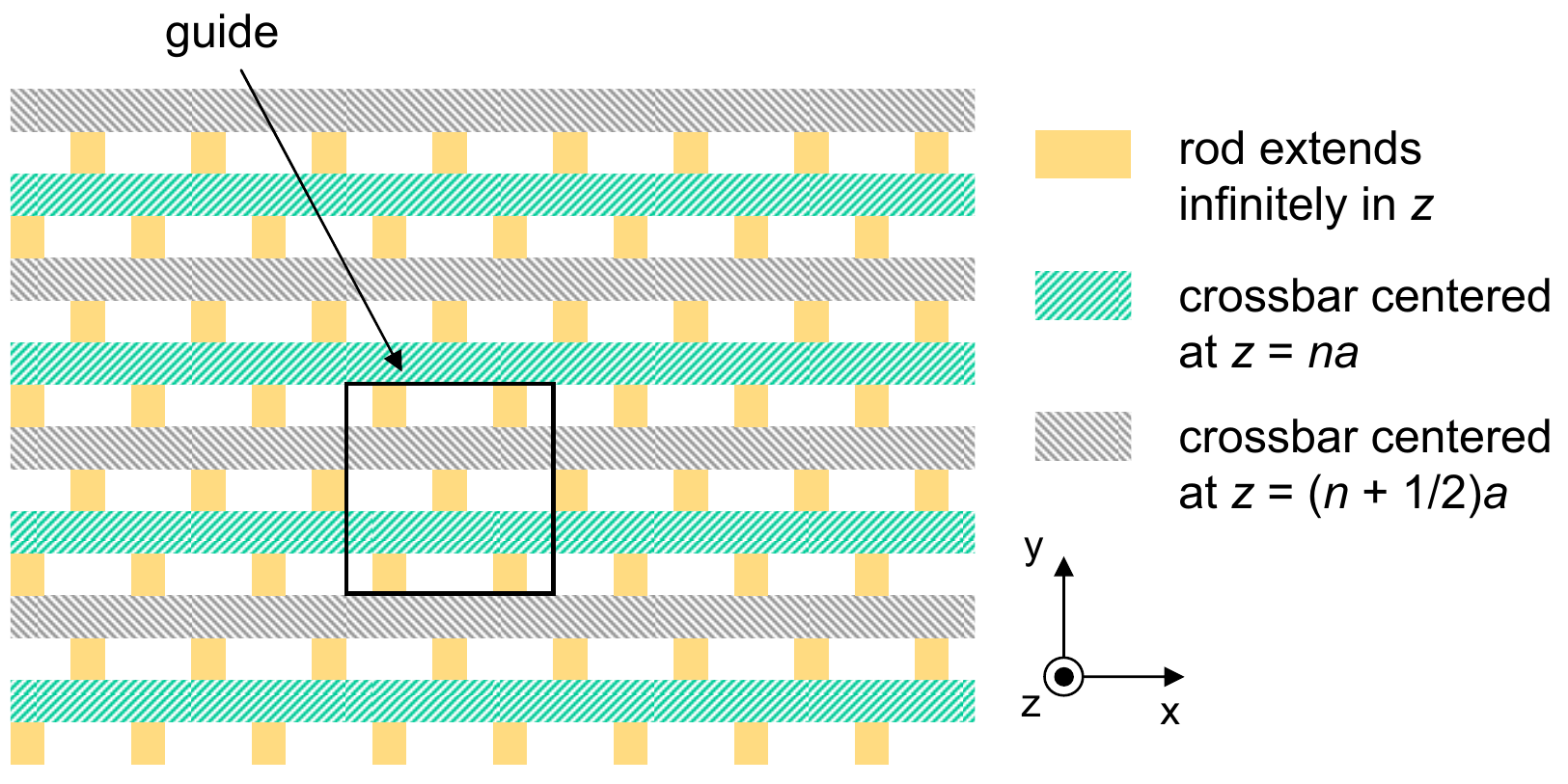}}
\caption{The geometry of a waveguide in a vertically asymmetric
  lattice.  The waveguide is formed by removing all dielectric
  material in the box shown, for all $z$.}
\label{fig:asymmetricguide}
\end{center}
\end{figure}

\begin{figure}
\begin{center}
\includegraphics{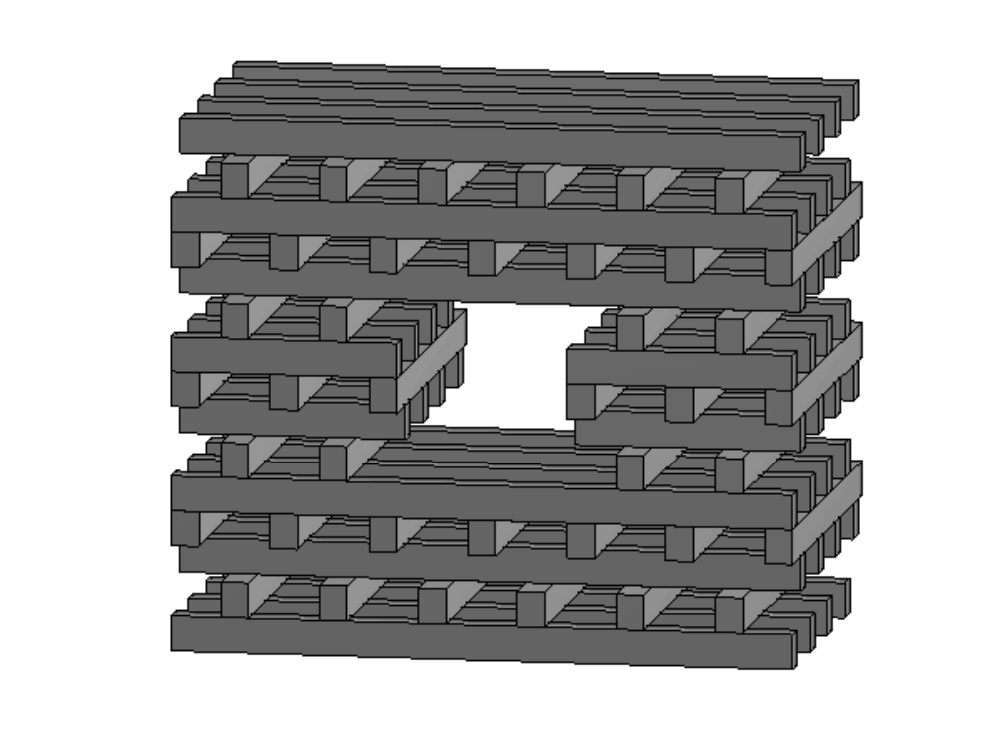}
\caption{An asymmetric waveguide.}
\label{fig:asymmetricmodel}
\end{center}
\end{figure}

Having established the presence of a photonic bandgap in the woodpile
lattice, we are now ready to introduce a defect to form a waveguide.
We do so by removing all dielectric material in a region which is
rectangular in the transverse $x$ and $y$ dimensions, and extends
infinitely in the $e$-beam propagation direction $z$, as shown
schematically in Fig.~\ref{fig:asymmetricguide} and visually in
Fig.~\ref{fig:asymmetricmodel}.

\begin{figure}
\begin{center}
\resizebox{!}{3in}{\includegraphics{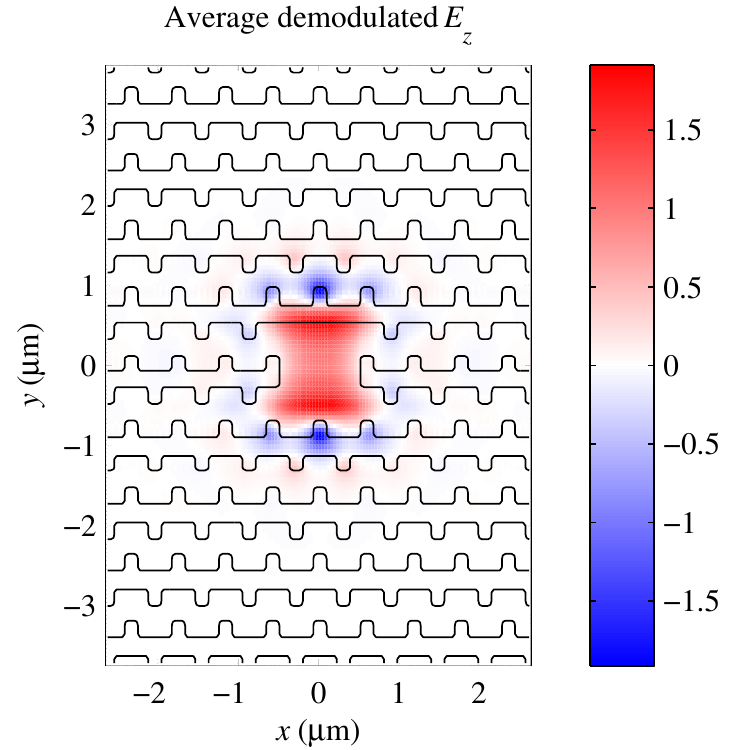}}
\resizebox{!}{3in}{\includegraphics{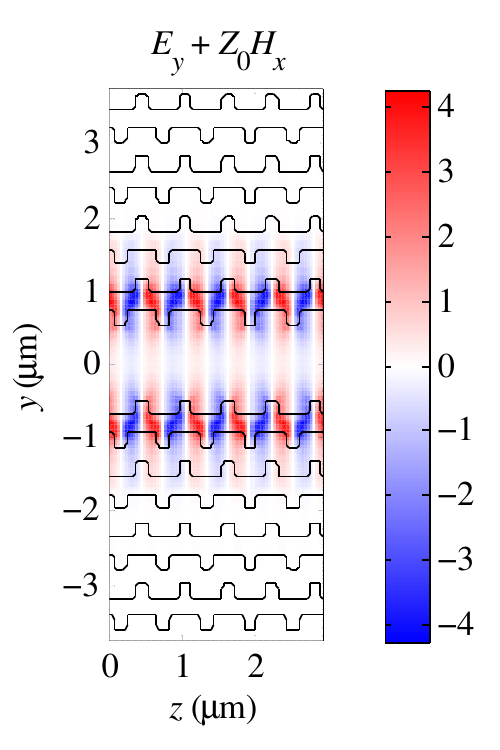}}
\caption{Left: The average accelerating field seen by a speed-of-light
  particle.  The fields were first demodulated by multiplying by
  $e^{ik_zz}$ in order to remove the period-to-period phase shift,
  then averaged over a period. The structure contours are shown for a
  transverse slice at $z = 0$.  Right: The vertical deflecting fields
  seen by a speed-of-light particle, shown with structure contours for
  a longitudinal slice at $x = 0$.  In both plots the fields are
  normalized to the accelerating field on axis.}
\label{fig:asymmetricmode}
\end{center}
\end{figure}

This waveguide supports an accelerating mode, that is, a mode with
speed-of-light phase velocity and nonzero longitudinal field $E_z$ on
axis.  For this mode, $a/\lambda = 0.379$, so setting $\lambda =
\unit[1550]{nm}$ determines $a = \unit[588]{nm}$. The accelerating
field is shown in the left plot of
Fig.~\ref{fig:asymmetricmode}.
Examining the fields, we find that there is a vertical deflecting
field, shown in the right plot, with magnitude comparable to the
accelerating field.  This is due to the fact that the structure is not
vertically symmetric.  In fact, this is an inherent property of the
photonic crystal: The lattice geometry itself is not symmetric under
reflection across any plane perpendicular to the $y$ axis.

However, the structure is symmetric under the transformation of
reflection in $y$ followed by translation of half a period in $z$.
Because of this symmetry, these vertical deflection fields average to
zero over a lattice period, so particles will see no net deflection.
Rather, they will see an undulator field with a period equal to the
lattice period of \unit[588]{nm}.  The transverse fields on axis will
cause energy loss due to synchrotron radiation.  Since synchrotron
loss scales as $E^2$ with beam energy $E$, the loss will be
significant at sufficiently high energies.

In examining the synchrotron loss in this structure, let us assume an
accelerating gradient on axis of $E\tsub{acc} = \unit[1]{GeV/m}$.
While it is unlikely based on the damage threshold studies presented
in Section~\ref{sec:damage} that such a gradient is achievable in
silicon at \unit[1550]{nm}, other materials and wavelengths may allow
such high fields.  Therefore this gradient value should be taken only
as an example used to examine synchrotron loss, not as the breakdown
limit of the structure.

In a high-energy collider, an accelerating field of \unit[1]{GeV/m}
will result in average incoherent synchrotron radiation loss of
$\ip{dE/dz} = (\unit[2.0\e{-4}]{eV/m})\gamma^2 = \unit[200]{MeV/m}$
for \unit[0.5]{TeV} electrons.  Therefore this structure would not be
appropriate for use in a high-energy linear collider.  To overcome
this problem, we must make the structure symmetric in both transverse
dimensions to suppress dipole fields.

Although it might be tempting to consider this structure as a
micro-period undulator for a radiation source, the extremely short
period, together with the field strength limitation, leads to such a
low undulator parameter that the photon yield would be insignificant.
For instance, even if the fields were strong enough that peak vertical
force were $F_0 = \unit[1]{GeV/m}$, the undulator parameter
\cite{Jackson:EM} would then be
\[ K = \frac{F_0 a}{2\pi mc^2} = 1.83\e{-4}. \]
The undulator parameter is a general measure of transverse field
strength.  An undulator causes a peak angular deviation of $K/\gamma$,
where $\gamma$ is the Lorentz factor.  In this case, $K$ is several
orders of magnitude smaller than in typical radiation devices, where
$K\sim 1$.  The number of photons per electron per period is $N_\gamma
= 2\pi\alpha K^2/3 = 5.13\e{-10}$, so only with large amounts of charge
propagating for meters within the structure would result in
significant photon yield.

\section{Mode in symmetric structure}
\label{sec:SymmMode}

\begin{figure}
\begin{center}
\resizebox{.8\textwidth}{!}{\includegraphics{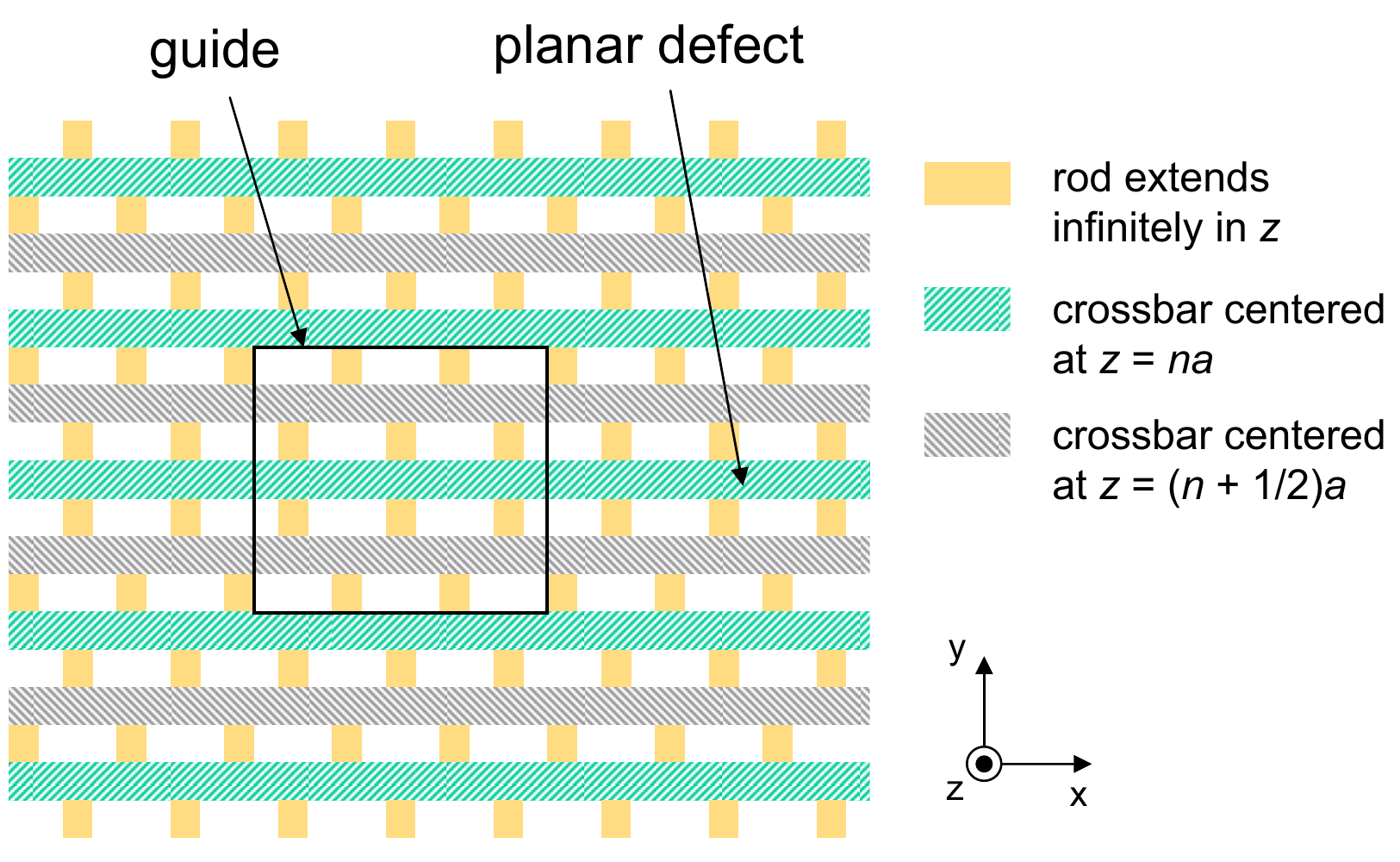}}
\caption{The geometry of a vertically symmetric waveguide structure.}
\label{fig:symmetricguide}
\end{center}
\end{figure}

\begin{figure}
\begin{center}
\includegraphics{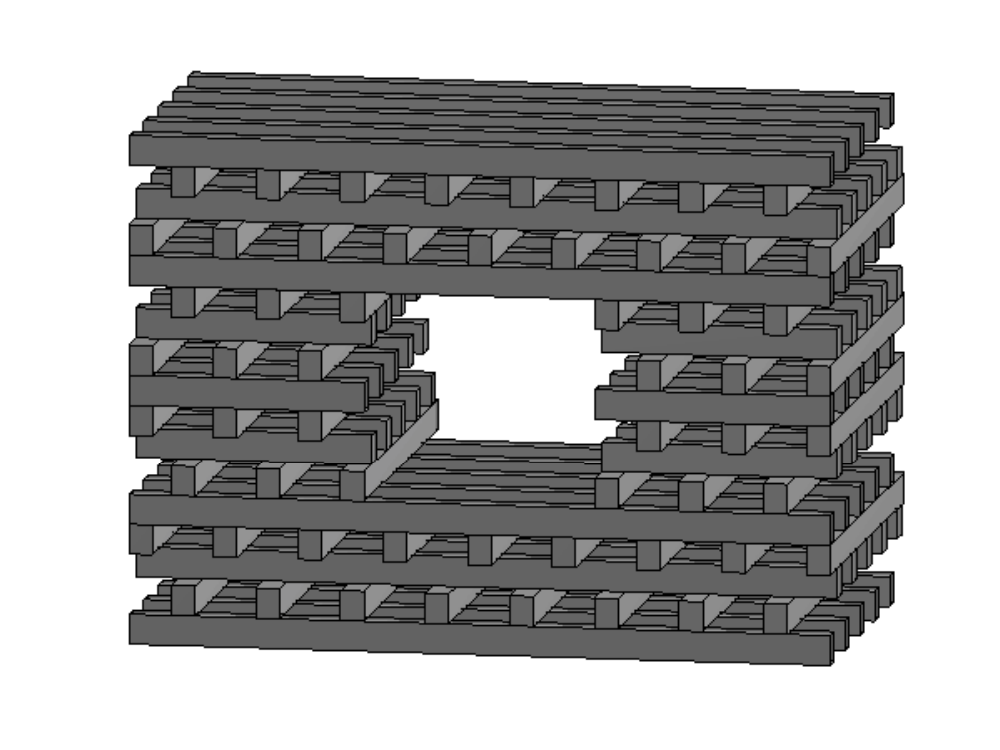}
\caption{A symmetric waveguide.}
\label{fig:symmetricmodel}
\end{center}
\end{figure}

In order to make the structure vertically symmetric, we invert the
upper half of the lattice so it is a vertical reflection of the lower
half.  The geometry, with a defect waveguide introduced, is shown
schematically in Fig.~\ref{fig:symmetricguide} and visually in
Fig.~\ref{fig:symmetricmodel}.

\begin{figure}
\begin{center}
\includegraphics{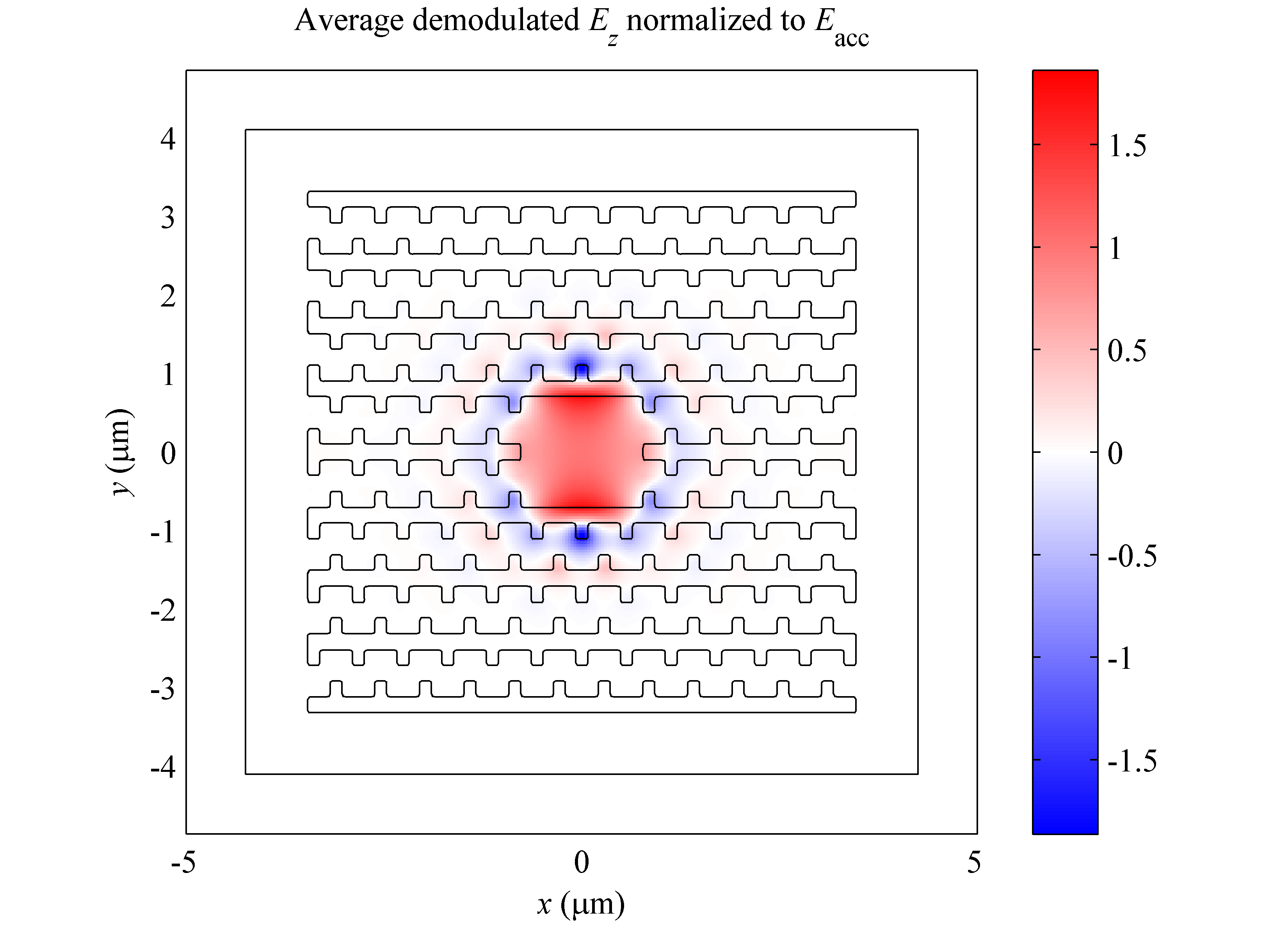}
\caption{The accelerating field seen by a speed-of-light particle,
  averaged over a lattice period, normalized to the accelerating field
  on axis, shown with structure contours for a transverse slice at $z
  = 0$.  The inner rectangle denotes the interface between free space
  and the absorbing boundary in the simulation.}
\label{fig:SymmetricMode}
\end{center}
\end{figure}

The inversion of half the lattice introduces a planar defect where the
two halves meet, but this waveguide still supports a confined
accelerating mode.  Indeed, the mode is lossless to within the
tolerance of the calculation, placing an upper bound on the loss of
\unit[0.48]{dB/cm}.  Its fields are shown in
Fig.~\ref{fig:SymmetricMode}; the computations are described in
Secs.~\ref{sec:FDTDSolver} and \ref{sec:ModeConvergence}.
In this case the dipole fields are suppressed by the vertical symmetry
of the structure.  For this mode $a/\lambda = 0.367$, so using a
\unit[1550]{nm} source determines $a = \unit[569]{nm}$.  The
individual rods are then \unit[159]{nm} wide by \unit[201]{nm} tall.

We can now explore the performance of this structure, according to the
parameters defined in Sec.~\ref{sec:Parameters}.  First, the damage
impedance is \unit[6.10]{\ohm}.  Damage threshold measurements of
silicon, described in Sec.~\ref{sec:damage}, have shown a maximum
sustainable energy density of $\unit[13.3]{J/cm^3}$ at $\lambda =
\unit[1550]{nm}$ and \unit[1]{ps} FWHM pulse width, resulting in an
unloaded accelerating gradient of \unit[221]{MV/m}.  Further
measurements have suggested that higher gradients could be achieved at
longer wavelengths and shorter pulse widths, but that \unit{GeV/m}
acceleration is unlikely in silicon for near-infrared pulses.

Second, we examine the efficiency of the woodpile structure.  The
phenomena which determine the efficiency are described in
Sec.~\ref{sec:Overview}.  The efficiency depends not only on the
electromagnetic mode itself, but also on the beam being accelerated.
Now, we quantitatively compute the efficiency, and the characteristics
of the particle beam needed to optimize the efficiency, following the
treatment in \cite{Na:Efficiency}.  For this mode the characteristic
impedance is $\unit[460]{\ohm}$ and the group velocity is $v_g =
0.253c$.  As described in Sec.~\ref{sec:Parameters}, the
\v{C}erenkov impedance can be estimated from the characteristic radius
of the waveguide.  In our case the aperture is rectangular, so we
define the parameter $R$ by $R = \sqrt{A}/2 = 0.476\lambda$, where $A$
is the aperture area.  This yields $Z_H = \unit[264]{\ohm}$.

With these parameters, we proceed to examine the efficiency of the
structure using the same assumptions as in
\cite{Na:Efficiency}.  The accelerator segment is inside an
optical cavity with a round-trip loss of 5\%, and the external laser
pulse is coupled into the cavity through a beamsplitter with
reflectivity $r$.  We consider a train of $N$ optical bunches, each
with the same charge, and assume that the duration of the train is
much less than the slippage time $\Delta\tau$ between the particles
and the laser pulse inside the structure.  We also assume that
$\Delta\tau = 3\sigma_\tau$, where $\sigma_\tau$ is the laser pulse
duration.

\begin{figure}
\begin{center}
\resizebox{!}{3in}{\includegraphics{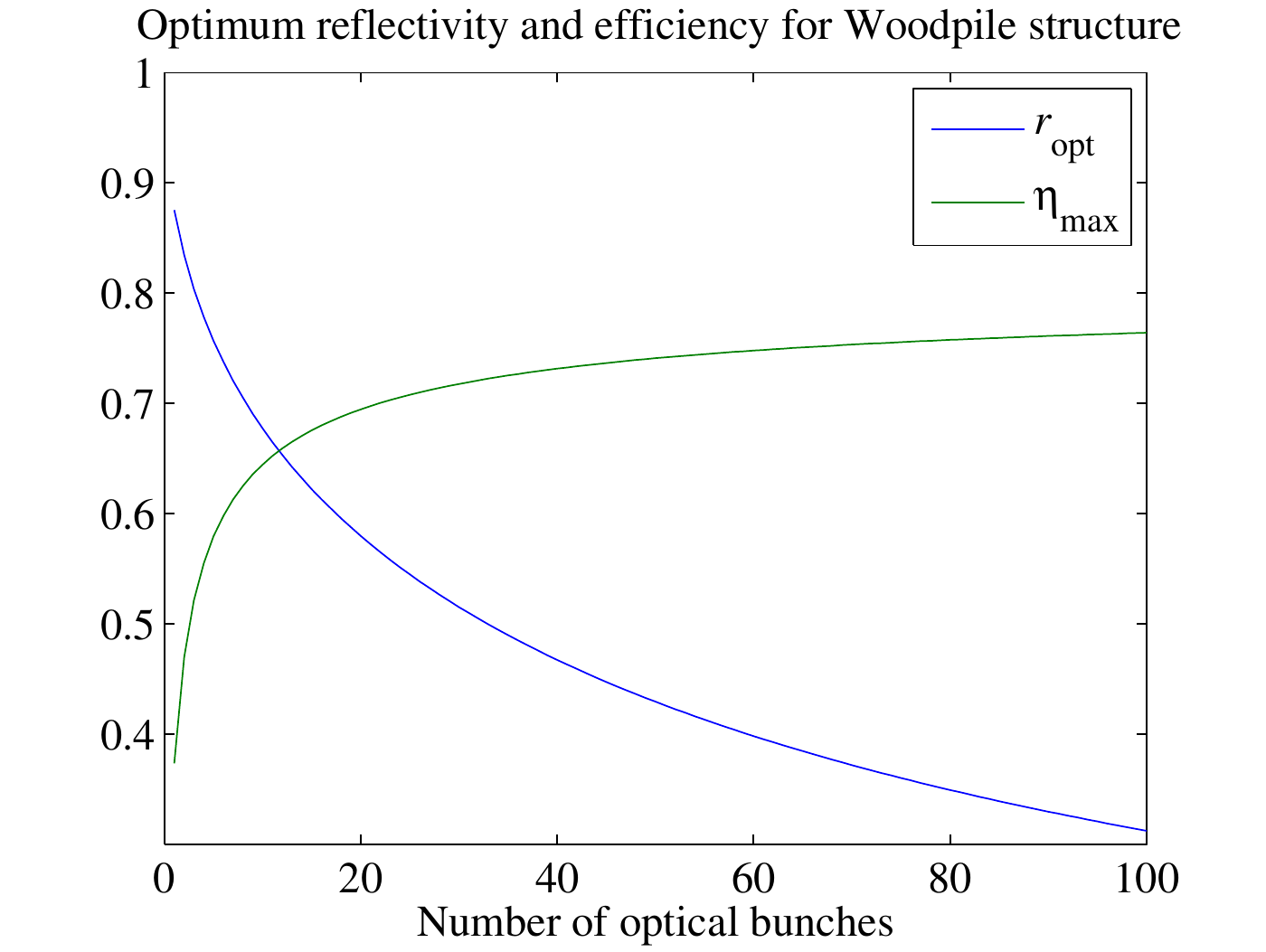}}
\caption{The optimum reflectivity and the corresponding maximum
  efficiency as a function of the number of optical microbunches in
  the bunch train.}
\label{fig:efficiency}
\end{center}
\end{figure}

\begin{figure}
\begin{center}
\resizebox{!}{3in}{\includegraphics{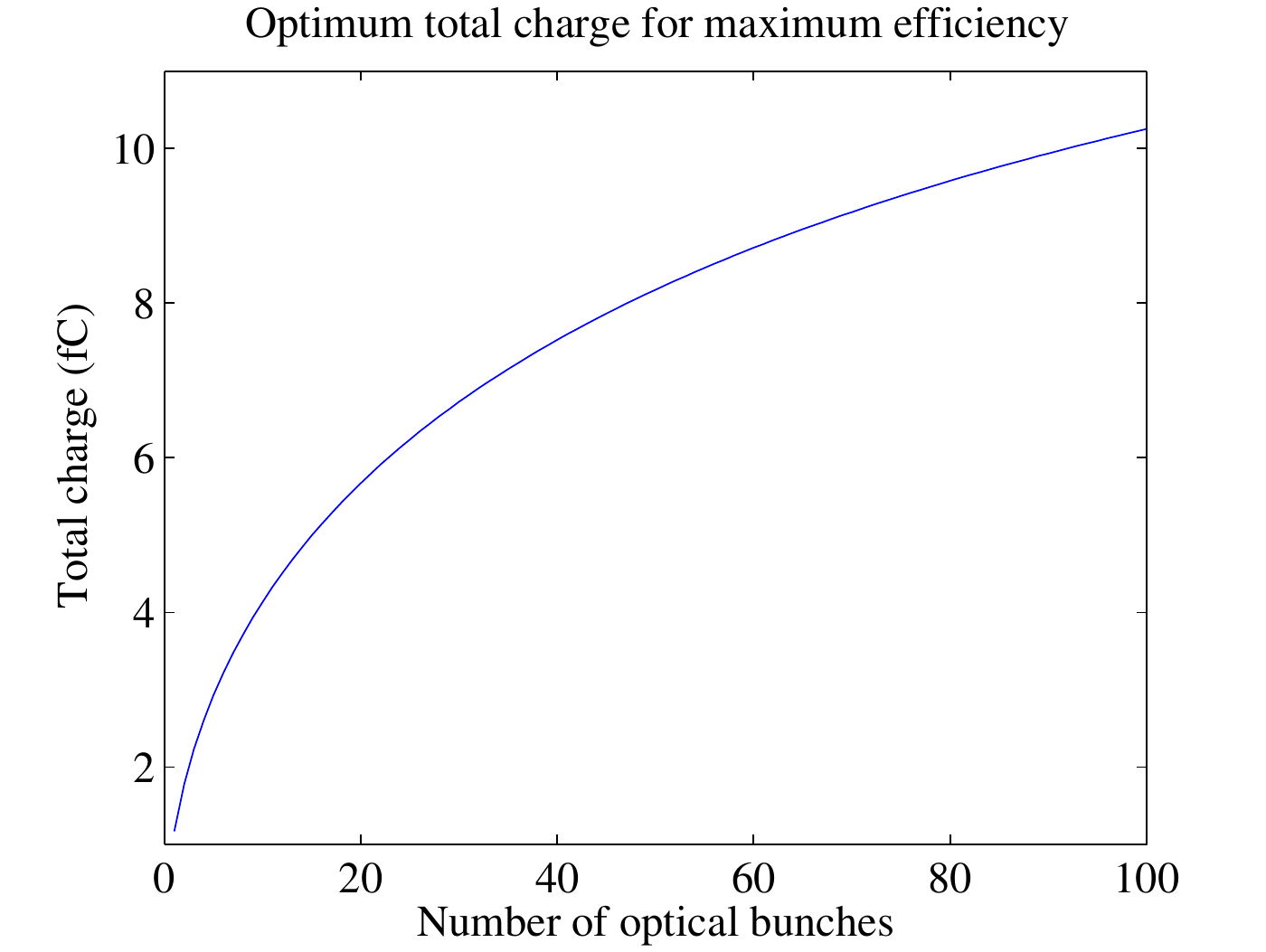}}
\caption{The optimum total charge of a bunch train as a function of
  the number of optical microbunches.}
\label{fig:charge}
\end{center}
\end{figure}

\begin{figure}
\begin{center}
\resizebox{!}{3in}{\includegraphics{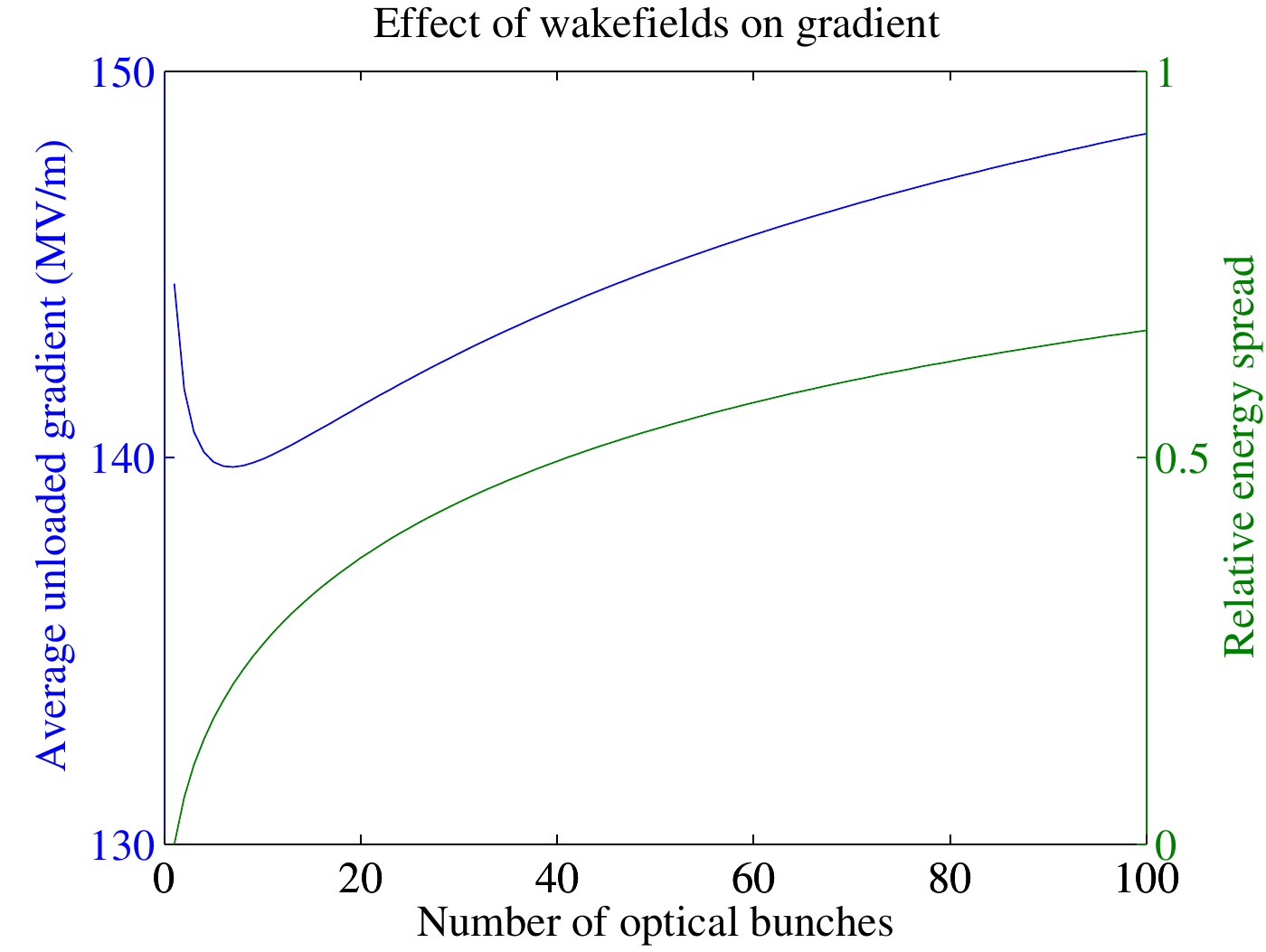}}
\caption{The average unloaded gradient and induced energy spread as a
  function of the number of optical microbunches.}
\label{fig:acceleration}
\end{center}
\end{figure}

As described in Sec.~\ref{sec:Overview}, trying to accelerate too much
charge can reduce the efficiency of an accelerator due to wakefield
effects.  Therefore, for a given $r$ and $N$, the structure has a
maximum efficiency $\eta\tsub{max}$.  For each $N$, we choose $r$ to
maximize $\eta\tsub{max}$.  These optimum values are plotted in
Fig.~\ref{fig:efficiency}.
From this we see that the efficiency can be made quite high.  Even
with just a single bunch, the efficiency reaches 37\%, while for a
train of 100 bunches, the efficiency is 76\%.  Once the optimum $r$ is
computed for each $N$, the total train charge $q_t$ and external laser
pulse amplitude are chosen so that they together satisfy two
conditions: First, that the peak unloaded gradient is equal to the
maximum sustainable gradient in the structure, which we take to be
\unit[221]{MV/m} from the above discussion.  Second, we require that
the charge maximize the efficiency.  We plot the optimum charge as a
function of the number of bunches in Fig.~\ref{fig:charge}.
As expected from the results in \cite{Na:Efficiency}, the
optimum total charge is low, only \unit[1.17]{fC} for a single bunch
and \unit[10.3]{fC} for 100 bunches.  With the charge having been
computed, we can then find the average unloaded gradient for each $N$.
We can also compute the induced energy spread on the beam as the
difference between the average accelerating gradient experienced by
the first and last bunches in the train, relative to the average
unloaded gradient.  These quantities are plotted in
Fig.~\ref{fig:acceleration}.
From this plot we see that wakefields have a serious effect on the
acceleration.  For just a single bunch, the average unloaded gradient
is reduced from \unit[160]{MV/m} (which is reduced from
\unit[221]{MV/m} due to the advancement of the electrons with respect
to the Gaussian laser pulse envelope) to \unit[145]{MV/m}.  The
minimum at around 7 optical bunches is due to two competing effects.
First, as the number of bunches increases from the single-bunch case,
the total charge increases, generating more wakefields which are
recycled within the cavity.  But we see from
Fig.~\ref{fig:efficiency} that as the number of bunches increases,
the reflectivity decreases, so a smaller fraction of these wakefields
are recycled.  More concerning is the effect on the energy spread of
the beam.  Even with only two bunches, the spread in gradient is
6.1\%, and with 5 bunches, the spread is 16.3\%.

\section{Coupling}
\label{sec:Coupling}

A significant advantage of planar structures which are amenable to
lithography is that a coupler from a laser source to the accelerating
waveguide can be integrated with the rest of the structure as part of
the same manufacturing process.  While investigation of such couplers
is currently underway, several possibilities have arisen.  For a
future accelerator we desire a coupler that is both efficient and
compact.  Because we wish to improve the overall optical-to-beam
efficiency by recycling the laser power in an optical cavity as
described in Sec.~\ref{sec:SymmMode}, avoiding coupling losses is
essential to the efficiency of the accelerator.  A compact coupler is
desirable because any space that is used for coupling is not used for
acceleration, so larger couplers reduce the average accelerating
gradient.  For the time being however, we have the simpler
intermediate goal of a coupler that has sufficient efficiency for use
in a proof-of-principle experiment, given that the input fields are
limited by the damage threshold of the material.  We have developed
several preliminary ideas that might meet this goal.

\begin{figure}
\begin{center}
\includegraphics{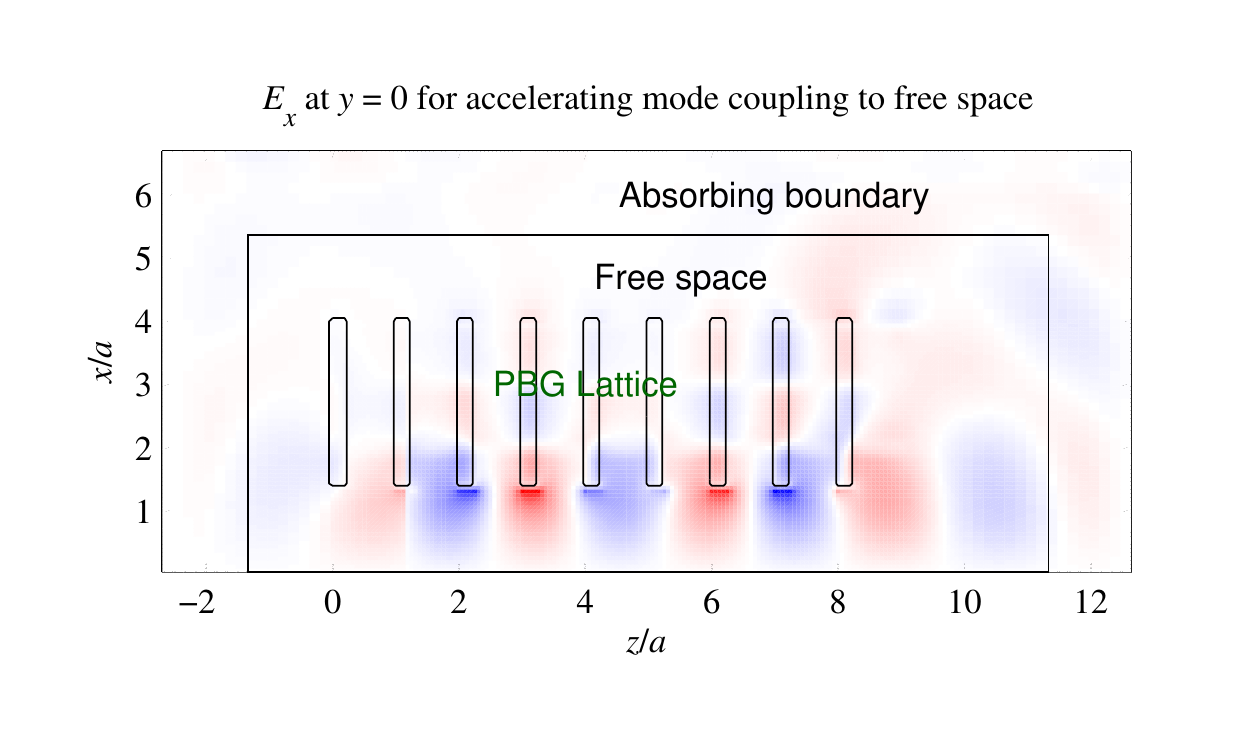}
\caption{Radiation from an accelerating waveguide into free space.
  Plotted here is the $E_x$ component at the $y = 0$ plane at a fixed
  point in time.  We show $E_x$ because the power corresponds most
  closely to the transverse fields.  The contours show the structure
  boundaries at the $y = 0$ plane.  The $yz$ plane is a magnetic
  boundary, used to exploit the symmetry of the structure to reduce
  the computational domain.}
\label{fig:FreeSpaceCoupler}
\end{center}
\end{figure}

The first possibility is simply to attempt to couple directly into the
accelerating waveguide from a free-space laser mode.  To investigate
this, we used the finite-difference time-domain (FDTD) technique (see
Sec.~\ref{sec:FDTD}) to simulate the reverse problem of radiation from
the woodpile waveguide described in Sec.~\ref{sec:SymmMode}.  We
launch the computed woodpile mode down a 10-period waveguide segment
using the total-field/scattered-field method, with the incident plane
being immediately after the first period.  The ends of the waveguide
are open to free space, and the simulation space is surrounded by a
uniaxial perfectly-matched layer.  To reduce computational effort, we
exploit the four-fold symmetry of the structure by placing magnetic
boundaries at the $x = 0$ and $y = 0$ planes.  We launch the mode as a
continuous excitation, with a gradual buildup in the form of an
error-function envelope in order to reduce the bandwidth.  Some of the
power is reflected at the end of the waveguide; some of the reflected
power is also reflected at the beginning of the waveguide.  We
therefore monitor $E_z$ at a point in the center of the waveguide and
continue the simulation until the amplitude at that point reaches
steady-state.  We then record the fields at points one quarter period
separated in time to reconstruct the complex field amplitudes.  We fit
the $E_z$ component on axis over one period to a sum of forward- and
backward-going accelerating modes.  The ratio of the backward power to
the forward power gives us the reflection coefficient for the exit of
the waveguide.  A plot of the fields for this simulation is shown in
Fig.~\ref{fig:FreeSpaceCoupler}.
We find that only 7\% of the power is reflected at the exit of the
guide, indicating that the guide is well-matched to free space.

\begin{figure}
\begin{center}
\includegraphics{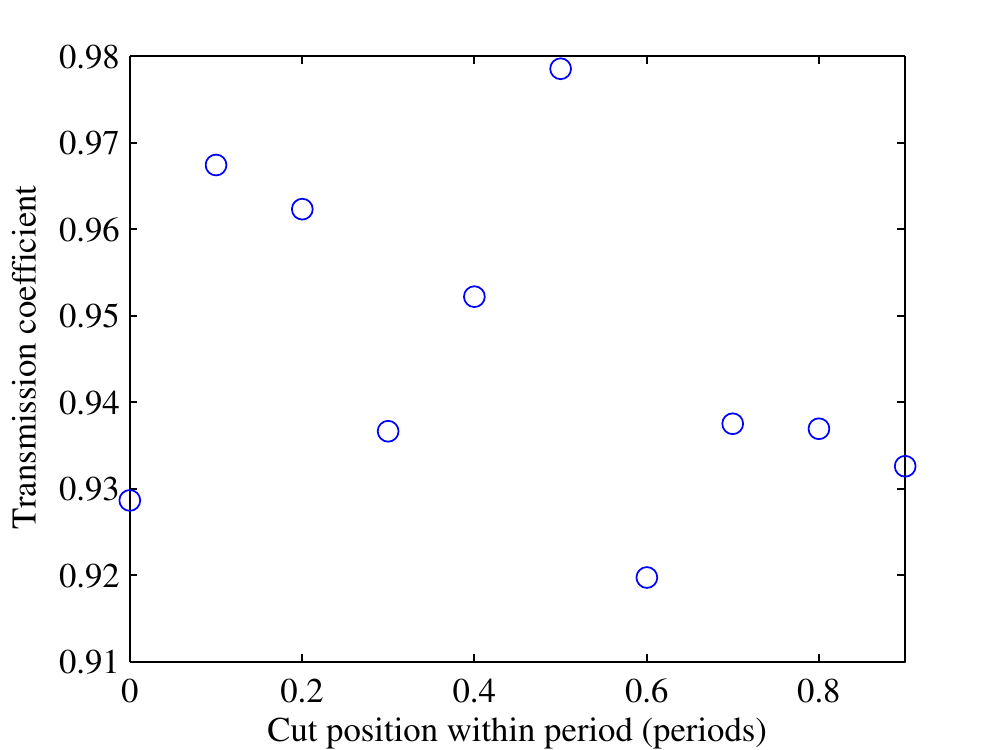}
\caption{Transmission coefficient from the woodpile waveguide to free
  space as a function of the longitudinal position within a lattice
  period of the end of the waveguide.}
\label{fig:FreeSpaceCouplerPosition}
\end{center}
\end{figure}

The reflection coefficient depends on the longitudinal position within
a period at which the waveguide is cut.  To compute this dependence,
we repeat the simulation, varying the longitudinal position of the end
of the waveguide.  The transmission from the waveguide to free space
remains quite high regardless of the end position.  The transmission
coefficients are plotted in Fig.~\ref{fig:FreeSpaceCouplerPosition};
we see that in all cases the transmission is above 90\%.
This is an encouraging result which suggests that we can couple a
significant fraction of the energy in a free-space pulse into the
waveguide.  To develop this idea further, one could take a Fourier
decomposition of the fields several wavelengths beyond the exit of the
guide to find the far-field mode pattern.  Then, one would determine
the field pattern of a free-space mode which matches that far-field
pattern as closely as possible while still being obtainable from a
laser source and the available optical transport.  By simulating,
using the same technique, the propagation of that mode into the guide,
one could determine the coupling efficiency from free space.

Another possible coupler involves using two identical accelerating
waveguides placed parallel to one another and offset by several
wavelengths, as has been explored for photonic crystal fibers
\cite{Mangan:DualCore,Saitoh:Coupling}.  Because of the symmetry of
such a structure the eigenmodes are the even and odd modes, which we
denote by $\psi_+$ and $\psi_-$ respectively.  We can choose the
phases of these modes so that the accelerating mode in a single guide
is very well approximated by the sum or difference of the eigenmodes.
We therefore let
\[ \psi_L = \psi_+ - \psi_-,\quad \psi_R = \psi_+ + \psi_-; \]
then $\psi_L$ and $\psi_R$ approximate accelerating modes in the left-
and right-hand guides respectively.  For a fixed frequency, the even
and odd eigenmodes have longitudinal wavenumbers which we denote $k_+$
and $k_-$ respectively.

Now suppose that an accelerating mode is coupled into the left-hand
waveguide.  After a propagation distance $z$, the state in the
waveguide will be
\[ \psi(z) = \psi_+e^{-ik_+z} - \psi_-e^{-ik_-z}. \]
If we let $\Delta z$ be half of the beat wavelength between the two
modes,
\[ \Delta z = \frac{\pi}{k_+ - k_-}, \]
we have that after propagating for $\Delta z$ distance,
\begin{align*}
\psi(\Delta z)
&= e^{-ik_+\Delta z}(\psi_+ - \psi_-e^{i(k_+ - k_-)\Delta z}) \\
&= e^{-ik_+\Delta z}(\psi_+ + \psi_-) \\
&= e^{-ik_+\Delta z}\psi_R.
\end{align*}
Thus after half a beat wavelength, the mode transitions entirely from
one waveguide to the other.  This would allow one to butt-couple the
mode to one waveguide, and run the particle beam through the other.

\begin{figure}
\begin{center}
\includegraphics{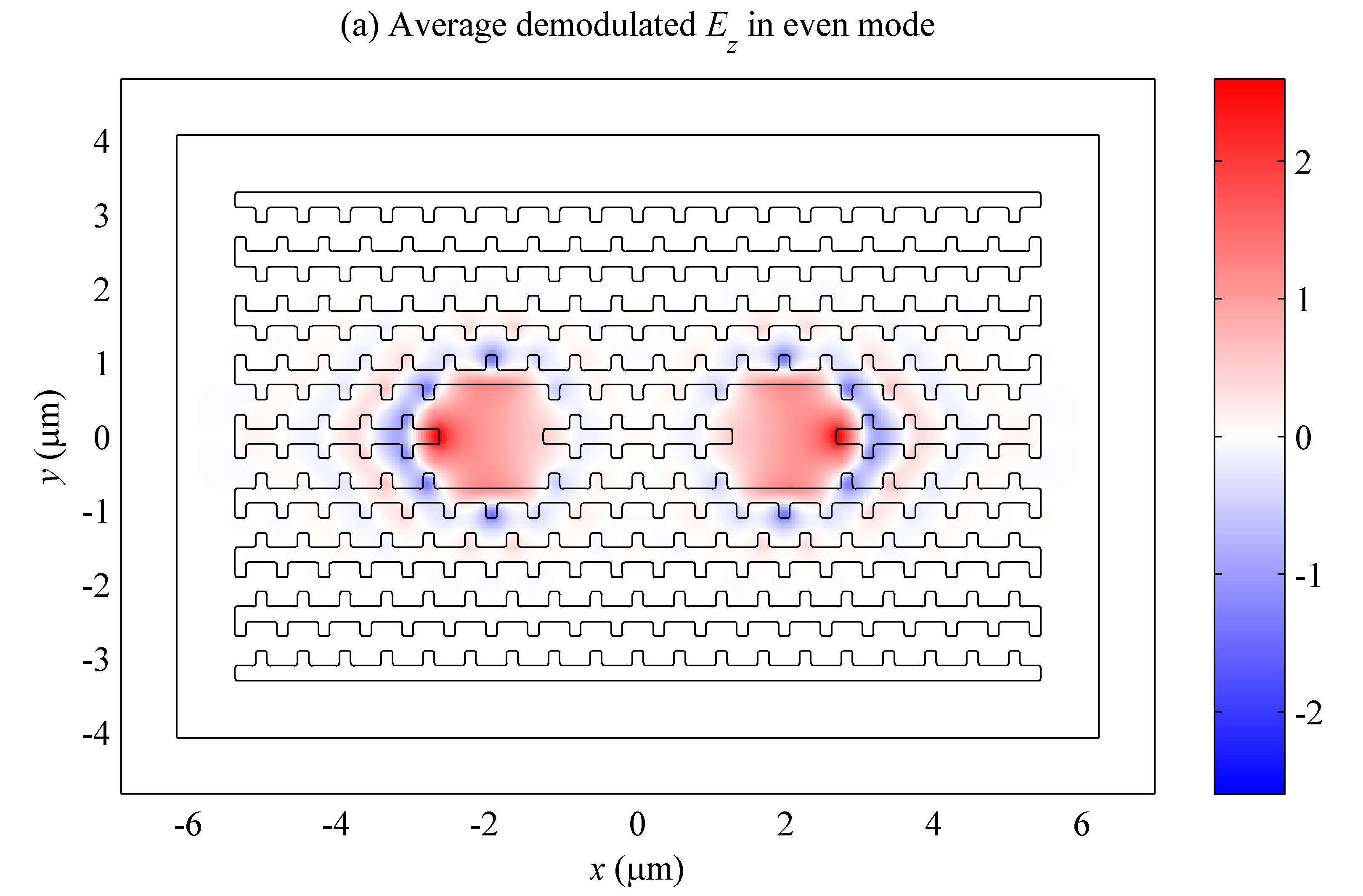} \\
\vspace{0.25in} \includegraphics{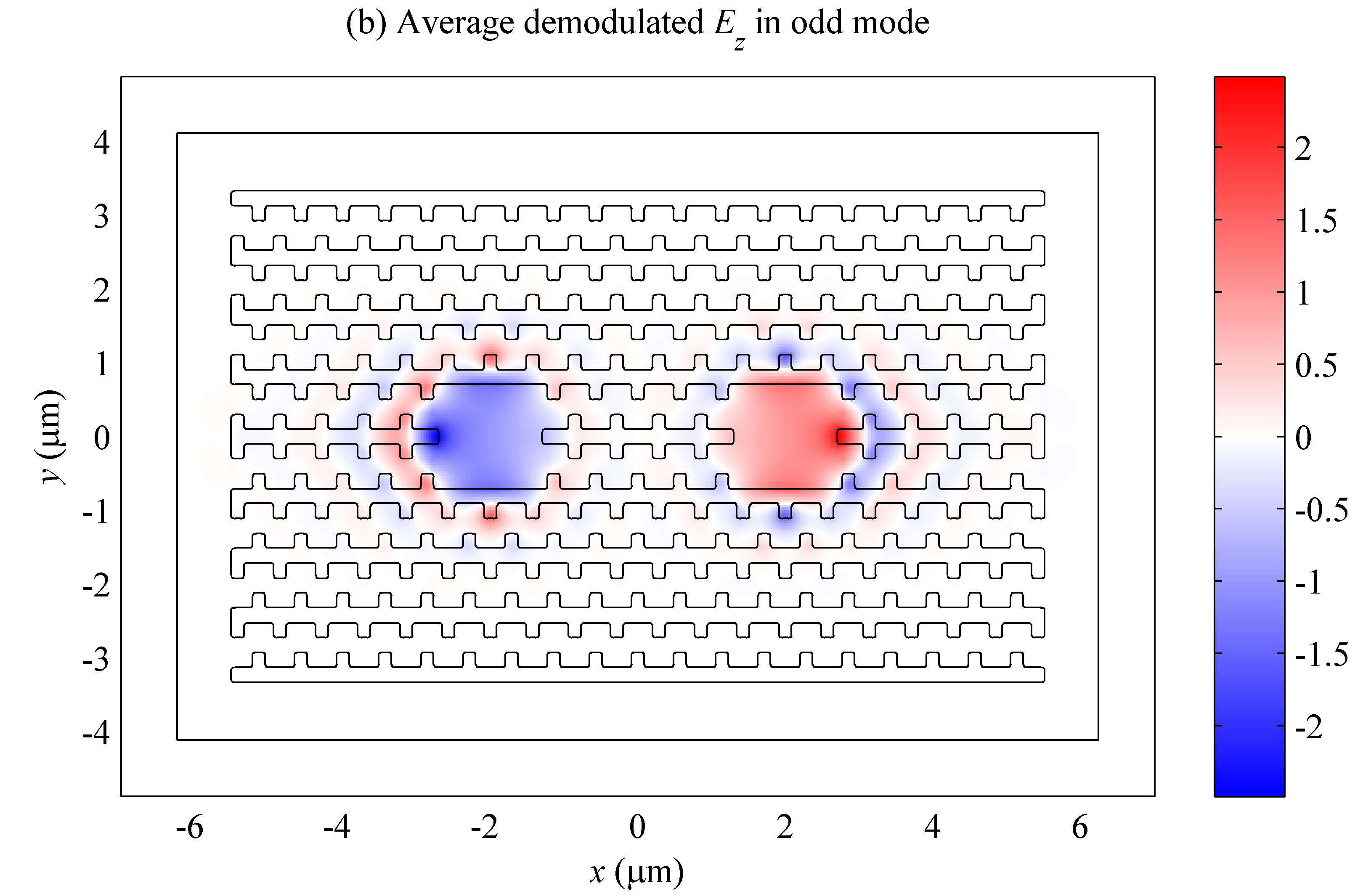}
\caption{The $E_z$ fields of (a) an even mode and (b) an odd mode in
  the parallel-coupler scheme.}
\label{fig:ParallelCoupler}
\end{center}
\end{figure}

A simulation of this scheme is shown in
Fig.~\ref{fig:ParallelCoupler}.
In this structure, the waveguide centers are separated by 7 lattice
periods, or \unit[4.0]{\micro m}.  We find that the beat length is
$\Delta z = 49.6\lambda$, or \unit[76.8]{\micro m}.  We see from the
figure that the fields are distorted due to the presence of the
neighboring waveguide.  The fields in each waveguide are not symmetric
in $x$ as the fields shown in Fig.~\ref{fig:SymmetricMode} are.  There is a
balancing act being performed here: If the waveguides are too far
apart, they effectively act independently, so the mode wavenumbers are
nearly equal and so the coupling length is very long.  However, the
coupling length cannot be made arbitrarily short, because if the
waveguides are too close together, they will disturb each other's
fields, and the sum and difference states will no longer resemble
accelerating modes.  To develop this idea further, one could simulate
this coupler design for a variety of waveguide separations.  By
finding the dependence on separation of both the beat wavelength and
accelerator parameters such as the damage and characteristic
impedances, one would be able to determine the suitability of this
technique and optimize the separation.

Ultimately, we do require a coupler that is both compact and
efficient, as remarked above.  This might be accomplished by
appropriately perturbing the geometry at a single position between the
guides, as has been demonstrated for a two-dimensional photonic
crystal waveguide \cite{Fan:ChannelDrop}.  There is also the
possibility of coupling directly from a waveguide perpendicular to the
accelerating guide, and this technique is currently under
investigation at SLAC.

\section{Particle beam dynamics}
\label{sec:BeamDynamics}

In Sec.~\ref{sec:AsymMode} we found we had to adjust the geometry
of the structure because of beam dynamics considerations, namely to
suppress dipole fields.  Here we examine the particle beam dynamics in
this structure in more detail.  Two concerns immediately arise.
First, the structure has an extremely small aperture, approximately
$\lambda$ in each transverse dimension, so confining the beam within
the waveguide presents a serious challenge.  Second, since the
structure is not azimuthally symmetric, particles out of phase with
the accelerating field will experience transverse forces.  As it turns
out, an investigation of the second problem will yield insight into
the first:  The optical fields in the structure provide focusing
forces which are quite strong compared to conventional quadrupole
magnets and can be used to confine a beam within the waveguide.
Therefore, we begin with a description of the transverse forces in
the woodpile structure.  We then propose a focusing scheme, and
compute the requirements on the phase space extent of the particle
beam.

In our analysis of the beam dynamics in this structure, we use the
average of the fields over one longitudinal period of the photonic
crystal waveguide.  This is justified because the submicron period
together with field intensities limited by damage threshold prohibits
particles from acquiring relativistic momentum on the scale of a
single period.  To see this, one need only examine a normalized vector
potential of the fields, defined by
\[ A = \frac{eEa}{mc^2}, \]
where $E$ is the electric field magnitude.  If $A\ll 1$, then the
variation of a particle's trajectory within a period is negligible, so
the time-averaged force on the particle is well approximated by the
force due to the longitudinally averaged fields.  In our case, even
for $E = \unit[1]{GeV/m}$, $A = 1.1\e{-3}$.  We therefore proceed to
define the longitudinally-averaged fields, as seen by a speed-of-light
particle beam.  We take the fields to have $e^{i\omega t}$ time
dependence, so that the real fields at spacetime coordinates $(t, x,
y, z)$ are
\[ \mathcal{E}(t, x, y, z) = \re[\vect{E}(x, y, z)e^{i\omega t}],
\quad \mathcal{H}(t, x, y, z) = \re[\vect{H}(x, y, z)e^{i\omega t}].\]
We also define the phase of the fields so that a particle with
trajectory $z = ct$ experiences the peak accelerating field.  The
electric field experienced by such a particle is then
\begin{align*}
\vect{E}_p(x, y, z) &= \re[\vect{E}(x, y, z)e^{i\omega z/c}] \\
&= \re[\vect{E}(x, y, z)e^{ik_zz}],
\end{align*}
where $k_z$ is the Bloch wavenumber, since the fields have
speed-of-light phase velocity.  We can then define the averaged
electric field by
\[ \bar{\vect{E}}(x, y) = \frac{1}{a}\int_0^a
\vect{E}(x, y, z)e^{ik_zz}\,dz, \]
and similarly the averaged magnetic field by
\[ \bar{\vect{H}}(x, y) = \frac{1}{a}\int_0^a
\vect{H}(x, y, z)e^{ik_zz}\,dz. \]

In order to better determine the nature of the forces a particle beam
will experience, we now wish to decompose the fields within the
waveguide into their azimuthal moments, that is, we wish to express
each field component $\psi$ in the form
\begin{equation}
\psi(\rho, \theta) = \sum_{m = -\infty}^\infty R_m(\rho)e^{im\theta}
\label{eq:AzimuthalDecomposition}
\end{equation}
in polar coordinates.  To perform this decomposition, we first notice
that since the fields have speed-of-light phase velocity, each field
component is harmonic in the transverse directions within the
waveguide, satisfying
\[ \nabla_\perp^2\psi = 0. \]
Let us now introduce the complex transverse coordinate
\[ w = \frac{x + iy}{\lambda} = \frac{\rho}{\lambda}e^{i\theta}. \]
We can then consider the averaged fields as functions of $w$.  A
result of complex analysis holds that any real harmonic function on an
open, simply connected region $U\subset\mathbb{C}$ is equal to the real part
of an analytic function on $U$ \cite{Marsden:Complex}.  It follows
that within the waveguide region, we can represent a field component
as $\psi = f + g^*$, where $f$ and $g$ are analytic.  Since an
analytic function can be represented as a power series, we can write
\[ \psi = \sum_{m = 0}^\infty (C_mw^m + D_m{w^*}^m) \]
for complex coefficients $C_m$ and $D_m$.  This forms a decomposition
as desired in Eq.~(\ref{eq:AzimuthalDecomposition}).

We can simplify the field expansions using the symmetries of the
mode.  We begin with the $\bar{E}_x$ component.  Let
\[ \bar{E}_x = \sum_{m = 0}^\infty (C_mw^m + D_m{w^*}^m). \]
Because of the even symmetry of the mode in the $y$ direction, we must
have $\bar{E}_x(w^*) = \bar{E}_x(w)$.  Then
\[ \sum_{m = 0}^\infty (C_m{w^*}^m + D_mw^m)
 = \sum_{m = 0}^\infty (C_mw^m + D_m{w^*}^m), \]
so $D_m = C_m$.  We then have
\[ \bar{E}_x = \sum_{m = 0}^\infty 2C_m\re(w^m). \]
The accelerating mode also has even symmetry under a 180\degree\ 
rotation around the $z$-axis.  Since $\bar{\vect{E}}$ is a vector
field, this means that $\bar{\vect{E}}_\perp(-w) =
-\bar{\vect{E}}_\perp(w)$.  For $\bar{E}_x$, this means that
\begin{align*}
\sum_{m = 0}^\infty C_m\re((-w)^m)
&= -\sum_{m = 0}^\infty C_m\re(w^m), \\
\sum_{m = 0}^\infty (-1)^m C_m\re(w^m)
&= \sum_{m = 0}^\infty -C_m\re(w^m).
\end{align*}
Thus $C_m = 0$ for $m$ even.  Finally, the system is symmetric under
the composition of time reversal and reflection in $z$.  Under time
reversal, $\bar{\vect{E}}\ad -\bar{\vect{E}}^*$ and
$\bar{\vect{H}}\ad\bar{\vect{H}}^*$, so under the composition we have
\begin{alignat*}{2}
\bar{\vect{E}}_\perp &\ad -\bar{\vect{E}}_\perp^*,
&\qquad \bar{E}_z &\ad \bar{E}_z^*, \\
\bar{\vect{H}}_\perp &\ad -\bar{\vect{H}}_\perp^*,
&\qquad \bar{H}_z &\ad \bar{H}_z^*.
\end{alignat*}
Since we are choosing the phase of the fields so that an on-crest
particle experiences peak acceleration, we have an even mode under
this symmetry.  Thus the $z$ components of the averaged fields are
pure real, while the transverse components are pure imaginary.  We can
therefore write
\[ \bar{E}_x = i\sum_{m = 0}^\infty A_{2m}\re\brckt{\paren{\frac{x +
	  iy}{\lambda}}^{2m + 1}} \]
for real coefficients $A_{2m}$.

We can similarly simplify the coefficients for $\bar{E}_y$.  In this
case, the even symmetry of the mode in the $y$ direction implies that
$\bar{E}_y(w^*) = -\bar{E}_y(w)$.  So if we now let
\[ \bar{E}_y = \sum_{m = 0}^\infty (C_mw^m + D_m{w^*}^m), \]
we must have $D_m = -C_m$.  Then
\[ \bar{E}_y = \sum_{m = 0}^\infty 2iC_m\im(w^m). \]
As was the case for $\bar{E}_x$, here we must also have $C_m = 0$ for
$m$ even due to the rotational symmetry around the $z$-axis.  Again
choosing the transverse field to be pure imaginary, we can write
\[ \bar{E}_y = i\sum_{m = 0}^\infty B_{2m}\im\brckt{\paren{\frac{x +
	  iy}{\lambda}}^{2m + 1}} \]
for real coefficients $B_{2m}$.

Using the Maxwell equations, all the other field components are
determined by $\bar{\vect{E}}_\perp$, and can thus be specified in
terms of the $A$ and $B$ coefficients.  If we let $k_0 = \omega/c =
k_z = 2\pi/\lambda$, then it follows from the fact that
$\del\cdot\vect{E} = 0$ in vacuum that
\begin{align*}
\bar{E}_z &= -\frac{i}{k_0}\del\cdot\bar{\vect{E}}_\perp
 = -\frac{i}{k_0}\paren{\frac{\ptl\bar{E}_x}{\ptl x} +
 \frac{\ptl\bar{E}_y}{\ptl y}} \\
&= \frac{1}{k_0}\sum_{m = 0}^\infty
\brce{A_{2m}\re\brckt{(2m + 1)\paren{\frac{x + iy}{\lambda}}^{2m}
\frac{1}{\lambda}} + B_{2m}\im\brckt{(2m + 1)
\paren{\frac{x + iy}{\lambda}}^{2m}\frac{i}{\lambda}}} \\
&= \frac{1}{2\pi}\sum_{m = 0}^\infty (2m + 1)(A_{2m} + B_{2m})
\re\brckt{\paren{\frac{x + iy}{\lambda}}^{2m}}.
\end{align*}
Thus the accelerating field on axis is given by $E\tsub{acc} = (A_0 +
B_0)/2\pi$.

The magnetic fields are then determined by the Maxwell curl equation
$\del\cross\vect{E} = -ik_0Z_0\vect{H}$, where $Z_0 =
\unit[376.73]{\ohm}$ is the impedance of free space, so that
\begin{align*}
Z_0\bar{H}_x
&= \frac{i}{k_0}\frac{\ptl\bar{E}_z}{\ptl y} - \bar{E}_y, \\
Z_0\bar{H}_y
&= -\frac{i}{k_0}\frac{\ptl\bar{E}_z}{\ptl x} + \bar{E}_x, \\
Z_0\bar{H}_z &= \frac{i}{k_0}\paren{\frac{\ptl\bar{E}_z}{\ptl x} -
  \frac{\ptl\bar{E}_z}{\ptl y}}.
\end{align*}
The average transverse forces on a speed-of-light particle of charge
$e$ moving parallel to
the $z$ axis are given by the Lorentz force equation $\vect{F} =
e(\bar{\vect{E}} + \zhat\cross Z_0\bar{\vect{H}})$.  We then have
\begin{align*}
F_x &= e(\bar{E}_x - Z_0\bar{H}_y)
 = \frac{ie}{k_0}\frac{\ptl\bar{E}_z}{\ptl x} \\
&= \frac{ie}{4\pi^2}\sum_{m = 1}^\infty (2m)(2m + 1)(A_{2m} + B_{2m})
 \re\brckt{\paren{\frac{x + iy}{\lambda}}^{2m-1}}
\end{align*}
and
\begin{align*}
F_y &= e(\bar{E}_y + Z_0\bar{H}_x)
 = \frac{ie}{k_0}\frac{\ptl\bar{E}_z}{\ptl y} \\
&= -\frac{ie}{4\pi^2}\sum_{m = 1}^\infty (2m)(2m + 1)(A_{2m} + B_{2m})
 \im\brckt{\paren{\frac{x + iy}{\lambda}}^{2m-1}}.
\end{align*}
Thus $E_z$ essentially forms a ``potential'' for the transverse force.
Since $E_z$ satisfies $\nabla_\perp^2E_z = 0$, we have that
$\del_\perp\cdot\vect{F}_\perp = 0$, so that focusing in both
transverse directions simultaneously is impossible.  Indeed, let us
compute the focusing gradients on axis: We have
\[ \evalu{\frac{\ptl F_x}{\ptl x}}_{x = 0, y = 0}
= \frac{3ie}{2\pi^2\lambda}(A_2 + B_2),\qquad \evalu{\frac{\ptl
F_y}{\ptl y}}_{x = 0, y = 0} = -\frac{3ie}{2\pi^2\lambda}(A_2 +
B_2). \]
Like in the case of a quadrupole magnet, we see here that the optical
fields provide an equal and opposite focusing gradient in the two
transverse dimensions.  The coefficient $A_2 + B_2$ specifies the
strength of the optical quadrupole field.  If we define the parameter
\[ F_1 = \frac{3}{2\pi^2\lambda}(A_2 + B_2), \]
then particles of energy $E$ ahead in phase by $\phi$ will experience
a focusing gradient in the $x$ direction, and a defocusing gradient in
the $y$ direction, equal to
\[ K_x = -K_y = \re\paren{\frac{ie}{E}F_1e^{-i\phi}}
= \frac{e}{E}F_1\sin\phi. \]
This is equivalent to a quadrupole magnet with gradient
$(F_1/c)\sin\phi$.

We now examine the implications of these forces for the structure we
examined in Sec.~\ref{sec:SymmMode}.  We first need to extract the
$A$ and $B$ coefficients from the computed fields.  We can express all
the averaged field components within the waveguide in terms of these
coefficients; the expressions for $\bar{\vect{E}}$ are given above,
while for $\bar{\vect{H}}$ we have
\begin{align*}
\bar{H}_x &= -i\sum_{m = 0}^\infty \brckt{B_{2m}
  + \frac{(2m + 2)(2m + 3)(A_{2m+2} + B_{2m+2})}{4\pi^2}}
\im\brckt{\paren{\frac{x + iy}{\lambda}}^{2m + 1}}, \\
\bar{H}_y &= i\sum_{m = 0}^\infty \brckt{A_{2m}
  - \frac{(2m + 2)(2m + 3)(A_{2m+2} + B_{2m+2})}{4\pi^2}}
\re\brckt{\paren{\frac{x + iy}{\lambda}}^{2m + 1}}, \\
\bar{H}_z &= -\frac{1}{2\pi}\sum_{m = 0}^\infty (2m + 1)(A_{2m} + B_{2m})
\im\brckt{\paren{\frac{x + iy}{\lambda}}^{2m}}.
\end{align*}
Since all the fields are linear in the $A$ and $B$ coefficients, we
perform a linear least-squares fit to the fields for $A_{2m}$ and
$B_{2m}$ for $m = 0,\ldots,6$.  From this fit, we find that for the
woodpile structure, we have $A_2 + B_2 = -0.907(A_0 + B_0)$.  Thus for
particles out of phase with the accelerating field by $\pi/2$,
experiencing peak transverse fields, the optical focusing force is
equivalent to that of a quadrupole magnet with $F_1/c =
\unit[411]{kT/m}$ gradient, for accelerating fields at the damage
threshold value of \unit[221]{MV/m}.

We can now see directly the problem caused by the two concerns of
small aperture and transverse forces in the structure.  Particles
which are out of phase with the accelerating field, even only
slightly, will be defocused in one of the two transverse dimensions
and rapidly driven out of the aperture.  For instance, a particle
just \unit[10]{mrad} out of phase and \unit[10]{nm} off axis will be
driven out of the waveguide within \unit[15]{cm} of propagation.
Clearly a means of overcoming this obstacle is necessary for such an
optical, azimuthally asymmetric structure to be viable as an
accelerator.

\begin{figure}
\begin{center}
\includegraphics{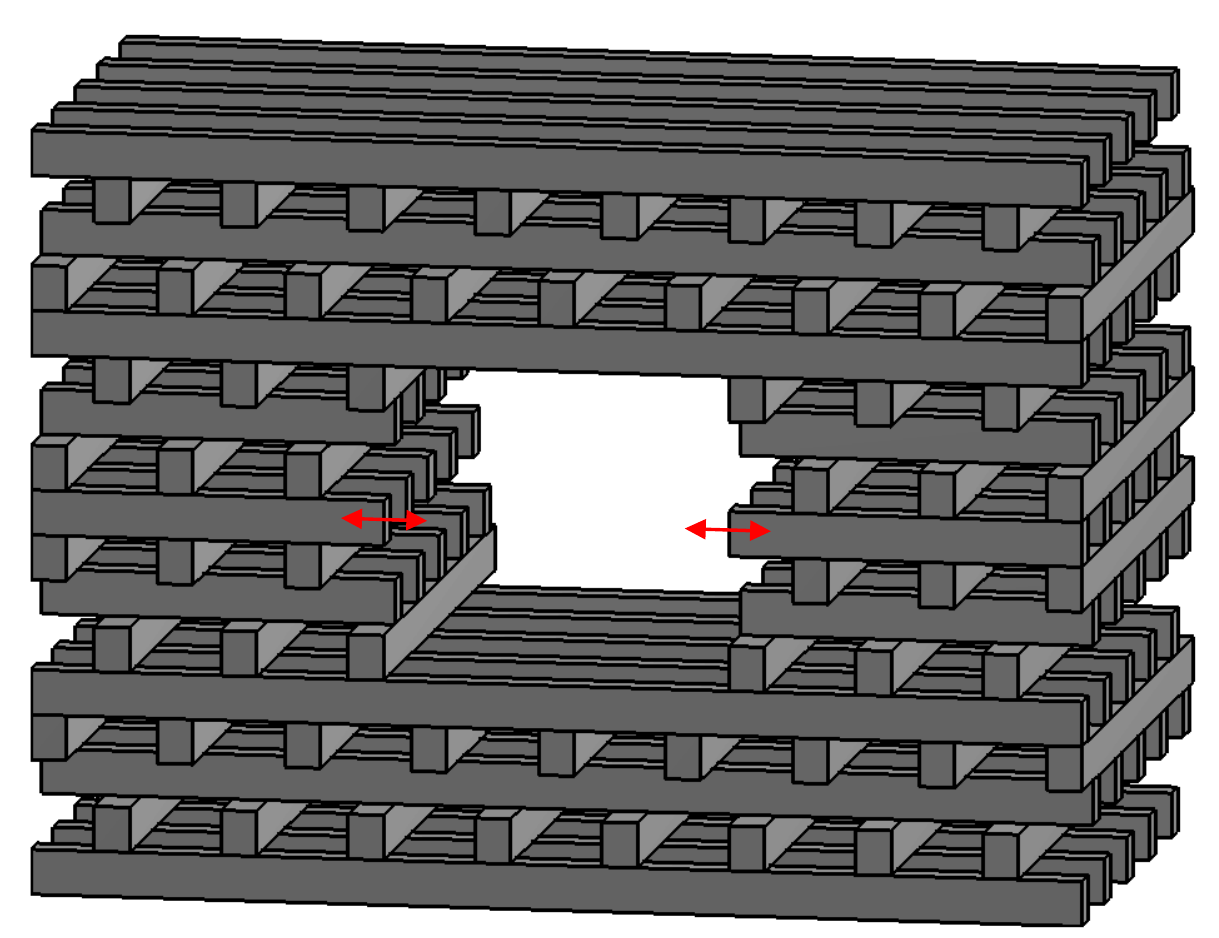}
\caption{The position of the end of the central bar is adjusted to
  suppress the quadrupole fields of the accelerating mode of the
  waveguide.  Both sides are adjusted symmetrically.  A positive
  displacement corresponds to removal of the bar from the guide; a
  negative displacement corresponds to insertion of the bar.}
\label{fig:BarOffset}
\end{center}
\end{figure}

\begin{figure}
\begin{center}
\includegraphics{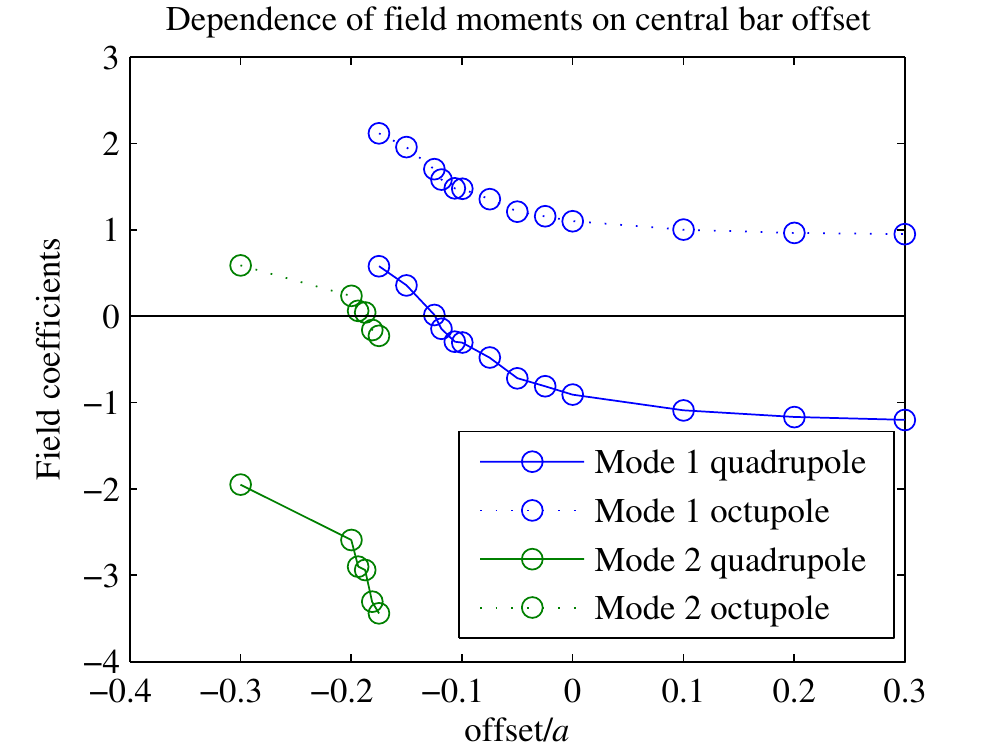}
\caption{The field moment coefficients as a function of central bar
  offset.  The solid lines show the quadrupole coefficients $(A_2 +
  B_2)/(A_0 + B_0)$, while the dotted lines show the octupole
  coefficients $(A_4 + B_4)/(A_0 + B_0)$.  The circles represent
  points where simulations were run.}
\label{fig:Coeffs}
\end{center}
\end{figure}

\begin{figure}
\begin{center}
\includegraphics{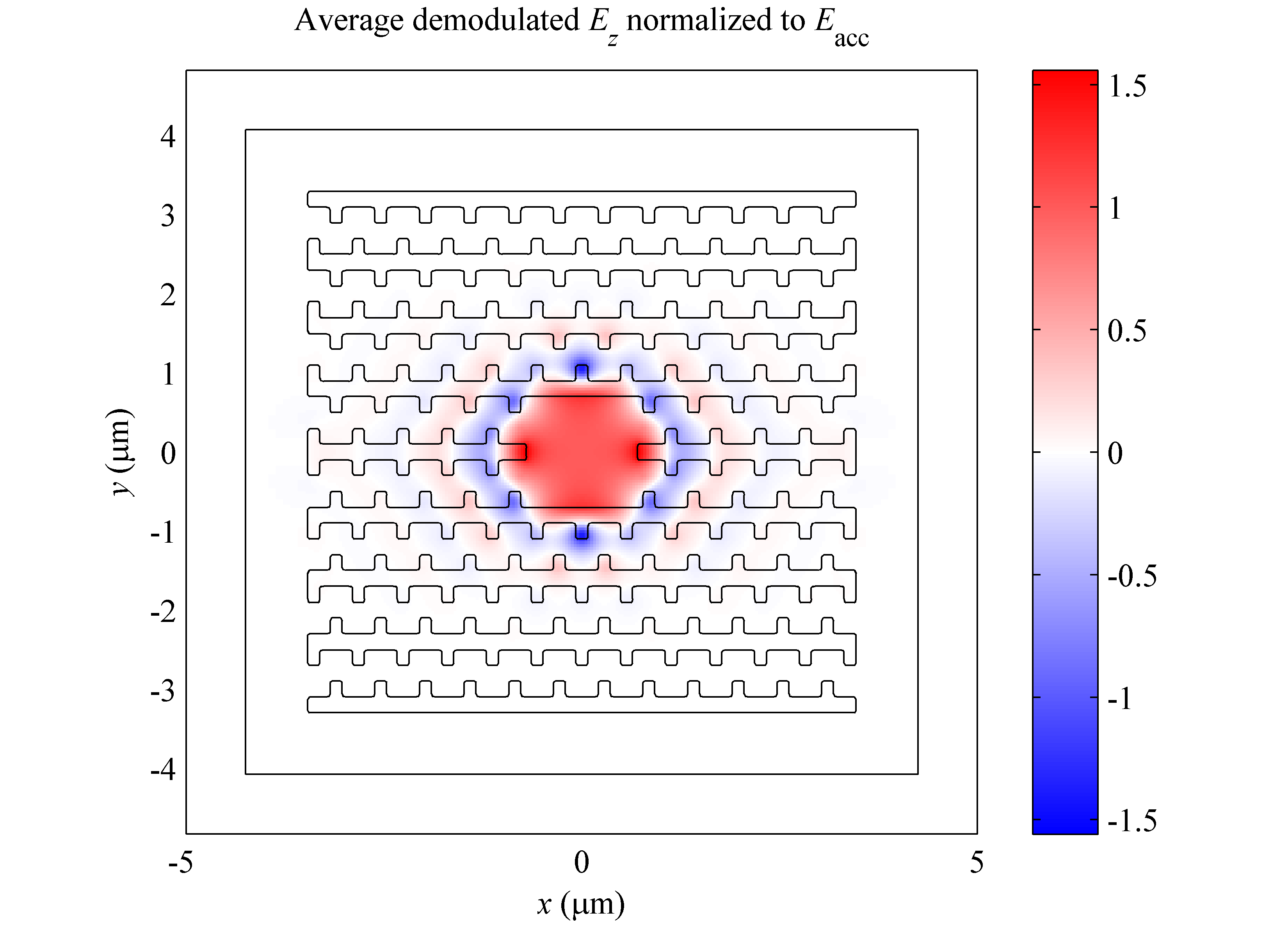}
\caption{The accelerating field in the structure modified to suppress
  the quadrupole moment.}
\label{fig:AccelMode}
\end{center}
\end{figure}

We can in fact overcome the problem of transverse defocusing by
altering the geometry of the structure to suppress the quadrupole
component of the fields.  To this end we adjust a single parameter of
the structure geometry, namely, the removal or extension of the
central bar a distance out of or into the waveguide, as shown in
Fig.~\ref{fig:BarOffset}.
In order to avoid dipole fields we adjust ends of the bars on both
sides of the waveguide symmetrically.  We define the offset to be the
distance the bars are removed from the guide, with a negative value
describing extension into the guide.  We can then examine the
quadrupole field coefficient $(A_2 + B_2)/(A_0 + B_0)$ as a function
of offset to see which offset value, if any, results in suppression of
the quadrupole field.  To this end we run mode computations with the
geometry adjusted with a variety of offset values.  We choose offset
values to converge on the point with zero quadrupole coefficient to
within the geometric resolution of the computation.  This is shown in
the solid blue curve of Fig.~\ref{fig:Coeffs} (the other curves in
the figure are discussed below).
We find that the quadrupole moment coefficient changes sign as the
central bar is extended into the guide, so that suppression of the
quadrupole field is possible by adjusting this parameter.  Indeed, the
quadrupole moment is suppressed by a factor of 57 with an offset of
$-0.125a$.  The accelerating field for this structure geometry is
shown in Fig.~\ref{fig:AccelMode}.
We notice that the field appears more symmetric under rotation by
90\degree, with the values increasing off axis equally in both
transverse directions.  It is also interesting to note that in this
geometry the dimensions of the aperture are nearly equal in the two
transverse directions.  If we take the horizontal aperture to be the
space between the central bars, then the horizontal and vertical
apertures differ by only 0.2\%.  This suggests that despite the
complex photonic crystal lattice geometry, the waveguide appears
``square'' to the fields.  Finally, we note that for this mode, the
damage impedance has increased from $\unit[6.10]{\ohm}$ for the
original mode to $\unit[11.34]{\ohm}$, improving the sustainable
gradient to \unit[301]{MV/m}.

By adjusting the geometry of the structure, we have addressed, at
least to first order, our initial concern of transverse forces during
acceleration.  The small waveguide aperture remains a concern, but the
above discussion of transverse forces presents a possible solution:
Use the extraordinarily strong focusing forces available from the
optical fields to confine the particle beam.  We expect that stronger
focusing forces will allow tighter confinement of the beam, and we now
proceed to explore this idea quantitatively.

\begin{figure}
\begin{center}
\includegraphics{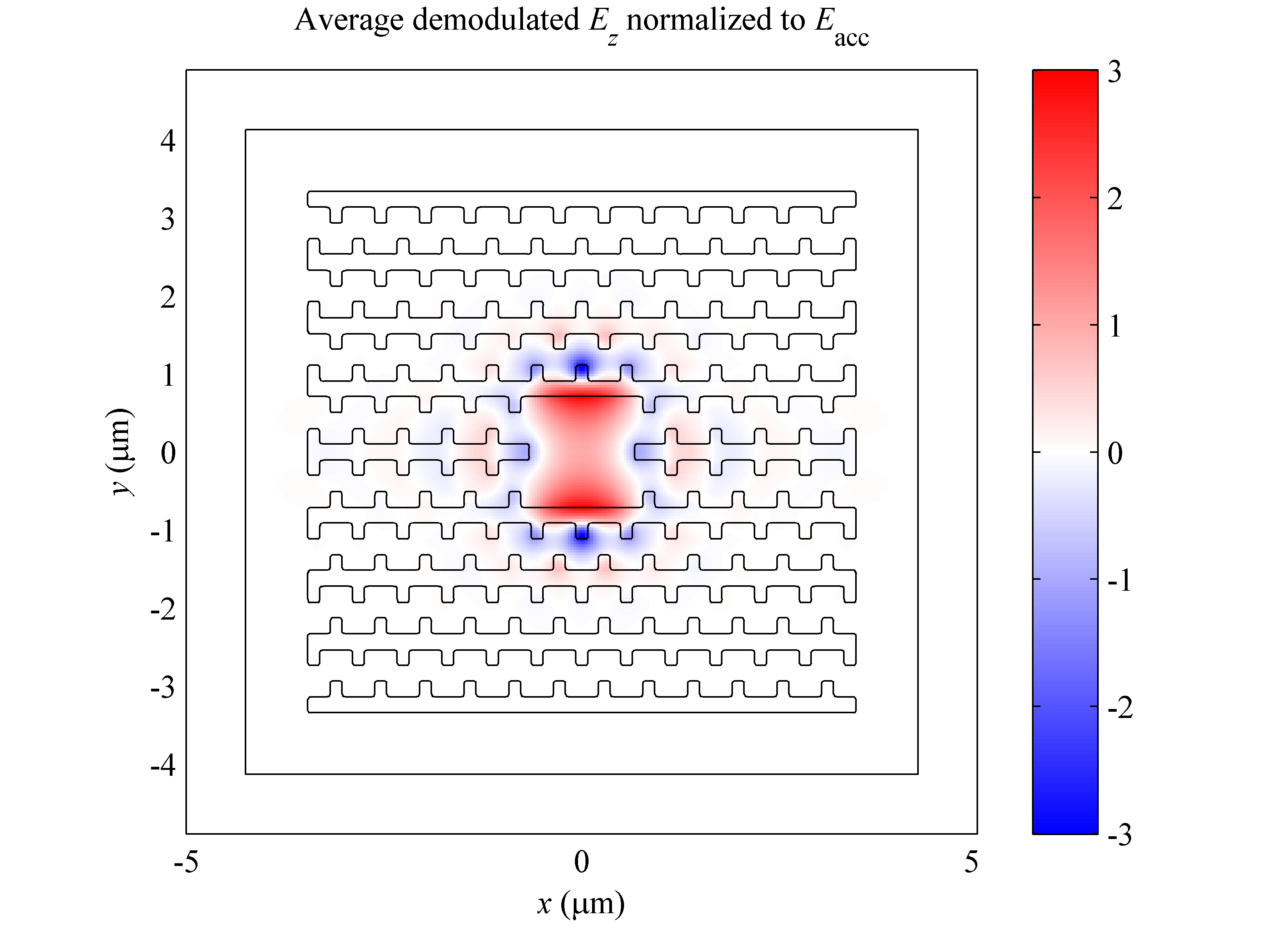}
\caption{The accelerating field in the focusing structure modified to
  suppress the octupole moment.}
\label{fig:FocusMode}
\end{center}
\end{figure}

Since we have suppressed the focusing field in the accelerating
structure, we must choose a different structure geometry for the
focusing structure.  While we could return to the original geometry
which provided a focusing field, we now have the opportunity to adjust
the central bar offset to optimize the structure for focusing.  As we
will see, nonlinear forces are responsible for instabilities in
particle trajectories.  Therefore we consider an optimal focusing mode
to be one in which the lowest-order nonlinear field, in this case the
octupole field due to the structure symmetries, is suppressed.
Returning to Fig.~\ref{fig:Coeffs}, the dashed blue line represents
the octupole field coefficient as a function of central bar offset,
normalized to the accelerating field coefficient; its value is $(A_4 +
B_4)/(A_0 + B_0)$.  We see from the dashed blue line that changes to
the central bar offset fail to suppress the octupole moment in the
accelerating mode.  However, when the central bar is inserted into the
waveguide by more than $0.175a$, a second mode appears which is
qualitatively different from the first.  The green lines in
Fig.~\ref{fig:Coeffs} show the quadrupole and octupole coefficients of
this mode.  From the dashed green line, we see that a appropriate
central bar offset can suppress the octupole field of this mode.
Indeed, for a central bar offset of $-0.188a$, the octupole moment is
suppressed by a factor of 26 relative to the original mode, while the
quadrupole moment is enhanced by a factor of 3.2.  The accelerating
field for this structure geometry is shown in
Fig.~\ref{fig:FocusMode}.

We have now found suitable optical modes for acceleration and
focusing.  The parameters of these modes, as well as the original
mode, are shown in Table~\ref{tab:ModeParams}, with one final minor
adjustment: Since the finite spatial resolution of the mode
computation prevented us from finding an accelerating mode with the
quadrupole moment entirely suppressed, we interpolate to find the
parameters of the ``ideal'' accelerating mode with quadrupole
coefficient identically zero.  We do this by linearly interpolating
the mode parameters, including multipole coefficients, offset,
$a/\lambda$, and damage impedance, between the mode with the smallest
positive quadrupole coefficient and the mode with the smallest
negative coefficient.  We similarly interpolate to find the focusing
mode with zero octupole coefficient.
\begin{table}
\begin{center}
\begin{tabular}{|l|c|c|c|}
\hline
Mode & Original mode & Accelerating mode & Focusing mode \\
\hline
Central bar offset & 0 & $-0.124a$ & $-0.186a$ \\
Quadrupole coefficient & & & \\ $(A_2 + B_2)/(A_0 + B_0)$
& \raisebox{1.5ex}[0cm][0cm]{$-0.907$}
& \raisebox{1.5ex}[0cm][0cm]{0}
& \raisebox{1.5ex}[0cm][0cm]{$-3.017$} \\
Octupole coefficient & & & \\ $(A_4 + B_4)/(A_0 + B_0)$
& \raisebox{1.5ex}[0cm][0cm]{1.100}
& \raisebox{1.5ex}[0cm][0cm]{1.689}
& \raisebox{1.5ex}[0cm][0cm]{0} \\
Damage impedance & $\unit[6.10]{\ohm}$
& $\unit[11.21]{\ohm}$ & $\unit[2.21]{\ohm}$ \\
\hline
\end{tabular}
\caption{Mode parameters for the original mode as well as the
  optimized accelerating and focusing modes.}
\label{tab:ModeParams}
\end{center}
\end{table}

With these parameters for the accelerating and focusing modes, we can
now describe a system of focusing elements designed to confine the
particle beam.  We consider a focusing-defocusing (F0D0)
lattice\footnote{For the remainder of this section, we refer to the
F0D0 lattice as simply the ``lattice,'' not to be confused with the
photonic crystal lattice.}, in which each cell consists of a focusing
segment of length $L_f$ run $\pi/2$ behind in phase for focusing in
the $x$ direction, followed by a an accelerating segment of length
$L_a$, a focusing segment of length $L_f$ run $-\pi/2$ behind in phase
for defocusing in the $x$ direction, and then another accelerating
segment of length $L_a$ \cite{Lee:AcceleratorPhysicsF0D0}.  We are
free to choose the parameters $L_f$ and $L_a$.  Let $K$ be the peak
focusing gradient for the focusing segment, given by $|eF_1/E|$, where
$E$ is the energy of the ideal particle.  If $\sqrt{K}L_f\ll 1$, then
the thin lens approximation applies and we can consider the focusing
and defocusing segments as lenses with focal lengths $f$ and $-f$,
respectively, where $f = 1/KL_f$.  In that case, the betatron phase
advance $\varphi$ per half cell is given by $\sin\varphi = L_a/2f$.
In addition, the maximum value of the $\beta$ function, which occurs
at the midpoint of the focusing segment, is
\[ \beta_F = 2f\frac{1 + \sin\varphi}{\cos\varphi}. \]
The maximum $\beta$ value is an important parameter of a lattice
because the RMS transverse size of a beam with geometric emittance
$\varepsilon$ is $\sigma = \sqrt{\beta\varepsilon}$, so the maximum
spot size over a lattice period is $\sigma\tsub{max} =
\sqrt{\beta_F\varepsilon}$.  The emittance must then be small enough
that $\sigma\tsub{max}$ is several times less than the waveguide
aperture.

We are now faced with a trade-off.  By increasing $L_f$, we can
shorten the focal length and thereby reduce $\beta_F$.  Since the same
maximum spot size can then be reached with a larger emittance, this
loosens the emittance requirement on the particle beam.  However,
assuming $\varphi$ is held constant, we have that $L_a\propto f$, so
that $L_a$ is reduced.  Then a smaller fraction of the accelerator
length is devoted to acceleration as opposed to focusing, so that the
average gradient is reduced.

Let us assume an initial ideal particle energy of \unit[1]{GeV}.  Then
if we run the focusing structure at damage threshold (with an on-axis
accelerating gradient of \unit[133]{MV/m}), we have $K =
\unit[2.49\e{5}]{m^{-2}}$.  For our lattice we choose $L_f =
0.2/\sqrt{K} = \unit[401]{\micro m}$, which gives $f =
\unit[10.0]{mm}$.  We also choose $\varphi = \pi/4$; then $L_a =
\sqrt{2}f = \unit[14.2]{mm}$.  The maximum beta function value is then
$\beta_F = \unit[48.4]{mm}$, the length of the unit cell is $L_c =
2(L_a + L_f) = \unit[29.1]{mm}$, and the betatron period is $4L_c =
\unit[117]{mm}$.  Note that the need to integrate both focusing and
accelerating segments imposes no additional fabrication burden: Since
the group velocity slippage of a laser pulse with respect to a luminal
particle beam over a distance $L_f$ is \unit[4]{ps}, coupling
structures will need to be integrated on a length scale shorter than
the smallest F0D0 lattice feature in any case.

The lattice described in the previous paragraph is stable under linear
betatron motion.  However, nonlinear transverse forces in the
accelerating and focusing modes may cause instabilities which limit
the dynamic aperture of the lattice and constrain the allowable
emittance of the particle beam.  To determine the effect of these
forces and compute the required emittance of a particle beam, we
perform full simulations of particle trajectories.  These simulations
include the dynamics of the particles in all six phase-space
coordinates, with field values taken from the fit $A$ and $B$
coefficients.  Note that the structure lengths are appropriately
scaled to the ideal particle energy.  Since $E$ is not constant,
neither is $K$, and both $L_a$ and $L_f$ are proportional to
$1/\sqrt{K}$.  The simulation techniques are described in detail in
Sec.~\ref{sec:BeamDynamicsComputations}.

\begin{figure}
\begin{center}
\includegraphics{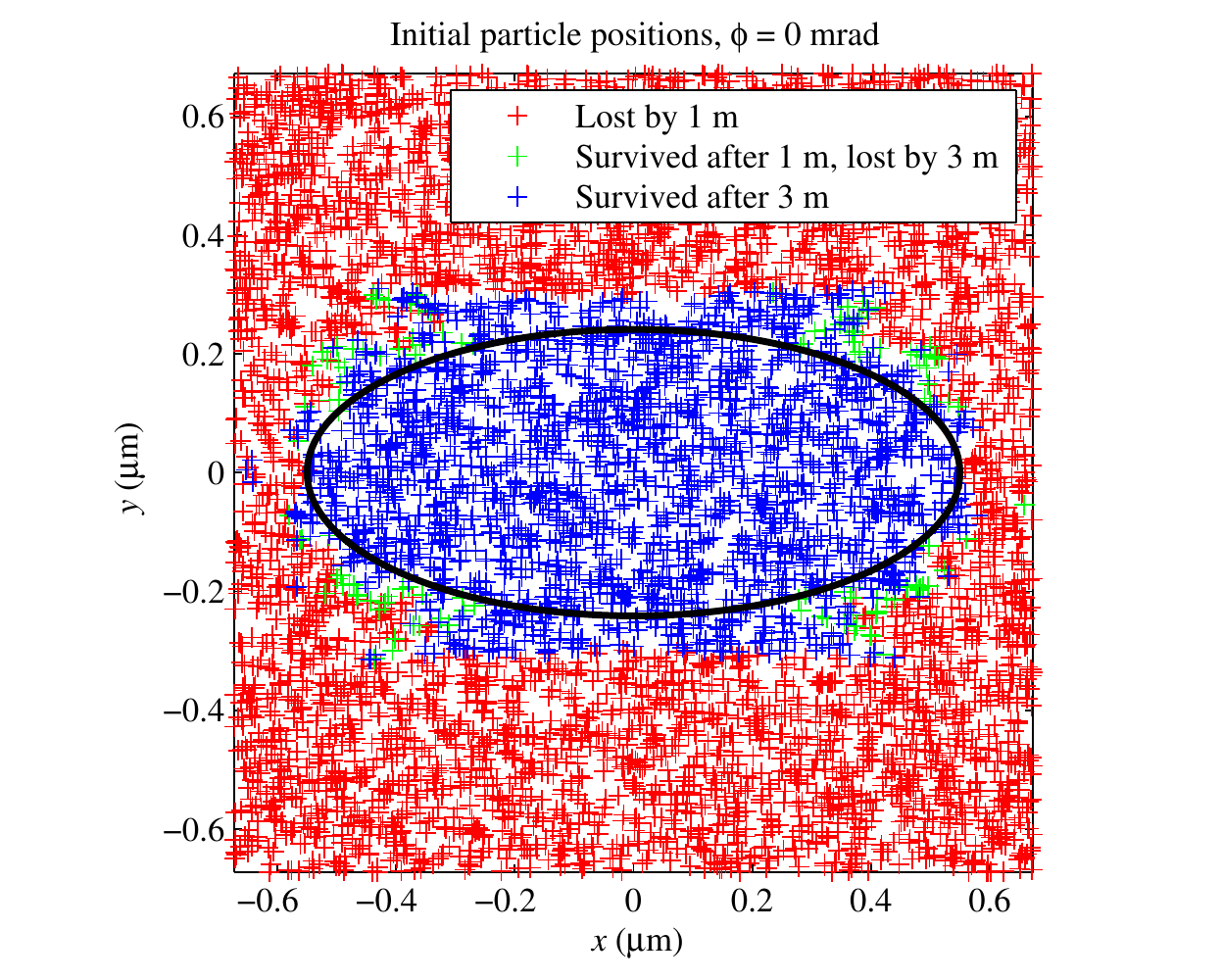}
\caption{Initial particle positions, with colors indicating whether,
  and when, each particle exited the waveguide aperture.}
\label{fig:Aperture0}
\end{center}
\end{figure}

To compute the dynamic aperture of the lattice, we simulate particles
with initial positions uniformly distributed throughout the waveguide
aperture.  The initial transverse momenta of the particles are taken
to be 0, as are the initial deviations of the particle energies from
the ideal case.  We first simulate particles which are exactly on
crest initially.  We propagate the particles for \unit[3]{m} through
the lattice and record whether or not each particle collides with the
waveguide edge.  The initial particle positions, color-coded as to
the particle's survival, is shown in Fig.~\ref{fig:Aperture0}.
In that figure, a red marker indicates that a particle initially at
that position exited the waveguide aperture within \unit[1]{m} of
propagation.  A green marker indicates that it remained within the
waveguide through \unit[1]{m}, but exited before \unit[3]{m}, while a
blue marker indicates that the particle remained in the waveguide
through all \unit[3]{m} of propagation.  We see from the plot that
most particles that are driven out of the waveguide exited within the
first meter.  The initial energy of the particles is \unit[1]{GeV},
and after \unit[3]{m} the energy of the ideal particle is
\unit[1.87]{GeV}.  The ellipse indicated in the figure is the ellipse
of largest area which still contains only particles which remain in
the waveguide.  If we take this ellipse to be the $3\sigma$ boundary
of the particle distribution, we can use the relation $\sigma^2 =
\beta\varepsilon$ to obtain the emittance requirement of the lattice.
We find invariant emittances of
\[ \varepsilon^{(I)}_x = \unit[9.2\e{-10}]{m},\qquad
\varepsilon^{(I)}_y = \unit[1.09\e{-9}]{m}. \]
While these emittances are several orders of magnitude smaller than
those produced by conventional RF injectors, because of the small
bunch charge as discussed in Sec.~\ref{sec:SymmMode}, the
transverse brightness (charge per phase space area) required in the
optical structure is similar to that of conventional accelerators.

\begin{figure}
\begin{center}
\includegraphics{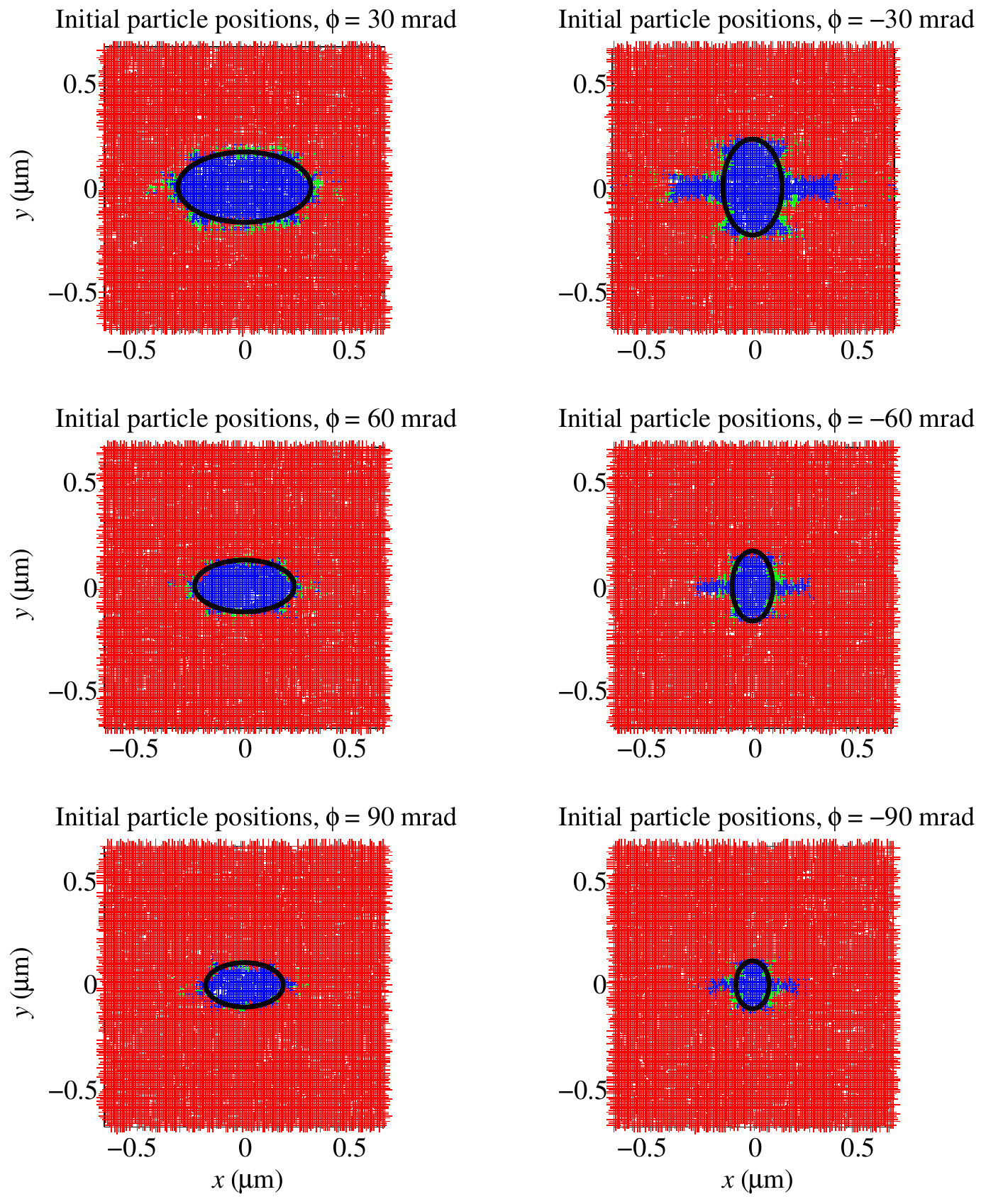}
\caption{Initial particle positions, with colors indicating whether,
  and when, each particle exited the waveguide aperture, for initial
  phases ranging from $\unit[-90]{mrad}$ to \unit[90]{mrad}.}
\label{fig:PhaseDep}
\end{center}
\end{figure}

We can also examine the beam dynamics for particles which are slightly
out of phase with the crest of the accelerating laser field.  Running
the same simulation as above, but with the particles given an initial
phase offset, we find that the dynamic aperture is reduced as
particles become further out of phase.  This is shown in
Fig.~\ref{fig:PhaseDep}.
For particles ahead in phase ($\phi < 0$), we can see in the plots a
pattern characteristic of a fourth-order resonance, indicating that
the octupole field in the accelerating structure becomes significant
as particles become off-crest.  The difference in aperture shape
between positive and negative phase shifts could be attributed to the
sign difference of that octupole field.

\begin{figure}
\begin{center}
\includegraphics{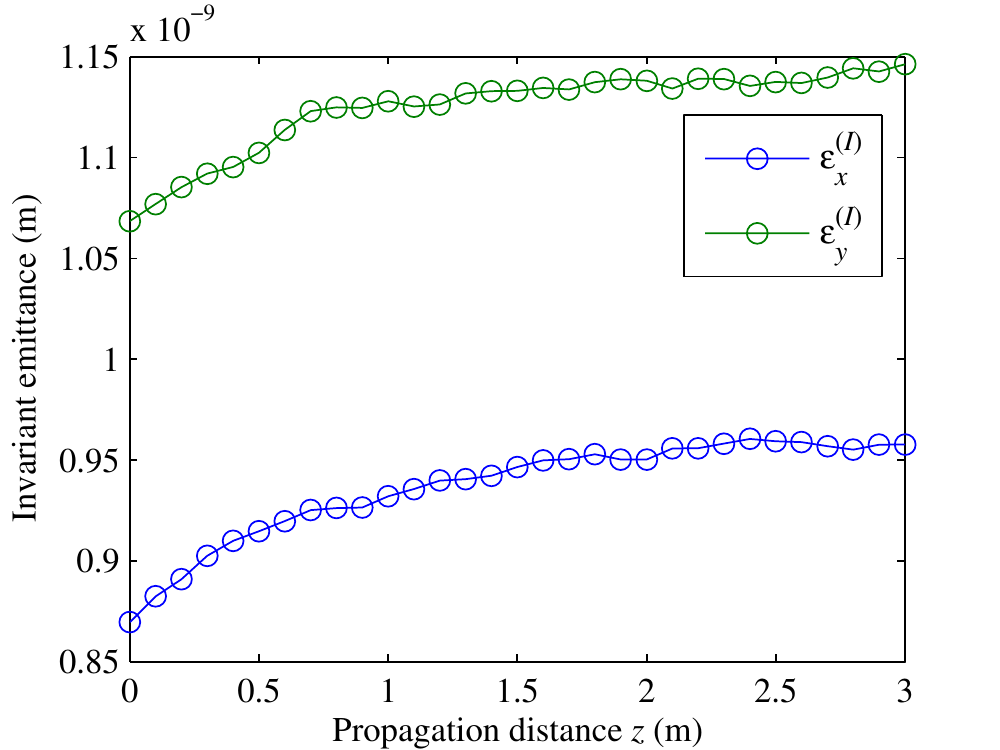}
\caption{Transverse emittances of the particle bunch as a function of
  propagated distance.}
\label{fig:TransverseEmittance}
\end{center}
\end{figure}

With simulations of the dynamic aperture of the lattice for both on-
and off-crest particles, we can now compute the dynamics of a
realistic particle bunch, with variation in all six phase space
coordinates.  We take the initial transverse emittances to be those
computed above for on-crest particles,
\[ \varepsilon^{(I)}_x = \unit[9.2\e{-10}]{m},\qquad
\varepsilon^{(I)}_y = \unit[1.09\e{-9}]{m}. \]
We also take the RMS phase spread and relative beam energy deviation
to be $\sigma_\phi = \unit[10]{mrad}$ and $\sigma_\delta = 10^{-3}$,
respectively.  We track the ensemble of particles for \unit[3]{m} of
propagation, as before, and record the phase space coordinates every
\unit[10]{cm}.  We find that 261 of the 13228 particles simulated, or
2.0\%, exited the waveguide before completing the \unit[3]{m}
propagation.  From the recorded phase space coordinates we can track
the invariant emittances as a function of propagated distance.  The
transverse emittances are shown in Fig.~\ref{fig:TransverseEmittance}.
We see that the emittances increase initially, but that this increase
slows as the propagation continues.  For the full \unit[3]{m}
$\varepsilon^{(I)}_x$ increases by 10.1\%, while $\varepsilon^{(I)}_y$
grows by 7.3\%.  Thus the growth in invariant emittance is much
smaller than the energy gain, so that the geometric emittance of the
beam will still be adiabatically damped.  This shows that particle
bunches can be propagated stably in a photonic crystal accelerator
without an extraordinary enhancement to beam brightness.

\chapter{Woodpile materials and fabrication}
\label{ch:Materials}
In Ch.~\ref{ch:Woodpile} we presented detailed simulations of
a photonic accelerator structure, exploring several aspects of its
operation.  Bringing this type of structure from computer model to
experimental reality involves several unique, difficult, and
interrelated challenges, however.  In this chapter we describe two of
those challenges, namely the choice of material and the fabrication of
the structure.

In the simulations presented in Ch.~\ref{ch:Woodpile}, we used
silicon as the model for our material parameters.  We made this choice
because silicon has several properties making it a very attractive
material for structure-based laser-driven acceleration.  It is
transparent in the mid-infrared \cite{Edwards:HOCSilicon}, in
particular in the telecommunications band at 1550 nm where many
promising sources exist (see for instance \cite{IMRAWeb}).  It has a
high index of refraction at those wavelengths, allowing for the
complete photonic bandgap modeled in Sec.~\ref{sec:WoodpileLattice}. It
is highly resistant to ionizing radiation damage
\cite{Colby:RadiationDamage2002}.  In addition, well-developed
microfabrication techniques exist for silicon due to its use in
integrated circuits.  In Sec.~\ref{sec:damage} we explore in detail
the suitability of silicon as an accelerator material with respect to
the key parameter of the optical breakdown threshold.

While advanced fabrication tools and techniques exist for creating
microstructures, the technology may not yet be available to
manufacture optical accelerator structures with adequate precision.
In Sec.~\ref{sec:fabrication}, we describe some possible
fabrication processes and the implications for choice of material.  We
then quantify the required fabrication tolerance of the woodpile
structure in Sec.~\ref{sec:Tolerance} by investigating the effect
of layer-to-layer misalignment on the accelerating mode.

\section{Damage threshold studies}
\label{sec:damage}

As mentioned in Sec.~\ref{sec:Parameters}, the accelerating
gradient in a structure is ultimately limited by damage to the
material.  The value of this threshold, which is a function of
wavelength and pulse duration, therefore has an overriding influence
on the choice of structure material and laser source.  It would then
be highly informative, as we embark on an exploration of
structure-based optical acceleration, to have a comprehensive set of
damage threshold data for a variety of materials and full range of
laser pulse parameters.  Such knowledge does not exist.  However, a
significant amount of research has taken place on optical damage of a
few materials, so that theoretical descriptions of dielectric
breakdown have been developed.  We discuss the prevailing theories
behind the optical breakdown process and their implications for
accelerators in the next subsection.  Then we will describe an
experiment to measure the optical breakdown threshold of crystalline
silicon for ultrafast pulses in the infrared and present the results.

\subsection{Theoretical background}

As discussed in \cite{Lenzner:Breakdown1998}, optical breakdown in
dielectric materials occurs in four general steps: (1) Seed conduction
electrons are generated by photoionization, (2) they are accelerated
in the laser field and generate an avalanche by impact ionization, (3)
the laser pulse heats the resulting plasma, and (4) the electron
energy is transferred to the lattice, resulting in ablation.  This
schematic description raises an important question: To what extent do
multiphoton effects drive the photoionization process?  According to
the Keldysh theory of photoionization \cite{Keldysh:Ionization}, there
are two limiting cases.  For long wavelengths and strong fields, the
phenomenon becomes that of tunnel ionization (TI), in which the field
reduces the Coulomb barrier sufficiently for a valence electron to
tunnel through it.  In that case the ionization rate depends only
weakly on the wavelength.  In the other extreme, for shorter
wavelengths and weaker fields, the phenomenon becomes that of
multiphoton ionization (MPI), in which the energy of several incident
photons ionizes an electron.  In that case the ionization rate depends
strongly on the photon energy and hence the wavelength.  As a
practical matter for accelerator structure design, can we
significantly increase the damage threshold by lengthening the
wavelength beyond a multiphoton threshold?  For instance, the bandgap
of silicon at room temperature is \unit[1.12]{eV}, corresponding to a
free-space wavelength of \unit[1107]{nm}.  By using a wavelength
longer than \unit[1107]{nm}, in the multiphoton regime seed electrons
can only be generated by two-photon absorption, which has a much lower
cross-section than single-photon absorption.\pagebreak\ Similarly,
operating at wavelengths\nopagebreak\ beyond the two-photon threshold
at \unit[2214]{nm} might yield a further increase in sustainable
gradient.

The experiments described in \cite{Lenzner:Breakdown1998} were
conducted at a wavelength of \unit[800]{nm} on fused silica (bandgap
energy $\Delta\approx\unit[9]{eV}$, requiring six-photon absorption)
and barium aluminum borosilicate (BBS, $\Delta\approx\unit[4]{eV}$,
requiring three-photon absorption).  The authors reached the
conclusion that multiphoton ionization dominates as the seed process
for FWHM pulse widths $\tau\geq\unit[20]{fs}$, while tunneling
dominates for shorter pulses.  Other experiments which have been
conducted at \unit[527]{nm} and \unit[1053]{nm} are consistent with
this conclusion \cite{Stuart:BreakdownPRB1996}.  However, another
experiment measuring breakdown thresholds as a function of
polarization has called into question the significance of MPI in the
breakdown process \cite{Joglekar:Damage2003}.  In that study the
authors showed that breakdown thresholds were independent of
polarization, and argued that the inefficiency of circularly-polarized
light for MPI indicates that MPI is not the dominant process.  Also,
evidence indicates that TI is the dominant ionization process in the
mid-infrared \cite{Simanovskii:Damage}.  Therefore, we may not be able
to arbitrarily increase the sustainable gradient by using longer
wavelengths.

\subsection{Experimental setup and procedure}

As samples for the damage study, we used undoped crystalline silicon
cut from a wafer with a (100) surface orientation.  We conducted
experiments starting at $\lambda = \unit[1550]{nm}$ and extending
longer in wavelength beyond \unit[2200]{nm}.  The pulse widths varied
between 0.66 and \unit[1.12]{ps}, depending on wavelength. 

\begin{figure}
\begin{center}
\resizebox{\textwidth}{!}{\includegraphics{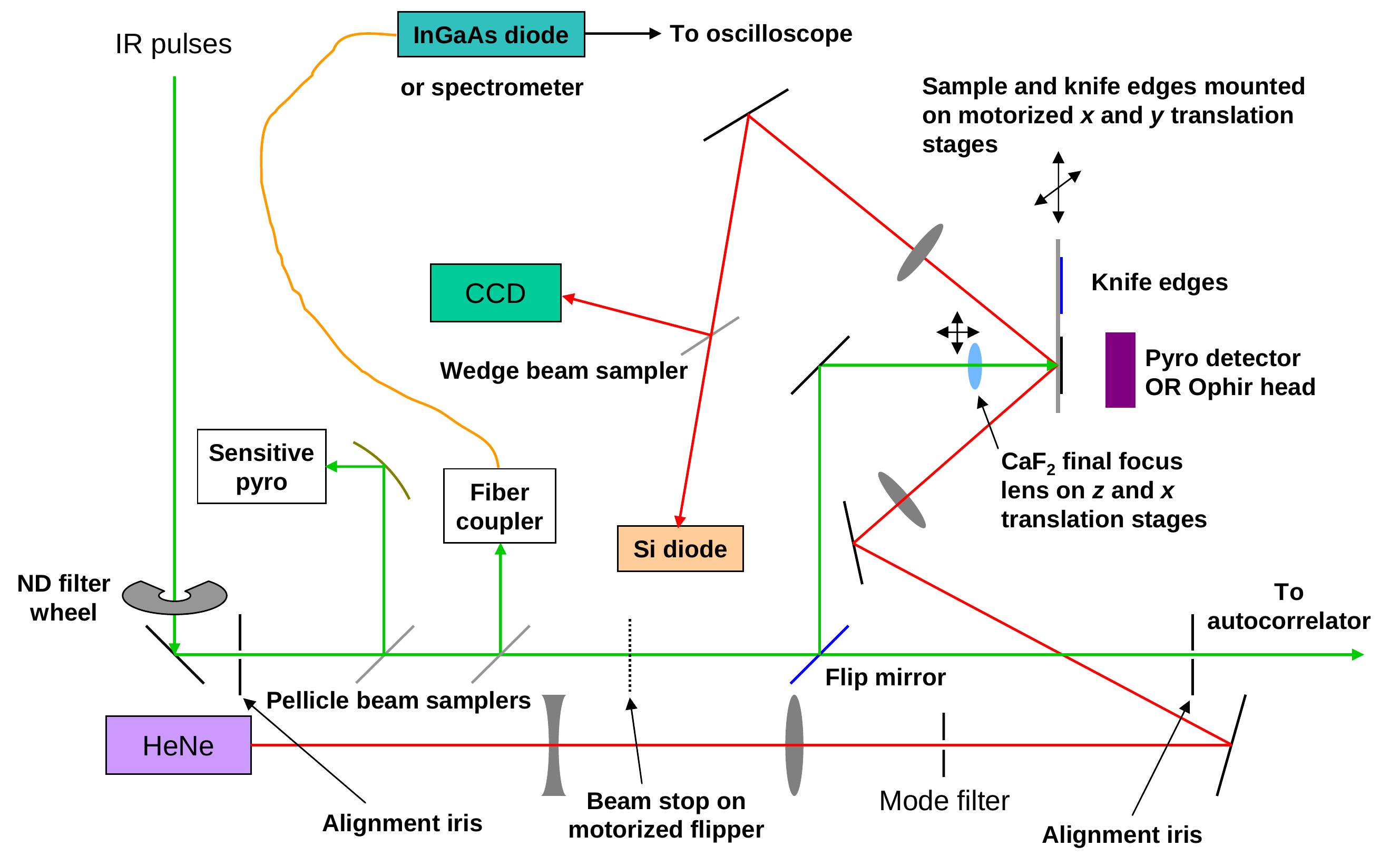}}
\caption{Schematic of the damage threshold measurement experiment.}
\label{fig:Schematic}
\end{center}
\end{figure}

This was a pump-probe measurement in which a CW helium-neon laser was
focused on the same spot on the sample as the infrared pulses and
damage was detected by observing a decrease in reflected HeNe
intensity.  A schematic of the experiment is shown in
Fig.~\ref{fig:Schematic}.
The experiment included the diagnostics necessary to measure both the
pulse energy and the extent of the pulse in all three spatial
dimensions, in order to obtain the energy density.  We also measured
the spectral content of the pulse to confirm that no residual light at
shorter wavelengths was present.

The infrared pulses were generated by a commercial Spectra-Physics
OPA-800 optical parametric amplifier (OPA) pumped by a Ti:sapphire
laser system.  The OPA produced pulses with energy
$\geq\unit[20]{\micro J}$ at a repetition rate of \unit[240]{Hz}; the
experiment detected multiple-shot damage.  The pulses first passed
through a neutral density filter to control the intensity.  They then
passed through several diagnostic pellicle beam samplers.  The first
of these directed a small fraction of the pulse energy into a
pyroelectric detector that served as the main pulse energy diagnostic.
An off-axis parabolic mirror was used to focus the pulse just enough
to ensure that all the energy was captured by the detector.  The
second beam pick-off directed a small amount of the pulse energy into
a multimode fiber.  This was connected to an InGaAs photodiode to
optimally align the fiber, and then to an optical spectrum analyzer
(OSA) to measure the spectrum.

\begin{figure}
\begin{center}
\includegraphics{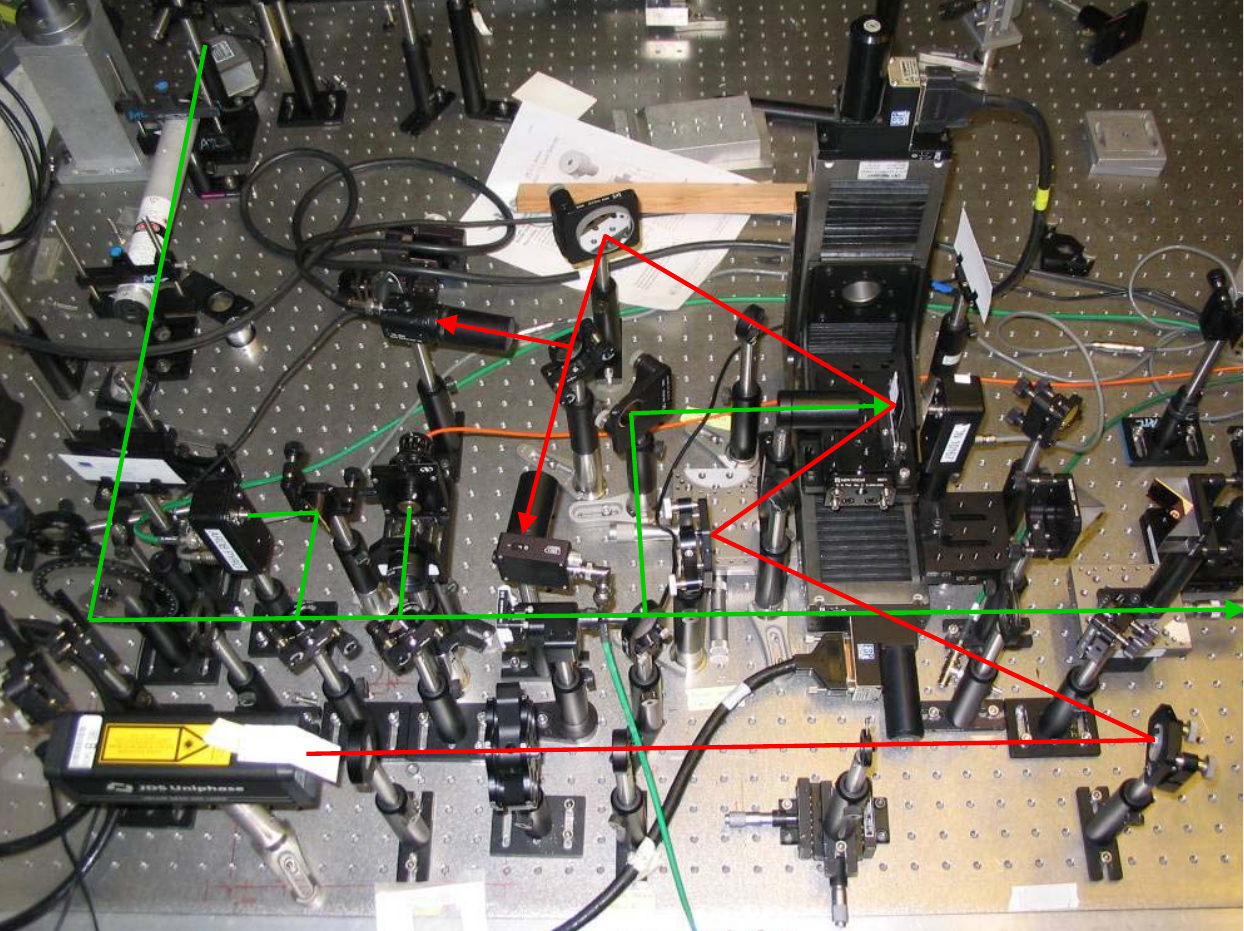}
\caption{Photo of the damage threshold measurement experiment.  The
path of the infrared pulses is drawn in green, and the HeNe path in
red.}
\label{fig:Picture}
\end{center}
\end{figure}

After the diagnostic pick-offs the pulses were directed by a flip
mirror toward the sample.  They were normally incident on the sample,
focused by a CaF\tsub{2} lens to minimize dispersion, while the HeNe
beam was incident at an angle.  The sample was mounted vertically on
motorized translation stages with motion in the plane of the sample,
and was oriented during initial setup so that it was parallel to the
directions of motion of the stages.  Razor blades were mounted in the
same plane as the sample to conduct knife-edge transverse spot size
measurements.  A photograph of the setup is shown in
Fig.~\ref{fig:Picture}.

\begin{figure}
\begin{center}
\resizebox{\textwidth}{!}{\includegraphics{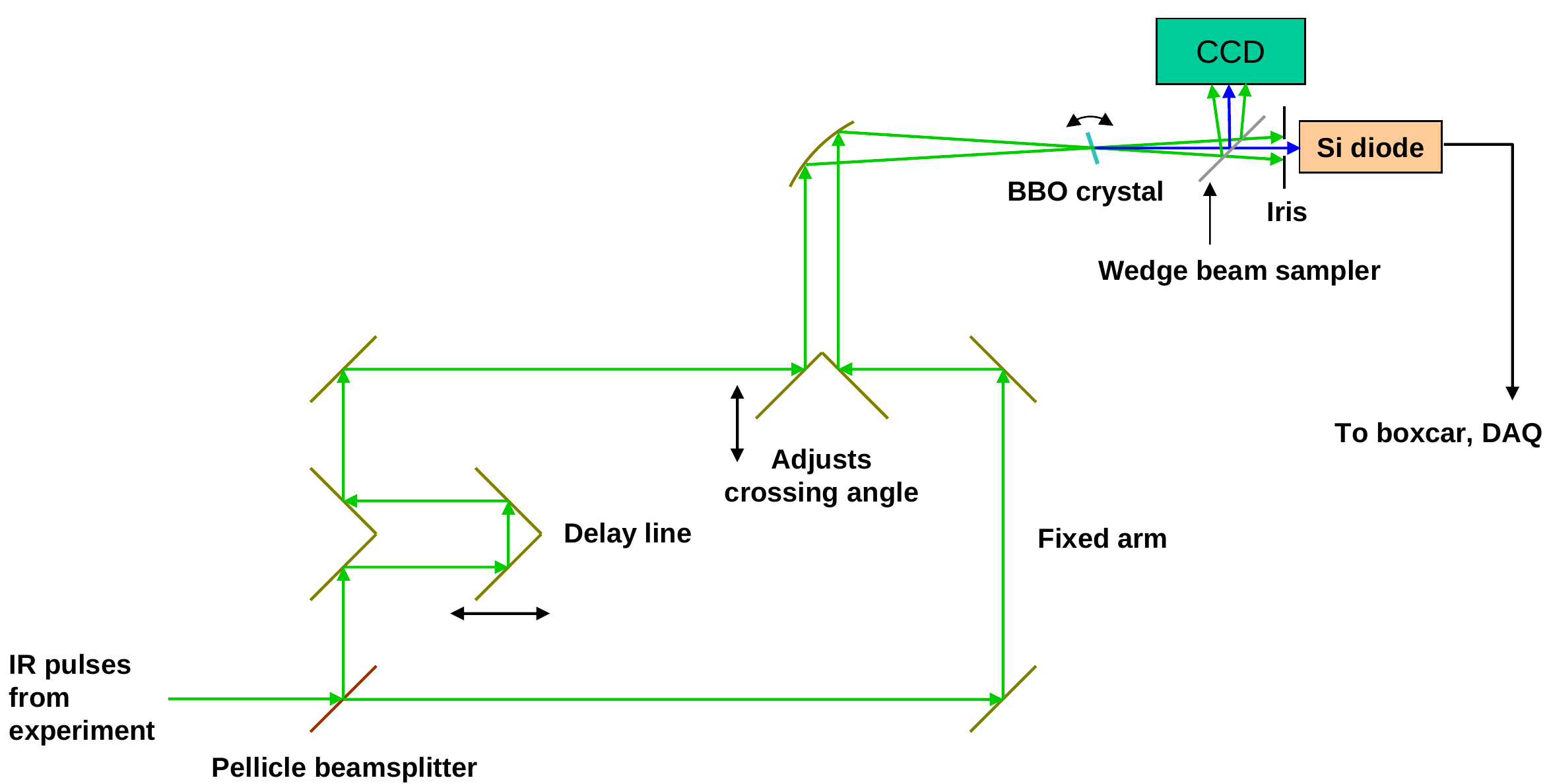}}
\caption{Schematic of the autocorrelator.}
\label{fig:AutocorrelatorSchematic}
\end{center}
\end{figure}

\begin{figure}
\begin{center}
\resizebox{!}{3.5in}{\includegraphics{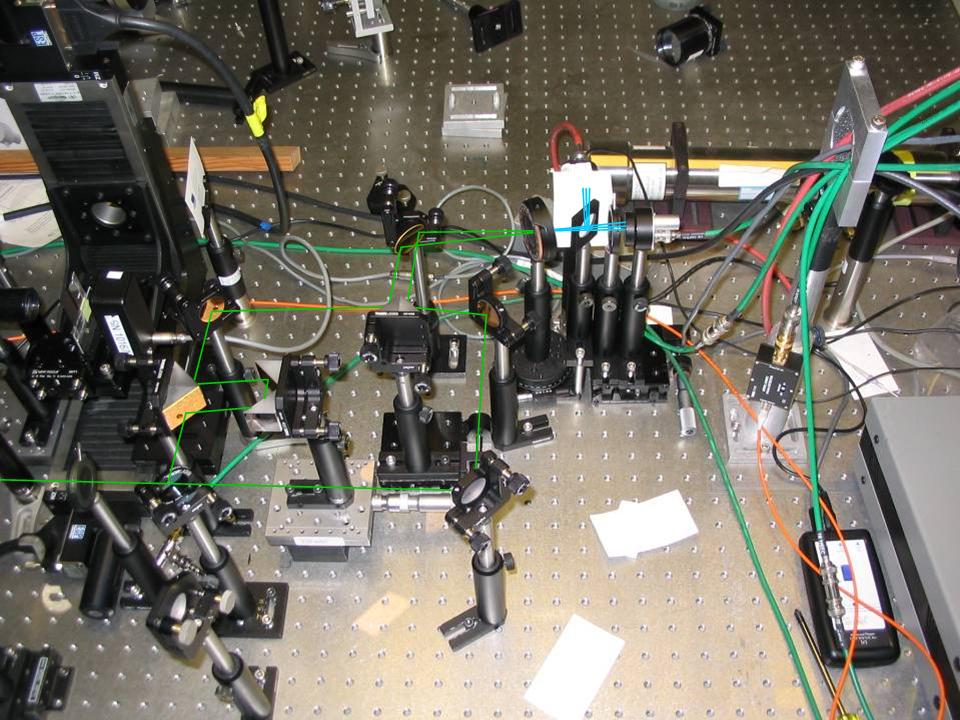}}
\caption{Photo of the autocorrelator setup.  The infrared pulses are
shown in green, while doubled light is shown in blue.}
\label{fig:AutocorrelatorPicture}
\end{center}
\end{figure}

With the flip mirror down, the pulses would proceed to an
autocorrelator setup used for measuring the pulse duration.  A
schematic of the autocorrelator is shown in
Fig.~\ref{fig:AutocorrelatorSchematic}.
The infrared pulses were split into two arms of an interferometer with
a pellicle beamsplitter, and one arm had an adjustable delay.  After
passing through their respective paths, the beams were directed
parallel to one another and then into an off-axis parabolic mirror.
This focused and crossed the beams at the same position on a nonlinear
barium borate (BBO) crystal, which was mounted on a rotation stage to
adjust the angle of its optic axis.  At the proper angle, pulses of
double the frequency (shown in blue in
Fig.~\ref{fig:AutocorrelatorSchematic}) were produced between the two
crossed beams, as long as the pulses were coincident on the crystal.
A beam sampler was used to direct some of the doubled light onto a CCD
to optimize the alignment.  The remainder was captured by a
photodiode, whose signal was integrated in a Stanford Research Systems
SR250 gated integrator, and then passed to the data acquisition system
(DAQ) for recording.  We obtained an autocorrelation trace by varying
the delay in one arm of the interferometer, and thus the temporal
overlap of the beams on the crystal, and observing the resulting
signal on the diode.  A picture of the autocorrelator is shown in
Fig.~\ref{fig:AutocorrelatorPicture}.

Our data acquisition system consisted of a Dell Precision 330 computer
with a \unit[1.8]{GHz} Pentium 4 processor and \unit[1]{GB} RAM.  It
contained a Newport ESP6000 motion controller card and a National
Instruments PCI-MIO-16E-4 data acquisition board.  The motion
controller was used to move, and read back position data from, the
translation stages on which the sample was mounted.  The DAQ board was
used to acquire analog voltage data from the pyroelectric detectors
and photodiodes, as well as control a motorized flipper used as a beam
stop to begin and end sets of damage data.  LabVIEW software was used
to automate the data taking and perform initial processing of the
data.  Our pyroelectric detectors output signals on the time scale of
\unit[100]{\micro s}, slow enough so that an entire trace could be
captured by the \unit[100]{kHz} ADCs in the DAQ board.  The signal
level was taken to be the maximum value of the pyro voltage during the
trace.

\begin{figure}
\begin{center}
\includegraphics{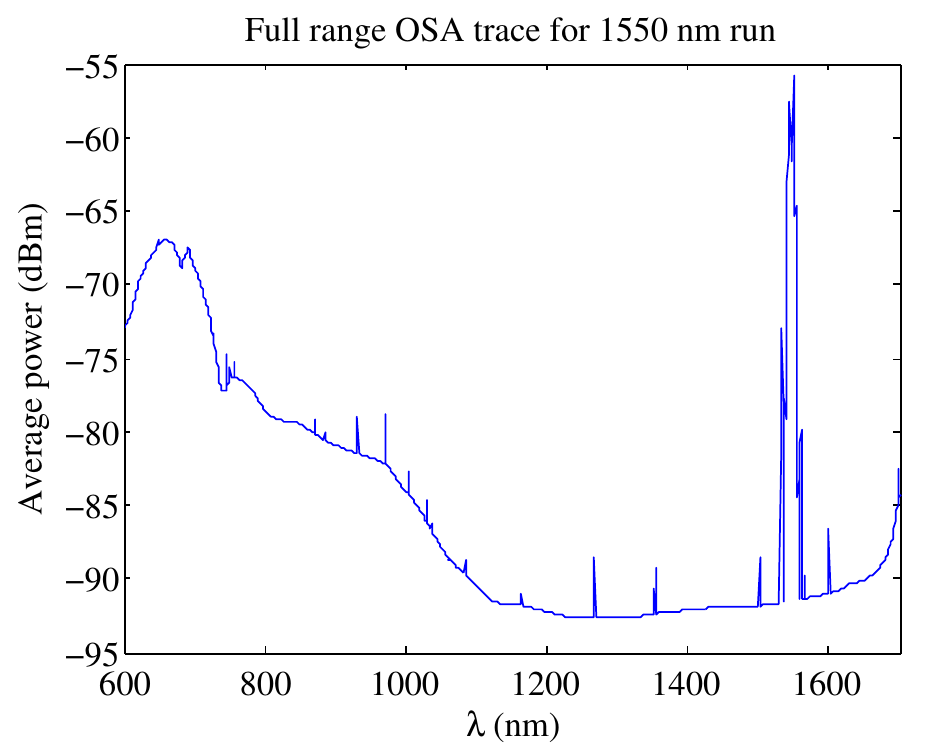}
\caption{An OSA trace for the \unit[1550]{nm} run covering the full
range of the spectrum analyzer.}
\label{fig:Spectrum}
\end{center}
\end{figure}

For each wavelength tested, we first measured the spectral content of
the pulses to confirm the absence of shorter wavelength light.  Such
light could have come from the Ti:sapphire pump at \unit[800]{nm}, the
signal wavelength when the idler was desired, or in the visible range
from the pump mixed with the signal or the idler within the OPA.  The
beam was coupled via fiber to an optical spectrum analyzer (OSA).  We
first acquired a spectrum over the entire \unit[600--1700]{nm} range
of the OSA; a sample spectrum from a \unit[1550]{nm} run is shown in
Fig.~\ref{fig:Spectrum}.
The noise floor in the spectrum shown is determined by the sensitivity
of the OSA.  For higher sensitivity, we also acquired spectra over
smaller ranges centered on wavelengths where shorter-wavelength
contamination might have been present.  Since we expect that such
contamination would have been narrow-band, centered only at the
wavelengths of light produced within the OPA, this was sufficient to
confirm the absence of additional wavelengths in the pulses.  In each
case we verified that the level of any shorter wavelength light was at
least \unit[30]{dB} less than the wavelength of interest.

\begin{figure}
\begin{center}
\includegraphics{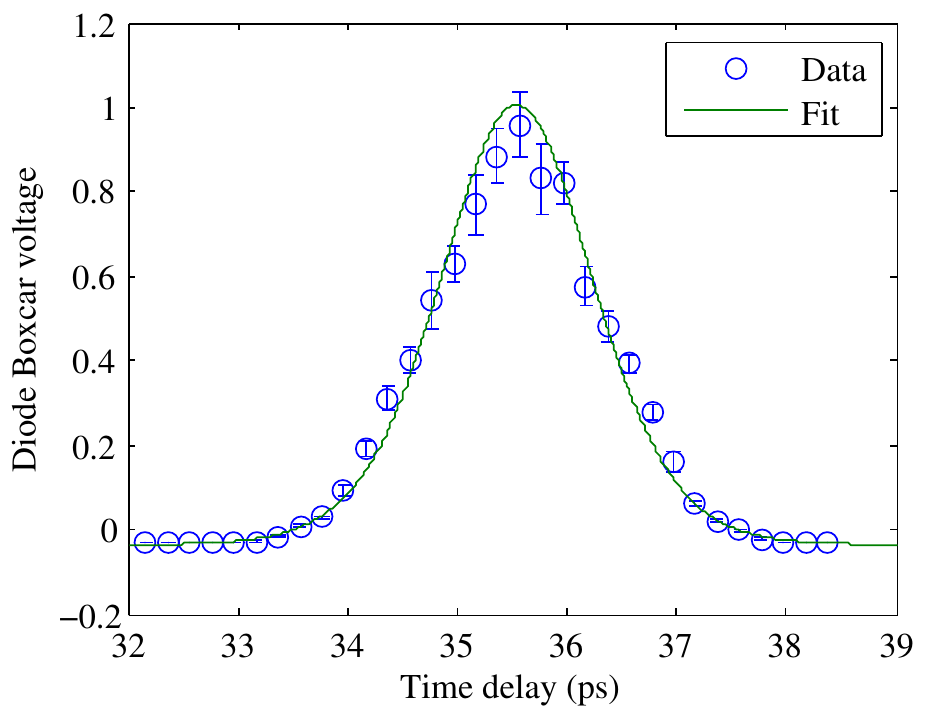}
\caption{The autocorrelation trace for the \unit[1550]{nm} run.}
\label{fig:AutocorrelationTrace}
\end{center}
\end{figure}

Next, we took an autocorrelation trace to measure the pulse duration.
We adjusted the delay arm of the autocorrelator in a pseudorandom
sequence in order to prevent slow pulse energy drifts from causing a
systematic error.  While the delay stage was manual, the DAQ prompted
the user to set each delay at the proper micrometer reading in order
to form the pseudorandom sequence.  For each delay, we acquired diode
levels for 1000 shots.  We then fit the data to an autocorrelation
profile, assuming a $\sech^2$ temporal profile of the incident pulse,
as is the case for passively modelocked lasers \cite{Siegman:Lasers}.
The autocorrelation trace for the \unit[1550]{nm} run, along with the
fit, is shown in Fig.~\ref{fig:AutocorrelationTrace}.
From the fit we deduce a FWHM pulse duration of $\unit[1.06\pm
0.01]{ps}$.

\begin{figure}
\begin{center}
\includegraphics{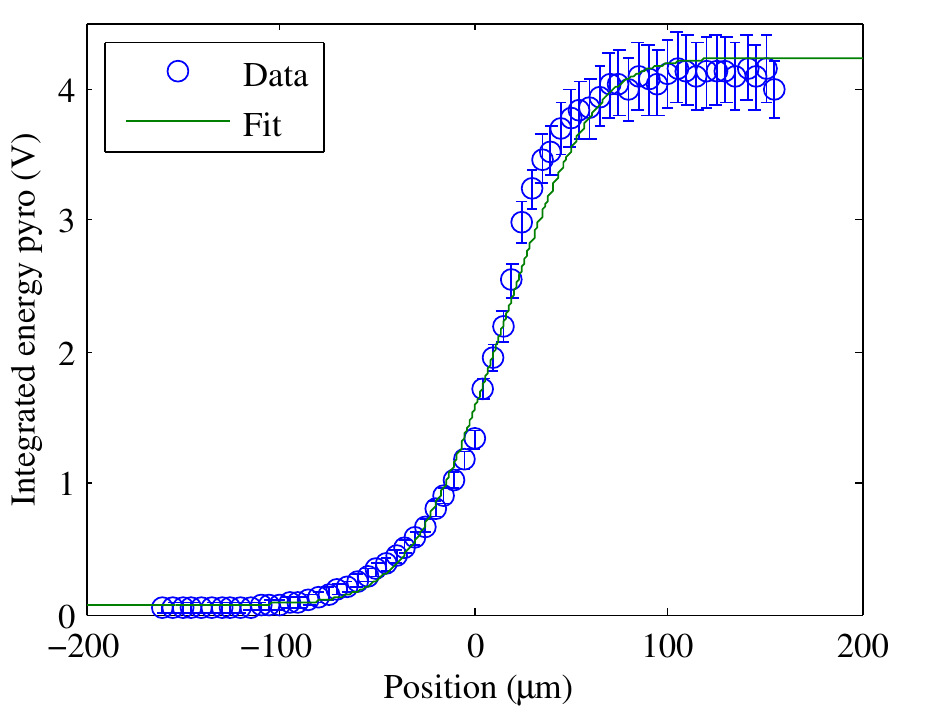}
\caption{The horizontal knife edge measurement for the \unit[1550]{nm} run.}
\label{fig:KnifeEdge}
\end{center}
\end{figure}

After the autocorrelation trace was taken, the final focus lens was
adjusted to place the beam waist just in front of the sample surface.
This was to ensure that maximum fluence occurred on the sample surface
rather than in the bulk, so that the transverse spot-size measurements
would yield relevant results.  Also, the HeNe beam was spatially
overlapped with the infrared pulses and one damage spot was created to
confirm that the pulse energy was sufficient and the focus was tight
enough for damage to occur.  Once this was complete, the transverse
spot sizes were measured using the knife-edge technique.  This was
accomplished using razor blades glued to the same microscope slide as
the silicon sample so that their edges were in the same plane as the
sample surface.  The platform holding the sample and the razor blades
was mounted on a pair of Newport M-UTM150CC1DD motorized linear stages
to allow for motion in both directions in the plane of the sample.  A
Molectron P1-45 pyroelectric detector was placed behind the knife
edges to capture any unblocked light.  A LabVIEW VI was used to
automatically move the stages, in one dimension at a time, to adjust
the position of a knife edge in pseudorandom order.  At each position,
we acquired pyro signal levels for 1000 shots.  The horizontal knife
edge measurement for the \unit[1550]{nm} run is shown in
Fig.~\ref{fig:KnifeEdge}.
We fit the data to an error function, which is the curve that results
from integrating a Gaussian spot intensity profile.  For this
knife-edge scan, we acquire a spot width of $w_x = \unit[74.3\pm
2.2]{\micro m}$.

As a final setup step, we calibrated the sensitive pyroelectric
detector used to measure pulse energy, also a Molectron P1-45.  This
was necessary for each wavelength because the reflectivity of the
pellicle beam sampler varied with wavelength.  We calibrated the
Molectron detector against an Ophir Optronics PE10 energy detector;
since the Ophir detector had an absolute calibration in our wavelength
range, this allowed us to have an absolute calibration of the
Molectron detector at each wavelength.  We placed the Ophir detector
behind the sample and used the stages to remove the sample from the
beam path.  This ensured that the Ophir detector was reading the pulse
energies after the beam passed through its transport through the
sample and therefore was measuring the actual incident pulse energies.
We acquired approximately 1200 shots on both the Ophir detector and
the Molectron sensor for a range of energies.  A linear fit was then
performed to obtain the calibration.

\subsection{Data analysis and results}

Once the setup was complete for a particular wavelength, damage data
were taken.  For each event, the pulses were allowed to illuminate the
sample by removing a beam stop, and infrared pulse energy and
reflected HeNe power were then acquired on a shot-to-shot basis.
Since the data acquisition rate was limited to
$\approx\unit[100]{Hz}$, less than the repetition rate of the laser,
not all samples were acquired.  The acquisition was stopped, and the
beam stop reinserted, when either the HeNe power decreased, indicating
damage, or a certain amount of time, usually $\geq\unit[100]{s}$, had
elapsed with no damage.  Each set of pulse energy and HeNe power data,
taken between the time the beam stop was initially removed and the
time the acquisition was stopped, constituted one ``event.''  Events
were taken both above and below the damage threshold, and the sample
was moved at least \unit[1]{mm} between each event to avoid geometric
deformities from one damage spot affecting subsequent measurements.

\begin{figure}
\begin{center}
\includegraphics{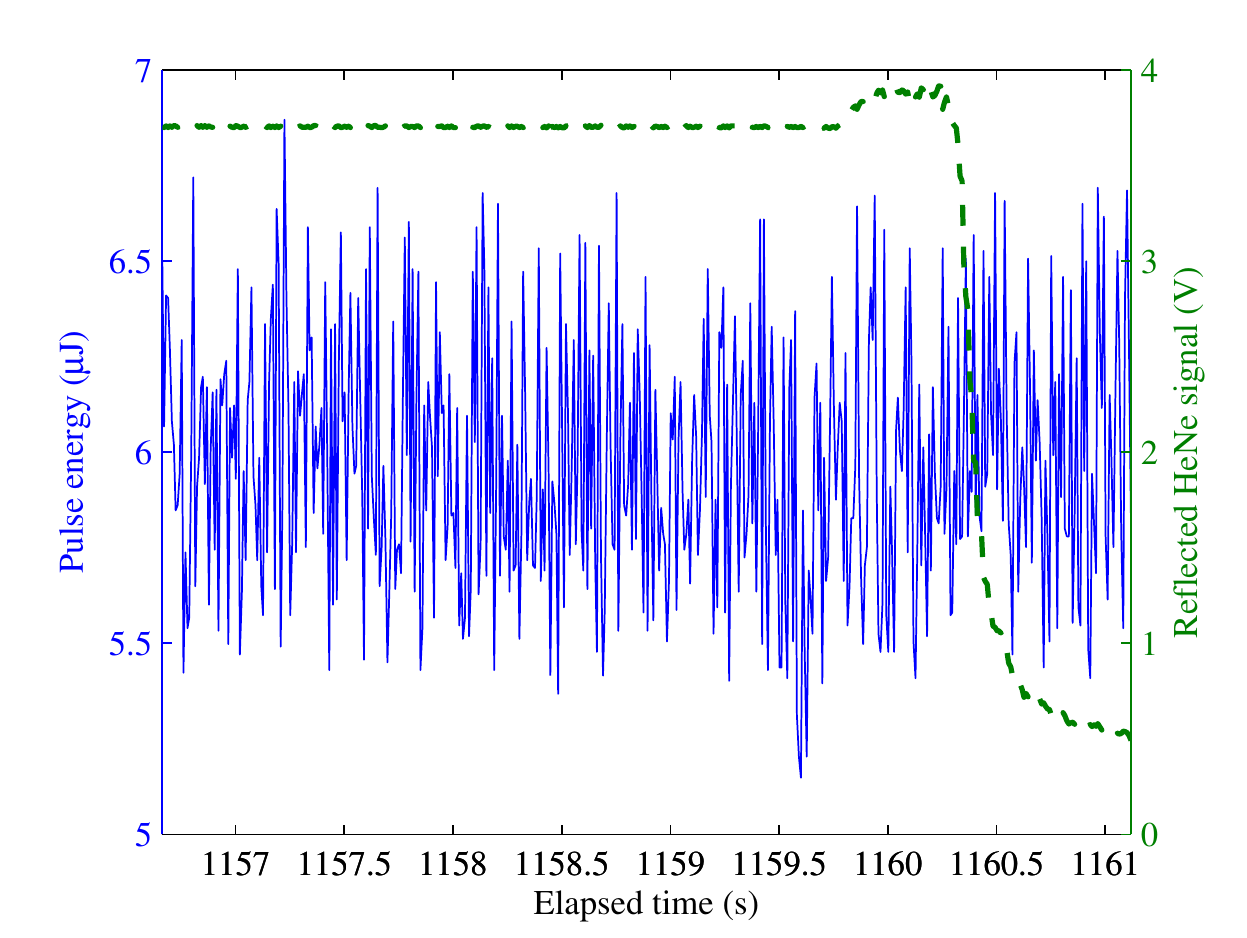}
\caption{A sample event.  Shown here are data from the last several
  seconds of an event in which damage occurred after about
  \unit[1160]{s}.  The solid line shows a trace of the acquired pulse
  energy, and the dashed line shows the reflected HeNe power.}
\label{fig:Event}
\end{center}
\end{figure}

A sample event is plotted in Fig.~\ref{fig:Event}.
We notice that the reflected HeNe power increases for a fraction of a
second before falling off.  This is ostensibly due to additional
focusing from the silicon surface as the damage morphology develops.
Indeed, we observed using a CCD image of the reflected HeNe light that
the mode pattern changes for about a second during damage before
finally disappearing.

From the multiple-shot data in this event and others like it, it is
not immediately clear how to compute the measured damage threshold.
We assume that damage was initiated by a single pulse, with further
damage occurring in each subsequent shot due to the field enhancement
resulting from the initial deformation of the surface.  That pulse may
or may not have been acquired, and the data do not tell us how long
before the visible onset of damage the pulse occurred.  However, we
can make a maximum likelihood estimate for the damage threshold based
on a few assumptions.  First, the damage process is deterministic, as
reported in \cite{Joglekar:Damage2003}, so damage did occur after a
pulse that exceeded the threshold and did not otherwise.  Second, at
least one pulse with energy above threshold therefore must have
occurred within the 1000 acquired shots before the visible onset of
damage.  Third, no such shot occurred more than 1000 shots before the
visible onset of damage, or more than 1000 shots before the end of an
event in which no damage was observed.

Using these assumptions, we form a maximum likelihood estimate of the
damage threshold as follows.  The data for each event include a
sequence of $N$ acquired pulse energies, $\{U_i\}_{i=1}^N$, as well as
a sequence of reflected HeNe diode voltages acquired simultaneously.
If damage occurred during the event, we define the ``damage index''
$i_d$ to be the index of the first shot in which the reflected HeNe
diode read more than $\unit[0.5]{V}$ below the initial reading at the
start of the event.  We then divide the event into ``blocks'' of 1000
acquired events as follows: If damage occurred, we have assumed that
the shot that initiated it occurred within the 1000 acquired shots
before $i_d$.  The index of the last shot where we know damage did not
occur is therefore $i_1 = \max(i_d - 1000, 0)$.  If $i_1 = 0$, then
there is no shot we can be sure did not cause damage.  We call the set
of pulse energies $\{U_i\}_{i = i_1 + 1}^{i_d}$ the ``damage block.''
We then divide the acquired pulse energies which we know did not cause
damage into ``no-damage blocks'' of 1000 pulses each, starting with
the first acquired pulse.  If no damage occurred during the event, we
let $i_1 = N$; then regardless of whether damage occurred there are
$\floor{i_1/1000}$ no-damage blocks in the event.

Because we did not acquire the energy of each incident pulse, we do
not use the acquired pulse energies directly to form our damage
threshold estimate.  Rather, we use the acquired pulse energies to
form a statistical description of each block, and use those
statistical parameters to estimate the threshold.  To obtain the
number of incident pulses in a block, we need to know the ratio $r$ of
incident to acquired pulses.  The DAQ records the total elapsed time
$T$ for the event; then for the known laser repetition rate
$f\tsub{rep}$ we have $r = f\tsub{rep}T/N$.  For each block $B$, we
fit the pulse energies in that block to an asymmetric Gaussian
distribution, with peak $\mu$ and RMS widths $\sigma_1$ and $\sigma_2$
for $U < \mu$ and $U > \mu$ respectively.  We ignore the pulse
energies with $U < \mu$, and consider the distribution of pulse
energies in a block to be a one-sided Gaussian with peak $\mu_B = \mu$
and RMS width $\sigma_B =
\sigma_2$.  We can justify this because if damage occurred, it is
unlikely that it was a pulse with such low energy that caused it, when
pulses with higher energies were present.  Conversely, if damage did
not occur, the fact that a relatively low-energy pulse did not cause
damage yields little information.  The probability density function
for pulse energy $U$ within block $B$ is therefore
\begin{equation}
f(U) = \begin{cases}
\sqrt{\frac{2}{\pi}}\frac{1}{\sigma_B}e^{-(U - \mu_B)^2/2\sigma_B^2}
& U\ge\mu_B, \\
0 & \text{otherwise}.
\end{cases}
\label{eq:PulseEnergyPDF}
\end{equation}
We estimate the number of relevant incident pulses within the block to
be $N_B = r|\{U\in B\mid U\ge\mu_B\}|$.

We can now form a likelihood function from these statistical
parameters.  For a given block $B$, let $P_1(U\tsub{th})$ be the
probability that a single incident pulse from block $B$ does not cause
damage if the pulse energy damage threshold is $U\tsub{th}$.  This is
the probability that a pulse energy $U < U\tsub{th}$.  From
Eq.~(\ref{eq:PulseEnergyPDF}) we have that $P_1(U\tsub{th}) = 0$ if
$U\tsub{th}\le\mu_B$, and that for $U\tsub{th}\ge\mu_B$,
\begin{align*}
P_1(U\tsub{th}) &= \int_{\mu_B}^{U\tsub{th}}
\sqrt{\frac{2}{\pi}}\frac{1}{\sigma_B}
e^{-(U - \mu_B)^2/2\sigma_B^2}\,dU \\
&= \erf\paren{\frac{U\tsub{th} - \mu_B}{\sqrt{2}\sigma_B}}.
\end{align*}
Then the probability $P_B(U\tsub{th})$ that no damage occurs within
the block is the probability that none of the incident pulses initiate
damage, or
\[ P_B(U\tsub{th}) = P_1(U\tsub{th})^{N_B} = \begin{cases}
\brckt{\erf\paren{\frac{U\tsub{th} - \mu_B}{\sqrt{2}\sigma_B}}}^{N_B}
& U\tsub{th}\ge\mu_B, \\
0 & \text{otherwise}.
\end{cases} \]
This equation makes physical sense, since not only is damage more
likely to occur if the mean pulse energy is increased, it is also more
likely to occur if the pulse energy jitter increases.  Increased
jitter makes it more likely that there will be a pulse above damage
threshold, underscoring the importance of stable laser sources for
accelerator applications.

To compute the likelihood function we use all the blocks acquired
during the course of a run at the given wavelength.  Let
$\mathcal{B}_d$ be the collection of damage blocks for the run, and
let $\mathcal{B}_{nd}$ be the collection of no-damage blocks.  Our
likelihood function is then
\[ L(U\tsub{th}) = \prod_{B\in\mathcal{B}_{nd}} P_B(U\tsub{th})
\cdot \prod_{B\in\mathcal{B}_d} \brckt{1 - P_B(U\tsub{th})}. \]
To find the value of $U\tsub{th}$ with maximum likelihood, we compute
the negative log likelihood, given by the function
\[ \chi^2(U\tsub{th}) = -\sum_{B\in\mathcal{B}_{nd}} \log P_B(U\tsub{th})
- \sum_{B\in\mathcal{B}_d} \log\brckt{1 - P_B(U\tsub{th})}. \]
We compute the $\chi^2$ function numerically, and use the \textsc{matlab}
minimization routine \texttt{fminbnd} to find the value of
$U\tsub{th}$ which minimizes $\chi^2$ as well as the minimum value
$\chi^2_0$.  We then use the routine \texttt{fzero} to find the pulse
energies $U_+$ and $U_-$ above and below $U\tsub{th}$, respectively,
for which $\chi^2(U_+) = \chi^2(U_-) = \chi^2_0 + 1$, and defined the
error on $U\tsub{th}$ to be $(U_+ - U_-)/2$.  For instance, for the
run at $\lambda = \unit[1550]{nm}$, the threshold was $\unit[(7.78\pm
0.05)]{\micro J}$.

We now have the information required to reconstruct the damage
threshold energy density in the material.  For a pulse with peak
intensity $I_0$, the energy density inside the material is given by
\[ u = \frac{I_0}{c}\frac{4n^2}{(n + 1)^2}, \]
where $n$ is the index of refraction; the factor of $4n^2/(n+1)^2$
takes into account Fresnel reflection at the surface and the slower
speed of light in the material.  For our pulses we have assumed an
intensity distribution in space and time given by
\[ I = I_0e^{-2(x^2/w_x^2 + y^2/w_y^2)}\sech^2\paren{\frac{t}{\tau}},
\]
and obtained fits to the parameters $w_x$, $w_y$, and $\tau$ from our
knife-edge and autocorrelation measurements.  The total pulse energy
is then $U = \pi w_xw_y\tau I_0$, so we have as our damage threshold
energy density
\[ u\tsub{th}
= \frac{U\tsub{th}}{\pi w_xw_y\tau c}\frac{4n^2}{(n + 1)^2}. \]

\begin{figure}
\begin{center}
\includegraphics{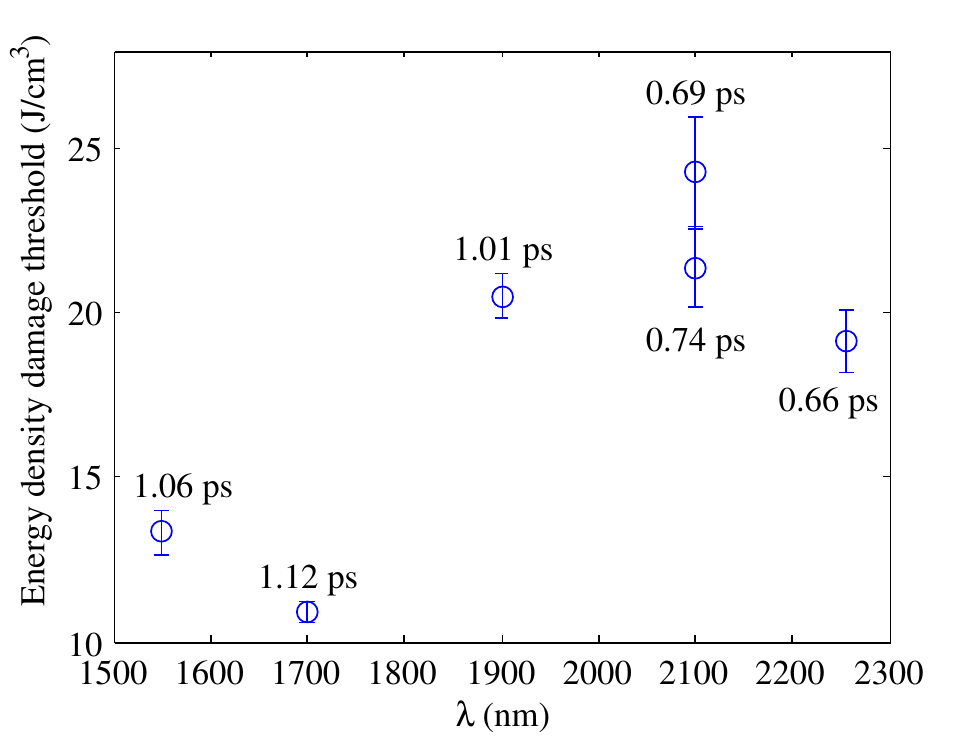}
\caption{Damage thresholds for a range of wavelengths.  The measured FWHM
  pulse width at each wavelength is shown next to the corresponding
  point.}
\label{fig:Thresholds}
\end{center}
\end{figure}

\begin{figure}
\begin{center}
\includegraphics{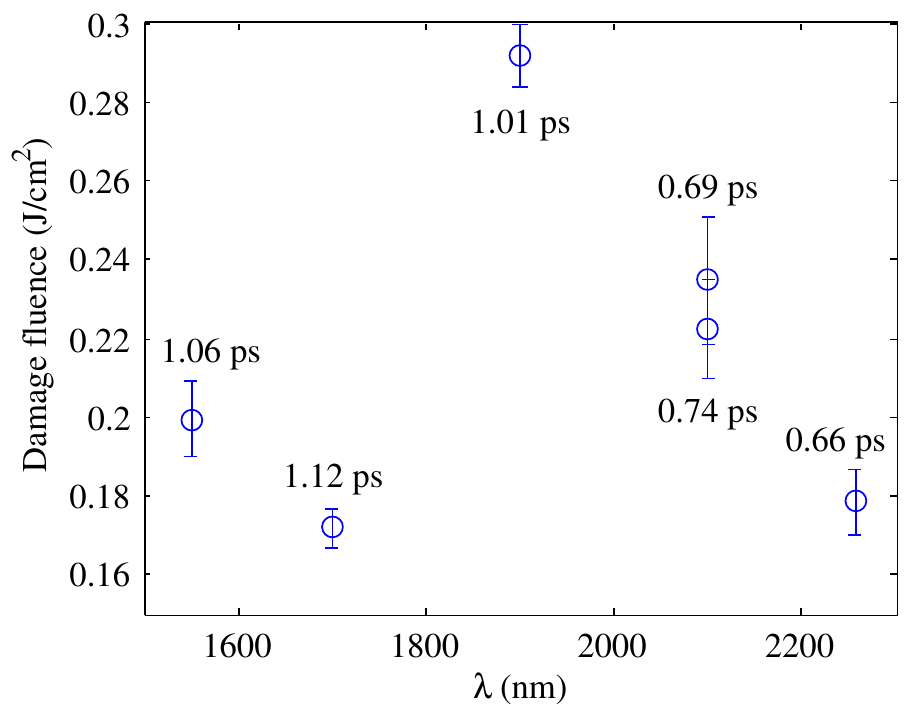}
\caption{The damage threshold values in terms of incident pulse fluence.}
\label{fig:DamageFluence}
\end{center}
\end{figure}

\begin{figure}
\begin{center}
\includegraphics{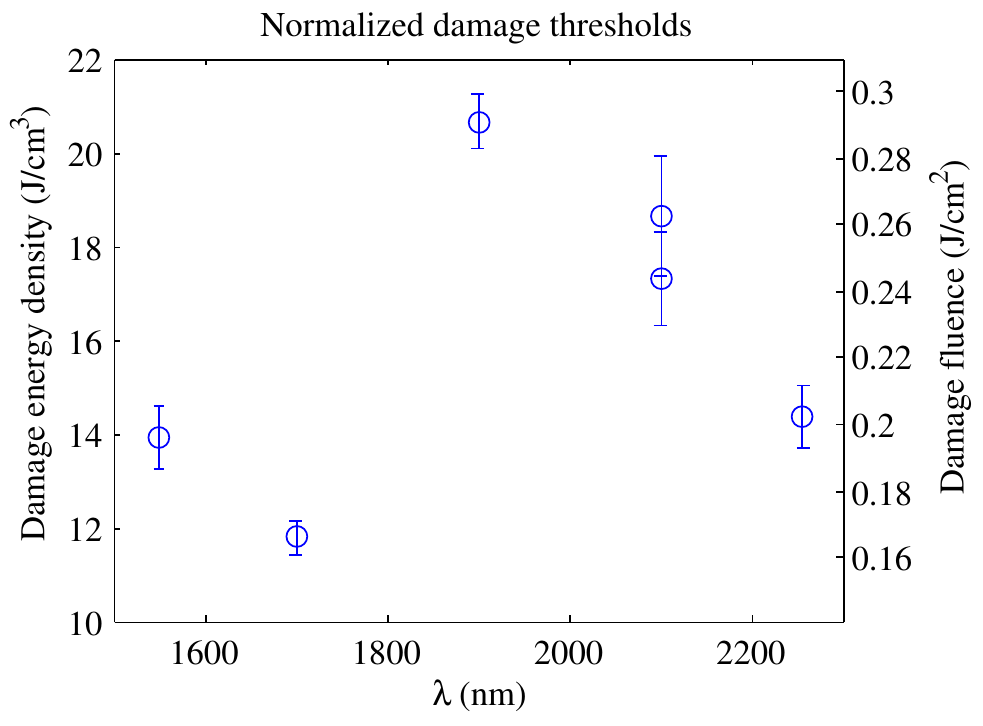}
\caption{Damage thresholds, normalized for pulses with
  \unit[1]{ps} FWHM duration.  Scales are shown for both energy
  density and fluence; with the normalization to a fixed pulse
  duration they are proportional to one another.}
\label{fig:NormalizedThresholds}
\end{center}
\end{figure}

We obtained data sets for $\lambda = \unit[1550]{nm}$,
\unit[1700]{nm}, \unit[1900]{nm}, \unit[2100]{nm}, and \unit[2256]{nm}.  The
damage threshold results are plotted in
Fig.~\ref{fig:Thresholds}.
The data point at \unit[2100]{nm} was repeated several months after
the initial point was taken, in order to check the consistency of the
setup; we found the two points to be within reasonable statistical
error of one another.  Because much of the damage literature gives
damage threshold values in terms of the incident pulse fluence, we
present those values as well here.  The damage fluence is computed
simply as $F\tsub{th} = 2U\tsub{th}/\pi w_xw_y$ and is plotted in
Fig.~\ref{fig:DamageFluence}.
Comparison among damage thresholds at different wavelengths is
complicated by the fact that each measurement was taken using a
different pulse duration.  We can attempt to remove the pulse
duration dependence using the empirical scaling law of
$F\tsub{th}\sim\tau^{0.3}$ reported in \cite{Mero:ScalingLaw}.
However, we should note that the result in \cite{Mero:ScalingLaw} was
determined for oxide thin films and might not be applicable to
silicon.  The damage energy density and fluence,
normalized for pulses with \unit[1]{ps} FWHM duration, are plotted in
Fig.~\ref{fig:NormalizedThresholds}.

We first notice that the damage threshold of silicon at these
wavelengths is low compared to larger-bandgap materials that have been
previously measured.  The normalized fluence thresholds of
160--$\unit[300]{mJ/cm^2}$ are an order of magnitude lower than those
for fused silica, CaF\tsub{2}, and other fluorides reported in
\cite{Stuart:BreakdownPRB1996}.  It is also several times lower than
the thresholds for the larger-bandgap semiconductors ZnS and ZnSe
reported in \cite{Simanovskii:Damage} for wavelengths of 400 and
\unit[800]{nm} and in the mid-infrared.

We also notice that the damage threshold does indeed increase as
the wavelength approaches the two-photon absorption threshold, between
\unit[1700]{nm} to \unit[1900]{nm}.  This is reasonable, since we
would not expect a sharp cutoff from an MPI-dominated process because
silicon has an indirect bandgap.  At wavelengths longer than
\unit[1900]{nm}, we then see a decrease in damage fluence.  This may be
due to the same mechanism as reported in \cite{Simanovskii:Damage}, in
which the damage threshold decreased at longer wavelengths because of
the increased probability of tunnel ionization.  Therefore, the
increase in damage threshold from \unit[1700]{nm} to \unit[1900]{nm}
is not dramatic enough to indicate that the damage process is highly
dominated by the MPI effect.  In addition, since the increase in
threshold would only increase accelerating gradient by $\sim 30\%$,
the decision whether to use longer wavelengths outside the telecom
band in an accelerator application might be guided by other
considerations.  And because of the low breakdown threshold overall,
the choice of material remains an open question, which we discuss
further in the next chapter.

\section{Fabrication possibilities}
\label{sec:fabrication}

A key consideration in the design of optical accelerators is the
challenge of fabricating structures at optical length scales.  It is
beneficial to choose geometries which capitalize on the many
microfabrication tools and techniques developed by the integrated
circuit and MEMS industries.  Indeed, the woodpile lattice stands out
among several possible three-dimensional PBG lattices in that the
methods of fabricating the lattice and defects within the lattice are
relatively straightforward.  Straightforward, to be sure, does not
necessarily mean easy, or even possible.  In fact, reliable,
economical production of the structures described in
Ch.~\ref{ch:Woodpile} at wavelengths in the near infrared might
not yet be possible.  However, a number of promising methods for
fabricating woodpile-based structures have been explored by the optics
community, and great progress has been made.  In this section we
review some of these techniques.  As the integrated circuit industry
continues to develop methods and equipment for constructing geometries
at ever-smaller feature sizes, we can expect that fabrication of
woodpile-based optical accelerator structures will become viable.

The woodpile structure is amenable to established microfabrication
techniques because of its layered structure.  Each layer of a device
can be constructed separately using a photolithographic process, and
the structure built up layer by layer.  Thus the fabrication process
divides naturally into two parts, the construction of each layer and
the composition of multiple layers into a working device.  As we will
see, the second part of the process presents a much greater challenge
than the first.

The photolithography process generally takes place as follows
\cite{Plummer:SiliconVLSI}.  It begins with a substrate in the form of
a circular wafer of solid material, typically crystalline silicon,
100--\unit[300]{mm} in diameter, and several hundred microns thick.
First, a thin film of the material to be patterned is deposited on the
substrate.  Then, liquid photoresist is placed on the wafer, which is
then spun at high velocity to form a thin layer.  Next, the wafer is
selectively exposed to ultraviolet light by placing a mask between the
UV source and the wafer, causing a chemical change in the photoresist
in the areas exposed.  The wafer is then immersed in a developer
solution which removes the resist in the exposed areas (for a positive
resist; negative resist behaves in the opposite manner).  Finally,
when the wafer is etched, the remaining resist protects the selected
areas of the wafer from the etch process, resulting in a patterned
layer.  Equipment exists which enables efficient, cost-effective
mass-production of patterned films.

The precision of photolithographic patterning is fundamentally limited
by the wavelength of UV light used to expose the resist.  The current
generation of tools uses \unit[193]{nm} light, which is greater than
the rod width of a woodpile structure operating at \unit[1550]{nm}.
While state-of-the-art equipment can produce subwavelength features
with sizes down to \unit[30]{nm}, a highly complex imaging system is
required \cite{JonesBey:ImmersionLitho}.  However, a patterning method
has been demonstrated which can accurately produce, with simpler
equipment, layers of the woodpile structure with feature sizes well
below the resist exposure wavelength.  This technique, called ``fillet
processing,'' is described in \cite{Fleming:WoodpileOL1999}.  It
proceeds by using adhered sidewalls from a thin film deposition as an
etch mask, rather than using photoresist directly.  First, a series of
lines are created in SiO\tsub{2} with period $2a$ and width $a - w$;
those features are large enough to be patterned with optical
lithography.  Then polysilicon is deposited over the SiO\tsub{2}
lines to form a layer of thickness $w$ on the entire surface,
including the sidewalls.  Next, the polysilicon is etched
anisotropically to remove just the material adhered to the tops of the
SiO\tsub{2} lines and in the trenches between them, but not the
sidewalls.  The SiO\tsub{2} is then dissolved, leaving only the
pattern of polysilicon sidewalls, with width $w$ and period $a$.
Those sidewalls are used as an etch mask for the material beneath.
Thus the width of the rods is controlled by the thickness of the
deposited material, rather than by a photomask, overcoming the
wavelength limitation of the feature size.  The required etch depth
does not pose a challenge, since the rod width and depth are $0.28a$
and $0.35a$ respectively, so the relevant aspect ratio of etch depth
to width is less than $1/2$.  It is also possible to use electron-beam
lithography to construct the small features of each layer, but that
method is highly time consuming and therefore unsuited to mass
production \cite{McCord:EbeamLitho}.

Previous investigations of woodpile structure fabrication have also
explored different means of stacking the layers into a photonic
crystal lattice.  In \cite{Fleming:WoodpileOL1999}, the structure is
built up layer-by-layer on a single substrate.  This is accomplished
by filling the vacuum regions of each layer with SiO\tsub{2} and
planarizing in order to have a flat surface on which to construct the
subsequent layer.  When the final layer is completed, the entire
structure is then wet-etched to remove the SiO\tsub{2}.  In
\cite{Noda:WoodpileScience2000}, the wafer fusion technique was used,
in which each layer is first constructed on a different wafer, and the
layers are then bonded together and one of the substrates removed.
This method is attractive because it allows all the photolithography
steps for a single structure to occur in parallel.

Both methods of stacking individual woodpile layers into a structure
present the considerable challenge of aligning the layers with respect
to one another.  In \cite{Fleming:WoodpileOL1999} and
\cite{Noda:WoodpileScience2000}, the investigators were only
interested in producing the woodpile lattice, so 4 or 8 layers were
sufficient, and an adequate structure could be produced even with
imprecise alignment.  However, for a woodpile accelerator structure,
20--30 layers are necessary because there must be enough layers of PBG
lattice to confine a mode, so a method of achieving highly precise
alignment is required if woodpile-based accelerators are to be
realized.  The fillet processing method which reduces the feature
sizes available from optical photolithography does not similarly
improve the alignment precision.  Both e-beam and state-of-the-art
optical lithography are capable of reaching the alignment tolerances
necessary to produce a working woodpile structure, but these methods
are either highly complex or time consuming
\cite{ITRSRoadmap,McCord:EbeamLitho}.  There are more exotic
techniques, such as interferometric alignment
\cite{Yamamoto:WoodpileAlignment} and stacking by micromanipulation
\cite{Aoki:WoodpileAssembly} which hold promise for efficient,
cost-effective fabrication.  However, alignment remains the most
challenging aspect of woodpile structure fabrication, so in the next
section we examine the effects of misalignment and evaluate the
tolerances required.

\section{Tolerance studies}
\label{sec:Tolerance}

To assess the effect of layer-to-layer misalignment in the woodpile
structure, we perform a set of simulations of the structure, with the
position of each layer randomly offset in the relevant horizontal
directions: Longitudinal rods are displaced transversely, while the
transverse crossbars are displaced longitudinally, and the crossbars
in the waveguide layers are displaced transversely as well.  Since we
are studying the effect on the accelerating mode, we do not include
angular misalignments: Angular misalignments would destroy the
longitudinal periodicity of the structure, so we would in effect have
a waveguide that slowly changes geometry with longitudinal position.
This would result in scattering or radiative loss as the fields
propagate down the waveguide, rather than a uniform alteration of the
accelerating mode.  The consequences of angular misalignments remain
to be studied.

\begin{figure}
\begin{center}
\includegraphics{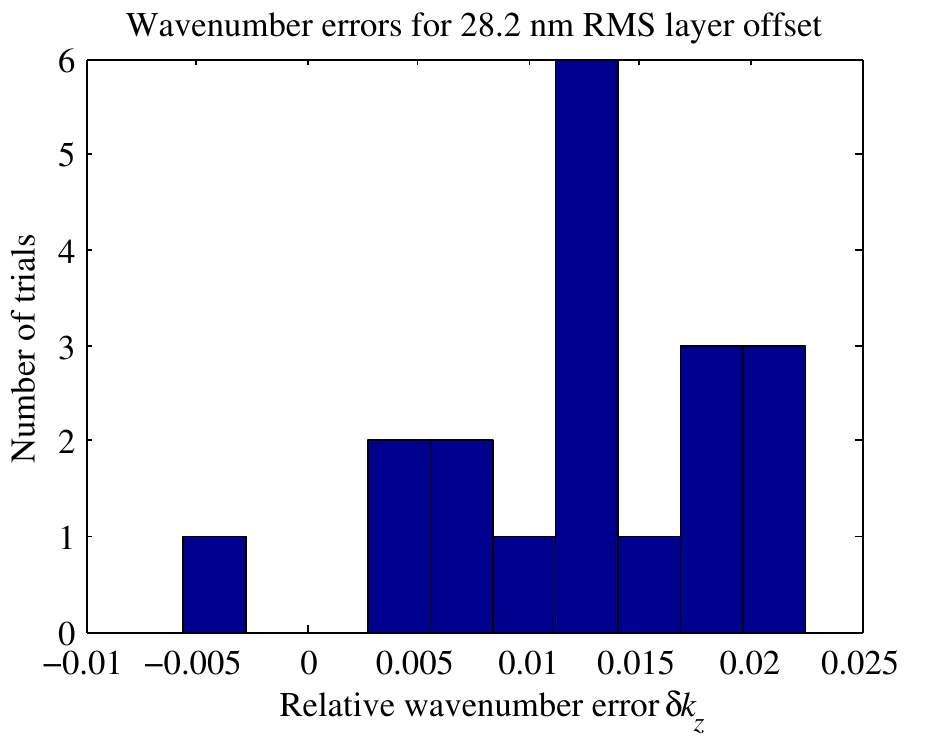}
\caption{The relative wavenumber errors for the 19 simulated
misaligned structures which supported accelerating modes.}
\label{fig:WavenumberErrors}
\end{center}
\end{figure}

We choose an RMS offset of $0.05a = \unit[28.2]{nm}$.  We performed 20
simulations, each with different offsets, using the \textsc{mpb} iterative
eigensolver described in Sec.~\ref{sec:IterativeEigensolver}.  Of
these simulations, one failed to support any accelerating mode.  For
the others, we found the accelerating mode at the speed-of-light
frequency of the perfectly aligned structure, as the source frequency
will be fixed by the bunching of the particle beam.  The key figure of
merit for a misaligned structure is the relative wavenumber error,
given by
\[ \delta k_z = \frac{k_z - k_z^{(0)}}{k_z^{(0)}}, \]
where $k_z^{(0)}$ is the wavenumber of the speed-of-light mode in an
ideal structure, and $k_z$ is the wavenumber of the accelerating mode
in the misaligned structure.  This is because if $\delta k_z$ is too
large, the fields will get out of phase by $\pi$ and become
decelerating fields.  A histogram of the relative wavenumber errors
is shown in Fig.~\ref{fig:WavenumberErrors}.
It is interesting to note that all but one of the wavenumber errors
were positive.  The cause of this is not clear, but may be related to
the second-order nature of effect of the transverse misalignments.  If
we follow the perturbation technique prescribed in
\cite{Johnson:Perturbation}, we find that the first-order frequency
shift from transverse horizontal displacements of the rods vanishes.

The fields will acquire a phase shift of $\pi$ relative to the
particle beam after length $L = \lambda/2|\delta k_z|$.  Let us
consider a structure to be acceptable if $L\ge\unit[100]{\micro m}$,
as this is the distance in which the laser pulse will slip by
\unit[1]{ps} with respect to the particle bunch, and is therefore the
characteristic length of a structure segment.  This requirement
corresponds to $|\delta k_z|\le 7.75\e{-3}$.  We find that 5 of the 20
simulated structures meet our requirement, giving a yield of 25\%.
This yield level is acceptable if we take advantage of economical
mass-production techniques for silicon microstructures.  We can
therefore quote a misalignment tolerance of the woodpile structure of
\unit[25--30]{nm}.

\chapter{Toward a photonic crystal accelerator}
In Ch.~\ref{ch:Woodpile} we described a structure suitable for
structure-based, laser-driven acceleration.  Its parameters are as follows:
\begin{itemize}
\item The underlying photonic crystal is a woodpile lattice, as shown
  in Fig.~\ref{fig:woodpileLattice}.  The rod spacing is $a =
  0.37\lambda$, the rod width is $w = 0.102\lambda$, and the rod
  height is $c/4 = 0.129\lambda$.  For a wavelength of
  \unit[1550]{nm}, this gives $a = \unit[565]{nm}$, $w =
  \unit[158]{nm}$, and $c/4 = \unit[200]{nm}$.  The lattice is
  constructed out of silicon with a relative permittivity of $12.1$.
\item The basic waveguide structure is formed by removing material
from the lattice in a region with rectangular transverse cross
section.  The waveguide dimensions are $3a - w = \unit[1.54]{\micro
m}$ horizontally by $7c/4 = \unit[1.40]{\micro m}$ vertically at
\unit[1550]{nm}; the geometry is shown in
Figs.~\ref{fig:symmetricguide} and \ref{fig:symmetricmodel}.
\item To improve particle beam stability in the structure, we modify
the geometry to suppress unwanted azimuthal moments, as described in
Sec.~\ref{sec:BeamDynamics}.  We modify the structure by inserting the
central bar into the waveguide, as shown in Fig.~\ref{fig:BarOffset}.
We obtain two structures; one for acceleration and the other for
focusing.  For the accelerating structure, the bar is inserted
$0.045\lambda = \unit[70]{nm}$ into the guide; for the focusing
structure, it is inserted $0.068\lambda = \unit[105]{nm}$.
\item The accelerating structure can sustain an unloaded gradient of
\unit[301]{MV/m} at $\lambda = \unit[1550]{nm}$.  Its characteristic
impedance is $\unit[483]{\ohm}$ and its group velocity is $0.269c$.
Therefore a laser pulse will lag the particle beam by \unit[1]{ps}
after \unit[100]{\micro m} of propagation, so that will set the length
of each accelerating segment.
\item Operated at damage threshold, the focusing structure has a
focusing gradient equivalent to an \unit[831]{kT/m} quadrupole magnet.
\end{itemize}

However, this is certainly not the only suitable photonic crystal
accelerator structure.  Indeed, there are many choices to be made to
optimize the accelerator.  These include not only the structure
geometry, but also the dielectric material and laser source.  In this
chapter we consider some of these choices.

\section{Materials and laser considerations}
\label{sec:Materials}

As we remarked in Ch.~\ref{ch:Materials}, silicon has several
properties desirable for a photonic crystal accelerator structure.
However, the damage threshold studies reported in that Chapter reveal
that silicon can sustain an accelerating gradient of only
$\sim\unit[300]{MV/m}$ at $\lambda = \unit[1550]{nm}$, improving to
$\sim\unit[400]{MV/m}$ at longer wavelengths.  We are therefore led to
question whether silicon is the most appropriate material, and what
some alternatives might be.  As described in Sec.~\ref{sec:damage},
our damage studies, combined with those in the literature, indicate
that materials with wider electronic bandgaps have a higher breakdown
threshold.  However, this involves a trade-off: A lattice with a
complete photonic bandgap generally requires a refractive index
contrast $\gtrsim 2$ \cite{Ho:Woodpile}, and high-index materials tend
to have narrow electronic bandgaps.  Silicon's high index is ideal for
photonic crystals but its narrow electronic bandgap yields a low
damage threshold.  On the other hand, fused silica, with an electronic
bandgap $\approx\unit[9]{eV}$ and a high breakdown threshold
\cite{Lenzner:Breakdown1998}, has a refractive index of only 1.44 at
\unit[1550]{nm} \cite{Malitson:FusedSilica}.  Other materials may
strike a better balance between refractive index and electronic
bandgap.  Possible candidates are rutile (birefringent; $n_o = 2.45$,
$n_e = 2.71$ \cite{DeVore:RutileIndex}, bandgap $\Delta =
\unit[3.05]{eV}$ \cite{Cronemeyer:RutileBandgap}), diamond ($n = 2.4$
\cite{Edwards:DiamondIndex}, $\Delta = \unit[5.45]{eV}$
\cite{Feldman:DiamondBandgap}), and silicon carbide ($n = 2.6$
\cite{Choyke:SiCIndex}, $\Delta = \unit[3.0]{eV}$
\cite{Shenai:SiCBandgap}).

\begin{figure}
\begin{center}
\includegraphics{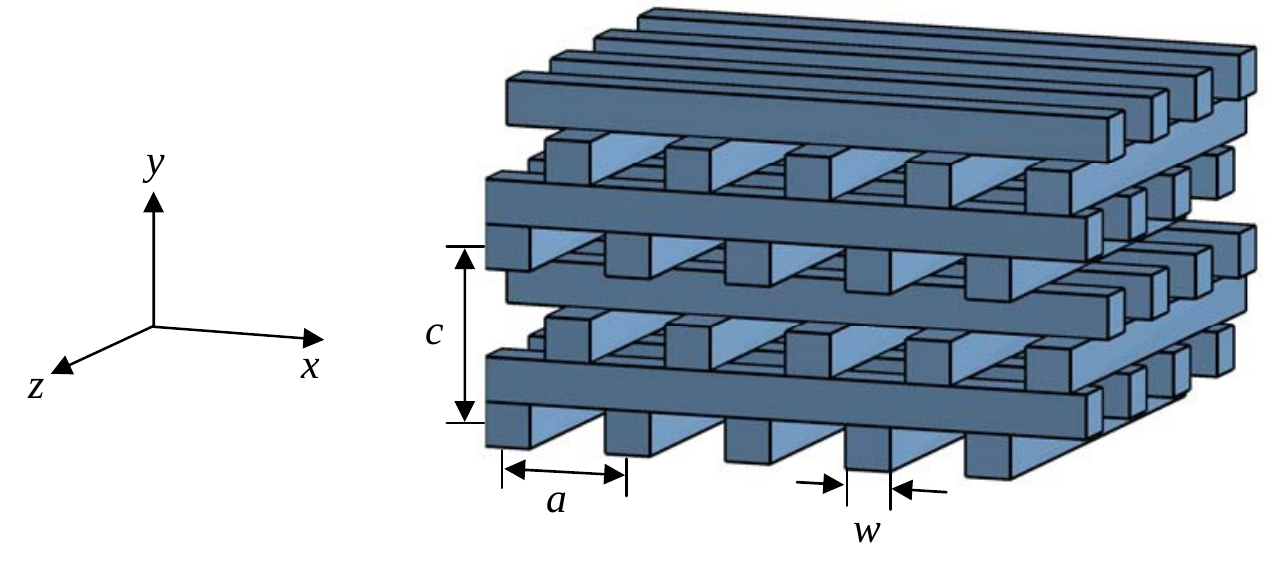}
\caption{The woodpile lattice constructed from diamond with maximal
  bandgap.}
\label{fig:DiamondLattice}
\end{center}
\end{figure}

To determine whether these lower-index materials might be suitable for
a photonic accelerator structure, we simulate a woodpile geometry
based on diamond, the lowest-index material mentioned with $n = 2.395$
at \unit[1550]{nm}.  The simulation proceeds similarly to that
presented in Ch.~\ref{ch:Woodpile}, starting with the
bandstructure computation.  The first step is to optimize the rod
width $w$ with respect to the lattice constant $a$ for maximal
bandgap.  We find that the optimum rod width occurs at $w/a = 0.37$.
This lattice with the wider rods is shown in
Fig.~\ref{fig:DiamondLattice}.

\begin{figure}
\begin{center}
\includegraphics{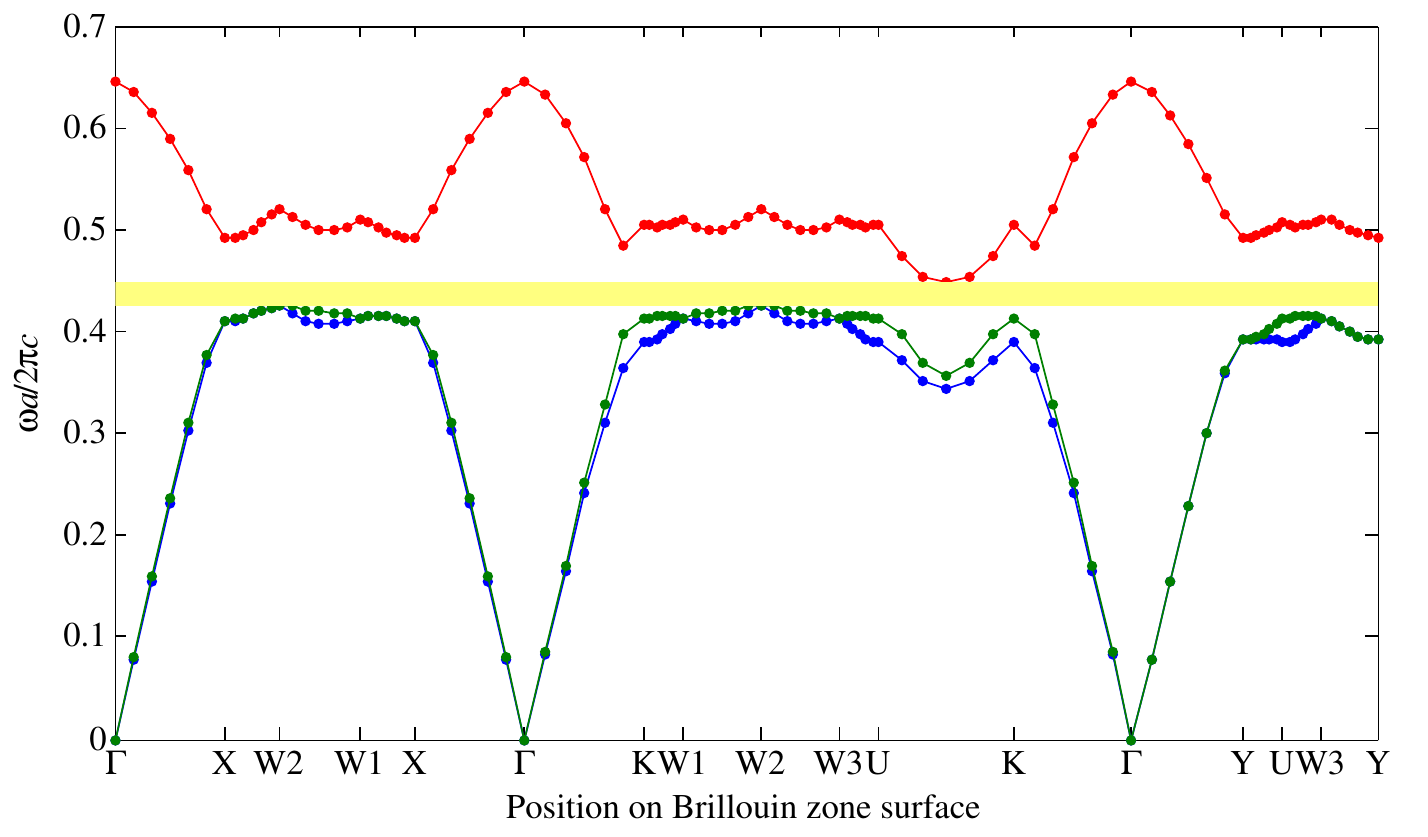}
\caption{Bandstructure of the optimal woodpile lattice constructed
  from diamond.}
\label{fig:DiamondBandstructure}
\end{center}
\end{figure}

As in Sec.~\ref{sec:WoodpileLattice}, we have computed the
bandstructure of this lattice, and we show it in
Fig.~\ref{fig:DiamondBandstructure}.
In that figure, the notation is the same as in
Fig.~\ref{fig:WoodpileBandstructure}, with Brillouin Zone points plotted in
Fig.~\ref{fig:WoodpileBZ}.  While the bandgap is not as wide as with
silicon, the diamond lattice does exhibit an omnidirectional bandgap,
shown in yellow in Fig.~\ref{fig:DiamondBandstructure}, with
width-to-center ratio of 5.4\%.

\begin{figure}
\begin{center}
\includegraphics{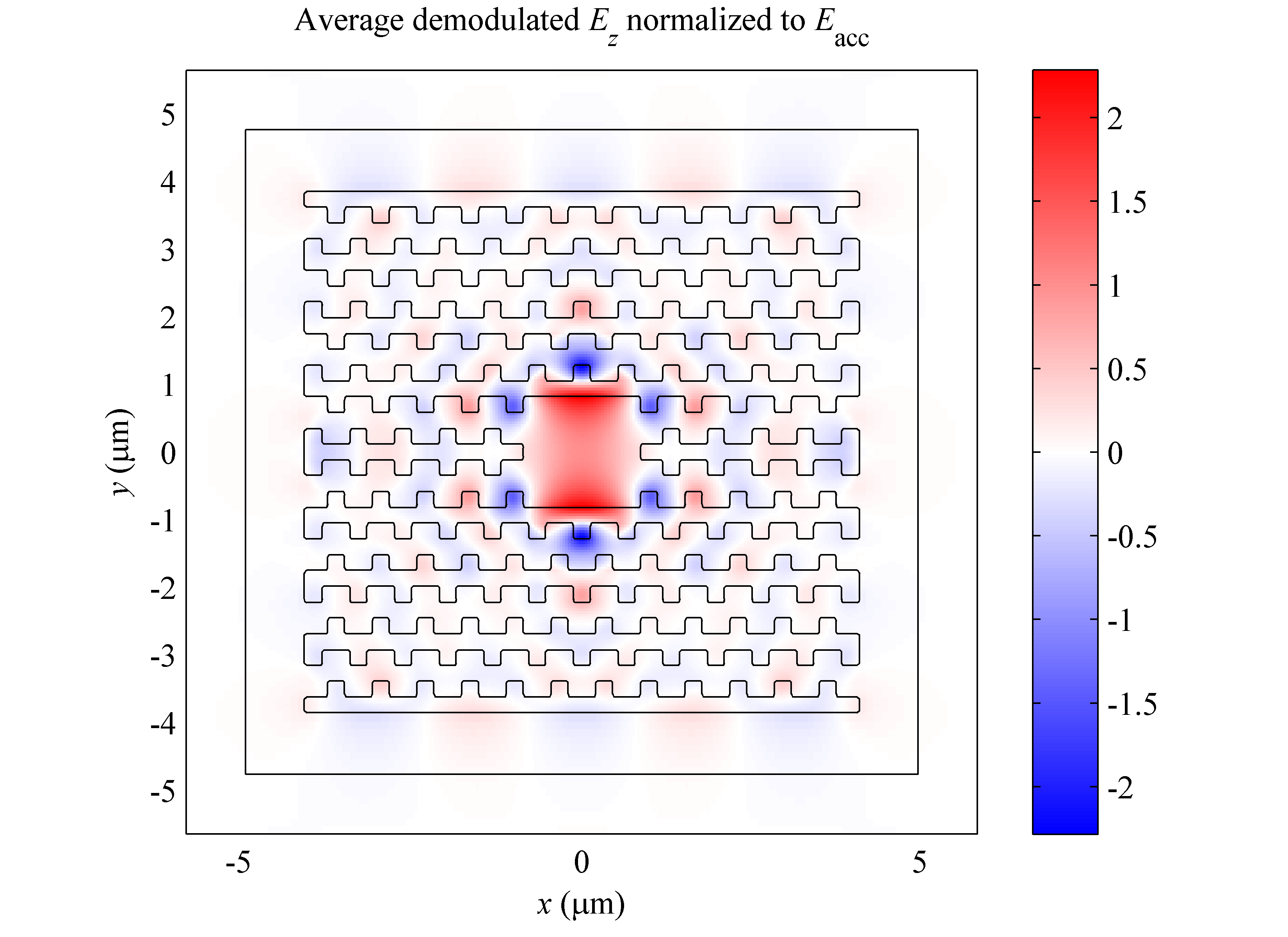}
\caption{Accelerating mode in a diamond-based woodpile structure.}
\label{fig:DiamondMode}
\end{center}
\end{figure}

Finally, we create a waveguide in this lattice and compute an
accelerating mode.  We use the same geometric parameters as in
Sec.~\ref{sec:SymmMode}; namely, we take the waveguide to be $3a - w$
wide by $7\sqrt{2}a/4$ (7 layers) tall.  We find that there is indeed
an accelerating mode in this waveguide; the accelerating fields are
plotted in Fig.~\ref{fig:DiamondMode}.
However, we notice that in comparison with the mode in the silicon
structure plotted in Fig.~\ref{fig:SymmetricMode}, there are
significant fields which extend throughout the PBG lattice surrounding
the waveguide.  Therefore this mode is not as well confined as the
mode in the silicon structure; indeed, instead of being radiatively
lossless, it has a loss of \unit[35.3]{dB/cm}.  Its impedance and
group velocity are also reduced, to $\unit[241]{\ohm}$ and $0.108c$
respectively.  These effects are likely due to the mode frequency of
$0.426c/a$ being close to the bandgap edge, and might be reduced by
perturbing the structure geometry slightly, as was done to affect the
quadrupole fields in Sec.~\ref{sec:BeamDynamics}.  More importantly,
however, the damage impedance is only reduced slightly from the
silicon case, to $\unit[5.56]{\ohm}$ from $\unit[6.10]{\ohm}$.
Therefore a gain in the damage threshold from an alternative material
would be largely preserved in the sustainable gradient of the
structure.  For instance, suppose the damage threshold in diamond is
10 times higher than that of silicon.  The damage field is then
$\sqrt{10} = 3.16$ times higher.  Then, since the sustainable gradient
is proportional to $\sqrt{Z_d}$, the slightly lower damage impedance
for this mode only reduces the gain in sustainable gradient to a
factor of 3.02.

This computation shows that larger-bandgap, lower-index materials
could be suitable for a photonic accelerator, and possibly yield
significant improvements in sustainable gradient.  However, those
improvements would need to be weighed against other advantages of
silicon, such as ionizing radiation hardness, thermal conductivity,
and amenability to microfabrication.  For the future, a comprehensive
study of damage thresholds and other material properties for a variety
of dielectrics would serve to inform the choice of accelerator
structure material.

Another possible mechanism for improving the damage threshold of an
accelerator structure material is to change the laser wavelength.  For
operation in the near infrared, structures such as the one presented
here are not limited by laser power availability.  Even if a material
were found that could sustain a \unit[1]{GeV/m} acceleration gradient
in the woodpile structure, the large characteristic impedance means
that at \unit[1550]{nm} the peak power required would be only $P =
E\tsub{acc}^2\lambda^2/Z_c = \unit[5.2]{kW}$.  Such powers are readily
available at high efficiencies from commercial fiber lasers; a system
is even available which produces pulses with \unit[2]{MW} peak power
at a \unit[60]{kHz} repetition rate \cite{CalmarOptcom}.  However,
because of the possible role of multiphoton ionization in the damage
process, especially for wide-bandgap materials, lengthening the laser
wavelength away from the near infrared may yield higher sustainable
gradient.  Efficient fiber sources exist in Er:glass at
$\unit[1.55]{\micro m}$ and Yb:glass at $\unit[1.05]{\micro m}$.
Degenerate optical parametric generation pumped by these sources to
double the wavelength would allow efficient generation of mid-infrared
wavelengths, and an effort to accomplish this using LiNbO\tsub{3} is
underway \cite{Wong:Degenerate}.  Ultimately, the challenge of
developing a mid-infrared source will need to be weighed against the
potential gain in gradient as well as the alternative challenge of
developing structures from higher-damage threshold materials.
 
\section{Structure considerations}

In addition to the choice of material and laser source, the geometry
of the structure itself is open to alternative candidates.  The most
significant drawback to the woodpile geometry described in
Ch.~\ref{ch:Woodpile} is the difficulty of fabrication, as discussed
in Ch.~\ref{ch:Materials}.  Other geometries may be easier to
fabricate by not requiring alignment of so many layers.

\begin{figure}
\begin{center}
\resizebox{\textwidth}{!}{\includegraphics{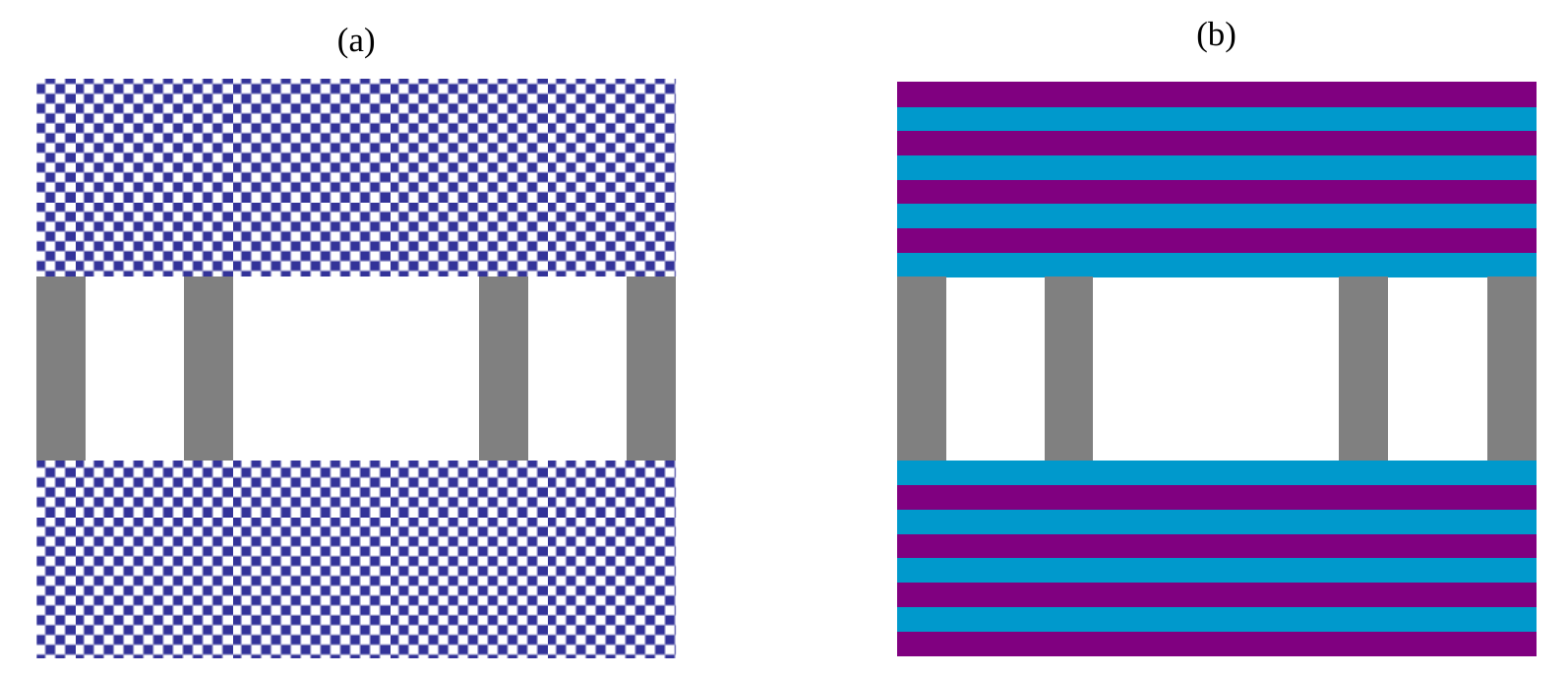}}
\caption{Schematic of a two-dimensional structure with vertical
confinement.  The vertical confinement might be provided by (a) a bulk
three-dimensional photonic crystal with a complete PBG, or (b) a
dielectric multilayer.}
\label{fig:SandwichSchematic}
\end{center}
\end{figure}

One possibility for simpler fabrication would be to return to the
two-dimensional geometries presented in Ch.~\ref{ch:2D}.  As mentioned
in that chapter, a method of vertical confinement is needed for these
structures to be practical.  We might therefore consider placing a
finite slab of 2D structure between other structures which provide the
vertical confinement.  This is shown schematically in
Fig.~\ref{fig:SandwichSchematic}.
One proposal for vertical confinement calls for using bulk
three-dimensional photonic crystals which can be assembled all at
once, rather than layer by layer
\cite{Chutinan:HeterostructuresPRL2003,Chutinan:HeterostructuresPRE2005}.
There are several such photonic lattices.  The inverse opal structure
\cite{Busch:InverseOpal,Vlasov:InverseOpalFab}
can be constructed by first creating a lattice of spheres by
self-assembly, depositing a second dielectric, and dissolving the
spheres.  Another option is the square spiral array geometry
\cite{Toader:SquareSpiralScience,Toader:SquareSpiralPRE,%
Kennedy:SquareSpiralFab}, which is constructed by rotating the
substrate during a deposition.  While these 3D lattices can be
fabricated economically, those same fabrication processes can not be
used to incorporate waveguide structures; those are incorporated only
in the two-dimensional slab.

An even simpler option would be to use a simple dielectric
multilayer stack for vertical confinement, as proposed in
\cite{Chen:DirectionalCoupler}.  This would involve only depositing
thin films of different materials in an alternating pattern; no
patterning of the films would be required.  The thicknesses of the
films would be chosen to provide the best vertical confinement;
however this a nontrivial task because a dielectric multilayer does
not exhibit a complete PBG.

In either case, the bandgap of the 2D and 3D structures would need to
be properly matched to minimize leakage in the heterogeneous
structure.  This is complicated, since the bandstructure of the 2D
slab depends on its placement within the vertically confining
structure.  In Ch.~\ref{ch:2D} we took the fields to be uniform in the
vertical direction; in the Fourier domain this was equivalent to
fixing $k_y = 0$.  The 2D bandstructure, though, will be different for
different values of $k_y$, and a slab within a larger structure will
exhibit a wide range of vertical spatial frequencies.  If the 2D and
3D structures shared the same periodicity in the horizontal
directions, the entire structure would then be periodic in both
horizontal dimensions.  It would therefore have a 2D bandstructure
which we might optimize for widest bandgap.

\chapter{Conclusion}
Significant improvement in accelerating gradient is necessary if
accelerator-based particle physics is to continue its pace of
discovery.  While lasers can provide extraordinary energy density, the
question remains of how to utilize properly the laser power to
accelerate a charged particle beam.

Photonic crystals have great potential for use in accelerator
structures.  By being constructed out of dielectric materials alone,
they can circumvent the problems of metals at optical frequencies.  We
have described a general procedure for the design of photonic crystal
accelerators, and then explored in detail a particular
three-dimensional geometry.  We found solutions to the fundamental
problems of supporting a speed-of-light mode in a photonic crystal
waveguide, and stably propagating a particle beam within the waveguide
aperture.  We also discussed the issues of coupling, material damage,
and fabrication.

There are many bridges yet to cross on the path to a practical
photonic accelerator, and this dissertation covers only the very
beginning of that journey.  One of the most important steps will be to
map the vast unexplored territory of material breakdown thresholds.
Indeed, it cannot be overstated just how much area there is to cover:
The optimal wavelength could occur in any regime from RF to optical
frequencies, and the choice of wavelength will inform power source and
fabrication issues.  A proof-of-principle demonstration of photonic
acceleration has yet to be done, and beyond that the challenge of
scaling to greater lengths must be considered.  Each step along the
way will present unique challenges, from initial fabrication to
maintaining optical-scale stability over a kilometer-scale
accelerator.

Significant improvement in accelerating gradient requires a
fundamentally new technique, and here we have demonstrated the
potential of photonic acceleration as one such technique.

\appendix
\chapter{Computation techniques}
\section{Iterative eigensolver}
\label{sec:IterativeEigensolver}

For the photonic crystal lattice computations we use the \emph{MIT
  Photonic-Bands} (\textsc{mpb}) package, a public-domain code using
an iterative eigensolver technique \cite{Johnson:MPB}.  For a given
Bloch wavevector $\vect{k}$, \textsc{mpb} solves the eigenproblem
given in Eq.~(\ref{eq:BlochEigenproblem}), $\Theta_{\vect{k}}\vect{u}
= (\omega^2/c^2)\vect{u}$, where $\Theta_{\vect{k}}$ is the Maxwell
operator defined in Eq.~(\ref{eq:MaxwellOperator}).  \textsc{mpb}
solves the eigenproblem in the planewave basis, and only the
transverse components of Bloch fields are used in the eigenvector.
Since $\Theta_{\vect{k}}$ is positive-definite, the eigenvector with
the smallest eigenvalue will be the one that minimizes the quantity
\[ \frac{\ip{\vect{u}, \Theta_{\vect{k}}\vect{u}}}{\ip{\vect{u},
    \vect{u}}}. \]
\textsc{mpb} finds the lowest-frequency eigenmodes by solving this
minimization problem using the preconditioned conjugate-gradient
method \cite{Press:NumericalRecipes}, with the modification that it
solves several eigenvectors at once using a block-iterative technique.

Representing continuous fields in the finite memory of a computer
necessarily involves discretization.  In \textsc{mpb}, space is
divided into cells of a uniform, but not necessarily orthogonal, grid.
Each grid cell therefore takes the shape of a parallelopiped.  For
grid cells in which $\epsilon_r$ is not constant, \textsc{mpb}
computes a tensor representing the average permittivity within the
cell, as described in \cite{Johnson:MPB}.

For the computation of the woodpile bandstructure described in
Sec.~\ref{sec:WoodpileLattice}, we used \textsc{mpb} with the FCC unit
cell and a resolution of 64 points per lattice constant in each
direction of the FCC basis vectors.  We computed 5 bands at each of 81
points in the irreducible Brillouin zone.  The computation took 4
hours, 12 minutes on a 1.8 GHz AMD Opteron computer with 2 GB RAM.

We also used \textsc{mpb} for several of the waveguide mode
computations, specifically for the two-dimensional waveguides
presented in Ch.~\ref{ch:2D}, the asymmetric woodpile waveguide in
Sec.~\ref{sec:AsymMode}, and the tolerance studies in
Sec.~\ref{sec:Tolerance}.  In that case, the computational domain
consisted of the waveguide surrounded by several layers of PBG
lattice.  Since \textsc{mpb} uses periodic boundary conditions, this
computation is not entirely physical, since we are actually simulating
an infinite array of waveguides rather than a single structure
surrounded by free space.  Because the size of the computational
domain is several wavelengths in each dimension transverse to the
waveguide, the confined waveguide mode is a high-order mode.  Consider
the scaling of computation resources with the number $n$ of PBG
lattice periods in each transverse dimension of the computational
domain.  The number of grid cells in the domain scales as $n^2$, and
the mode number of the accelerating mode is also $O(n^2)$.  In
addition, at each iteration of the eigensolver, the fields being
computed must be orthogonalized against all the lower-order
modes. This can dominate the computation time for large numbers of
modes, adding another factor of order $n^2$.  Thus the required memory
is $O(n^4)$ and the computational time $O(n^6)$.  This quickly becomes
prohibitive when more lattice periods are required, for instance when
computing the modes of a structure with a lower refractive index
material and thus poorer confinement.  Finding just one accelerating
mode of a perturbed structure for the tolerance study required $\sim
10^6$ CPU seconds on a \unit[2.4]{GHz} Xeon processor with 16 grid
points per lattice period.

\textsc{mpb} has the option of operating in a targeted frequency mode
to find eigenmodes with a frequency close to a target frequency
$\omega_0$.  In this mode, instead of finding the eigenvectors of
$\Theta_{\vect{k}}$ directly, \textsc{mpb} finds the eigenvectors of
the operator $A = (\Theta_{\vect{k}} - \omega_0^2/c^2)^2$.  The lowest
eigenvalues of $A$ correspond to the eigenfrequencies closest to
$\omega_0$.  However, critical to the efficiency of \textsc{mpb} is
the use of a preconditioner, an operator designed to approximate the
inverse of the operator in the eigenvalue problem.  An effective
preconditioning technique exists for $\Theta_{\vect{k}}$, and is
described in \cite{Johnson:MPB}.  On the other hand, the
preconditioner for $A$ is not nearly as effective, and computations of
waveguide modes using this method fail to converge.  Preconditioning
is described in \cite{Johnson:MPB} as a ``black art,'' so we do not
attempt to use the conjugate-gradient method by improving the
preconditioner for $A$.  Instead, we use an entirely different
technique, based on time domain methods, for computing the other
waveguide modes---those described in Secs.~\ref{sec:SymmMode},
\ref{sec:Coupling}, \ref{sec:BeamDynamics}, and \ref{sec:Materials}.
We describe time domain methods in the next section, followed by a
discussion of the related eigensolver technique.

\section{The finite-difference time-domain method}
\label{sec:FDTD}

The finite-difference time-domain (FDTD) method is a general and
robust technique for propagating the Maxwell equations forward in time
from a set of initial and source conditions \cite{Taflove:FDTD}.  The
method works simply by stepping the magnetic and electric
fields in time according to the Maxwell equations
\[ \frac{\ptl\vect{H}}{\ptl t} = -\frac{1}{\mu}\del\cross\vect{E},
\qquad \frac{\ptl\vect{E}}{\ptl t} =
\frac{1}{\epsilon}\del\cross\vect{H}, \]
where the time derivatives are discrete differences between subsequent
steps, and the spatial derivatives are represented as finite
differences on an orthogonal grid.  Key to the stable operation of
FDTD is that the field components are not co-located in space or time.
Instead, they are located on a Yee grid \cite{Yee:Grid}, in which for
each coordinate $i$, the component $E_i$ is displaced half a grid cell
in the $i$ direction but not in the other directions, while the $H_i$
component is displaced half a grid cell in the other directions but not
in the $i$ direction.  In addition, the $\vect{E}$ field is displaced
half a step in time from the $\vect{H}$ field.  This allows both the
space and time derivatives to be central differences.  For instance,
the $x$ component of the update equation for $\vect{H}$ is
\[ \frac{\ptl H_x}{\ptl t}
= \frac{1}{\mu}\paren{\frac{\ptl E_y}{\ptl z} - \frac{\ptl E_z}{\ptl
	y}}. \]
If we let $(i, j, k)$ denote spatial grid coordinates and $n$ denote
	the time step, this becomes
\begin{multline}
\frac{\evalu{H_x}^{n+1}_{i,j+1/2,k+1/2} -
	\evalu{H_x}^n_{i,j+1/2,k+1/2}}{\Delta t} \\
= \frac{1}{\mu}\paren{\frac{\evalu{E_y}^{n+1/2}_{i,j+1/2,k+1} -
	\evalu{E_y}^{n+1/2}_{i,j+1/2,k}}{\Delta z}
- \frac{\evalu{E_z}^{n+1/2}_{i,j+1,k+1/2} -
	\evalu{E_z}^{n+1/2}_{i,j,k+1/2}}{\Delta y}},
\label{eq:FDTDUpdateHx}
\end{multline}
where $\Delta t$ is the time step and $\Delta y$ and $\Delta z$ are
the grid spacings in the $y$ and $z$ directions respectively.
Thus the $H_x$ component is located between the components of
$\vect{E}$ which need to be differentiated.  In this scheme, it can be
shown that this algorithm is stable as long as the time step satisfies
the Courant condition,
\[ \Delta t < \frac{1}{c\sqrt{\frac{1}{(\Delta x)^2} +
	\frac{1}{(\Delta y)^2} + \frac{1}{(\Delta z)^2}}}, \]
as described in Ch.~4 of \cite{Taflove:FDTD}.

There are two important additional features of FDTD we implemented in
order to make our time-domain simulations effective.  The first is an
absorbing boundary layer.  We simulated structures surrounded by free
space, but it is not possible to include an infinite volume of space
in a finite simulation.  Instead, we surrounded the simulation space
with a \emph{uniaxial perfectly-matched layer} (UPML), which is a
non-physical material which is perfectly matched to free space but
which absorbs electromagnetic radiation.

The UPML is based on a simple yet sufficient criterion for perfect
transmission from one material to another.  Consider an isotropic
dielectric with permittivity $\epsilon_1$ and permeability $\mu_1$
adjacent to an anisotropic dielectric with permittivity and
permeability tensors $\bar{\bar{\epsilon}}_2$ and $\bar{\bar{\mu}}_2$
respectively, and suppose that the interface between them is normal to
the $x$ axis.  It can be shown that electromagnetic waves propagate
without reflection from the first material to the second as long as
$\bar{\bar{\epsilon}}_2 = \epsilon_1\bar{\bar{s}}$ and
$\bar{\bar{\mu}}_2 = \mu_1\bar{\bar{s}}$, where
\[ \bar{\bar{s}} =
\begin{pmatrix}
s_x^{-1} & 0 & 0 \\
0 & s_x & 0 \\
0 & 0 & s_x
\end{pmatrix} \]
for any scalar $s_x$.  In particular, we can introduce loss into the
anisotropic material by making $s_x$ complex:
\[ s_x = 1 + \frac{\sigma_x}{i\omega\epsilon_0}, \]
where $\omega$ is the frequency.  Since this anisotropic material is
perfectly matched to free space (or any isotropic lossless dielectric)
and is lossy, it functions as an absorbing boundary condition.  The
frequency dependence of $s_x$ can be efficiently implemented in the
time domain by storing auxiliary fields $\vect{B}$ and $\vect{D}$
along with $\vect{E}$ and $\vect{H}$.  A detailed description of the
UPML implementation as well as the three-dimensional formulation is
described in Ch.~7 of \cite{Taflove:FDTD}.

The second additional feature is the incident source condition called
the total-field/scattered-field technique.  In this technique, the
computational domain is divided into a total field and a scattered
field region, located downstream and upstream, respectively, of the
incident field.  In the total field region, the total electric and
magnetic fields are stored as usual.  However, in the scattered field
region, we store only the scattered fields, that is, the total fields
minus the incident fields.  As long as the incident fields satisfy the
Maxwell equations in the scattered field region, the fields can be
updated as usual in both the total field and scattered field regions.
The only required modification is at the interface between the
regions, when updating a field component in one region requires using
a component from the other.  For instance, consider the update
equation, Eq.~(\ref{eq:FDTDUpdateHx}), for $H_x$ given above.  Suppose
that the boundary between the scattered field and total field regions
is normal to the $z$ direction and occurs between grid lines at $k$
and $k + 1/2$, with the total field region being on the $+z$ side.
Then in that equation, all components are in the total field region
except $\evalu{E_y}^{n+1/2}_{i,j+1/2,k}$.  Since that component lies
in the scattered field region, we must add the incident field to it to
obtain the total field.  Thus we must add to the updated value
$\evalu{H_x}^{n+1}_{i,j+1/2,k+1/2}$ derived from
Eq.~(\ref{eq:FDTDUpdateHx}) the quantity
\[ -\frac{\Delta t}{\mu\Delta
  z}\evalu{E_{y,\text{inc}}}^{n+1/2}_{i,j+1/2,k}. \]
The modified updates of other components proceed similarly.  A
detailed description of the total-field/scattered-field technique is
given in Ch.~5 of \cite{Taflove:FDTD}.

With the FDTD technique with these features, we are able to include an
incident field in our coupling simulations, propagate the fields
forward in time, and allow outgoing radiation to exit the simulation
space without reflection.

To discretize the permittivity we use the averaging scheme described
in \cite{Kaneda:EpsilonAverage}.  Since the components of $\vect{E}$
are not co-located, the values of the averaged $\epsilon$ can be
different for every component.  We compute the average permittivity as
follows.  Let $\bar{\epsilon}_x$ be the average permittivity at the
location of an $E_x$ field component.  Suppose the component is
located at coordinates $(x_0, y_0, z_0)$ and that the cell has dimensions
$\Delta x\times\Delta y\times\Delta z$.  We define the average
permittivity by
\[ \frac{1}{\bar{\epsilon}_x} = \int_{x_0 - \Delta x/2}^{x_0 + \Delta x/2}
\brckt{\int_{y_0 - \Delta y/2}^{y_0 + \Delta y/2} \int_{z_0 - \Delta
z/2}^{z_0 + \Delta x/2} \epsilon(x, y, z)\,dy\,dz}^{-1}\,dx. \]
This is computed numerically by evaluating the permittivity at a
$5\times 5\times 5$ mesh within each grid cell.  The average
permittivities at the locations of the other $\vect{E}$-field
components are computed similarly.

\section{An FDTD-based mode solver}
\label{sec:FDTDSolver}

We can use the FDTD method to compute eigenmodes of photonic crystal
waveguides.  The typical technique for this is to use complex fields
and periodic boundary conditions with the desired Bloch wavenumber.
We can then excite the waveguide with a point source of the correct
polarization and finite duration, and then allow the fields to ring
down.  We can examine the time structure of one component of the
fields at a single point and use a procedure such as harmonic
inversion
\cite{Mandelshtam:HarmonicInversion,Mandelshtam:HarmonicInversionErratum}
to extract the eigenfrequencies.  We can then repeat the process with
a narrow-band source at one of the eigenfrequencies to isolate a
single mode.

However, this approach has a serious problem, but one that we can
overcome with a few modifications.  Specifically, it does not
converge to the desired eigenmode.  When the fields ring down, those
modes which are most confined remain.  Because of numerical dispersion
in the FDTD grid, plane waves at the Nyquist frequency of the grid
approach zero group velocity, and are therefore stationary and do not
decay away.  Thus the desired eigenmode is polluted with
high-spatial-frequency components which are not physical.  Indeed,
such numerical parasites grow in amplitude compared to the desired
mode.  Since we wish to get a highly accurate evaluation of the fields
at particular points (on axis especially), this numerical effect
detracts from the validity of our results.

We can overcome these numerical artifacts using three key observations.
The first is that if we consider the vector space of all the fields in
the simulation---including the $\vect{B}$ and $\vect{D}$ fields used
in the UPML region---then the action of a single timestep, in the
absence of sources, is a linear operator.  We can then use concepts
from linear algebra to address this problem.  Let $S$ denote the
timestep operator and $\psi$ denote a field vector.  If $\psi$
represents an eigenmode of the system, then $\psi$ is an eigenvector
of $S$ with eigenvalue $e^{i\omega\Delta t}$, where $\omega$ is the
mode frequency and $\Delta t$ is the time step.

The second observation is that if $\psi$ is an eigenvector of $S$ with
eigenvalue $\lambda$ and $P$ is any polynomial, then $\psi$ is also an
eigenvector of $P(S)$ with eigenvalue $P(\lambda)$.  The third
observation is that we have a wide choice of sources and temporal
excitations, and that the choice of source corresponds to an initial
vector, while the choice of temporal excitation corresponds to a
polynomial $P$, as we now show.  Addition of a source vector $v$
during a timestep corresponds to the transformation
\[ \psi\mapsto S\psi + v. \]
Now consider a source $v$ which is added to the simulation with
temporal profile $\{a_n\}_{n=0}^N$.  Let $\psi_n$ denote the state
vector at timestep $n$.  We can then compute $\psi_{n+1}$ according to
the recursion relation $\psi_{n+1} = S\psi_n + a_{n+1}v$:
\begin{align*}
\psi_0 &= a_0v, \\
\psi_1 &= S\psi_0 + a_1v = (a_0S + a_1)v, \\
\psi_2 &= (a_0S^2 + a_1S + a_2)v, \\
&\vdots \\
\psi_n &= \paren{\sum_{n=0}^N a_nS^{N-n}}v.
\end{align*}
Thus a source added to the simulation with an $N+1$-step temporal
excitation corresponds to applying $P(S)$ to the source vector, where
$P$ is a degree $N$ polynomial.

Given a state vector $\psi$, we can decompose $\psi$ into a
superposition of eigenvectors $\psi^{(m)}$,
\[ \psi = \sum_m b_m\psi^{(m)}. \]
If for each $m$ the vector $\psi^{(m)}$ has eigenvalue $\lambda_m$,
then applying the operator $P(S)$ to $\psi$ results in the
transformation
\[ \sum_m b_m\psi^{(m)} \mapsto \sum_m P(\lambda_m)b_m\psi^{(m)}. \]

To converge to our desired mode, we wish to find a polynomial $P$ for
which $|P(\lambda)|$ is large for the desired eigenvalue relative to
$|P(\lambda_m)|$ for the other eigenvalues.  To find a suitable
polynomial, we can use our knowledge of the physical properties of the
mode.  We can isolate an eigenvalue with a narrow-band excitation.
Consider a lossless mode, with eigenvalue $\lambda = e^{i\omega\Delta
t}$ with $\omega$ real.  Then for an excitation $\{a_n\}_{n=0}^N$, we
have
\[ P(\lambda) = \sum_{n=0}^N a_ne^{i\omega(N-n)\Delta t}
= e^{i\omega N\Delta t}\sum_{n=0}^N a_ne^{-i\omega n\Delta t}. \]
This corresponds to the Fourier amplitude of the sequence ${a_n}$ at
frequency $\omega$.  Thus we can isolate a particular mode by
tailoring the excitation in the frequency domain.  The traditional
method described above does this initially by starting with a
narrow-band excitation at the frequency of interest.  However, the
following polynomial is simply $P(S) = S^N$, which amplifies the
modes with largest magnitude eigenvalues, which are unfortunately the
numerical dispersion artifacts.  We modify this method by using
narrow-band excitations throughout the simulation.

Our method for solving guided modes in photonic crystal waveguides
proceeds as follows: Since we know the desired mode lies in the
bandgap, and we know the bandgap frequencies from a lattice
simulation, we start with a point excitation whose bandwidth covers
the bandgap.  We then let the fields ring down, and from the time
structure we can extract the guided mode frequencies as usual.  We
choose one of these modes to solve for.  We then compute an excitation
at the frequency of the desired mode with a Gaussian envelope.  We
choose the envelope to be sufficiently narrow-band in the frequency
domain to avoid the following unwanted trapped modes:  First, there
are the numerical dispersion artifacts discussed above; these are at
high frequencies.  Second, there are field divergences, which
correspond to charges on the FDTD grid.  These charges are static, so
they have zero frequency and do not ordinarily decay.  Finally,
the group velocity of a photonic crystal lattice mode approaches zero
as the frequency nears the edge of the bandgap, so such modes do not
decay.  Therefore we must make the excitation narrow-band enough that
the amplitude at the bandgap edges is very small relative to the
amplitude at the desired frequency.

Once we have computed the desired excitation, we apply it repeatedly,
using the fields at the end of one iteration as the sources for the
next.  Algebraically, if $P$ is the polynomial corresponding to our
excitation, then each iteration takes $\psi\mapsto P(S)\psi$.  This
differs from the traditional approach in which the iteration is
$\psi\mapsto S\psi$.  With our method, the procedure converges to the
desired eigenmode.

\begin{figure}
\begin{center}
\includegraphics{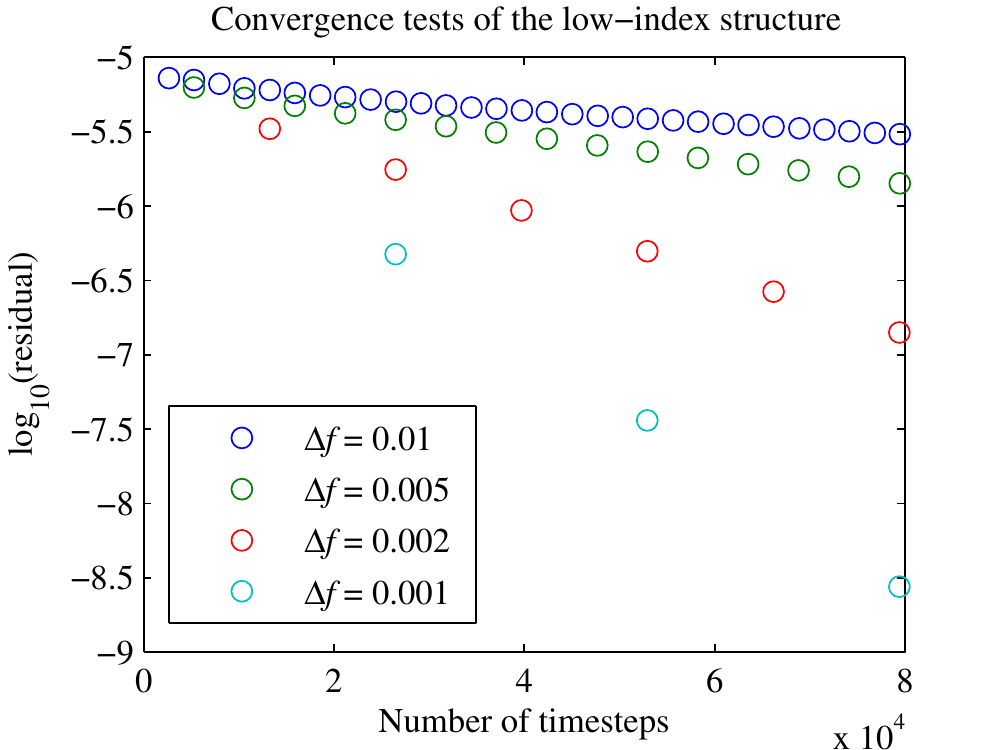}
\caption{Convergence of the FDTD-based mode solver algorithm for
  several excitation bandwidths.}
\label{fig:ConvergenceTest}
\end{center}
\end{figure}

To assess convergence we compute the residual, which is defined for
field state $\psi$ as $r = |S\psi - \lambda\psi|^2$.  For a given
state, the value of $\lambda$ which minimizes the residual is
\[ \lambda = \frac{\ip{\psi, S\psi}}{\ip{\psi, \psi}}. \]
For that value of $\lambda$ the residual is then
\[ r = \abs{S\psi}^2 - \frac{\abs{\ip{\psi,
      S\psi}}^2}{\abs{\psi}^2}. \]
We consider a mode to be converged when the residual has dropped below
a certain threshold, typically $10^{-6}$.  We show the convergence of
this algorithm in Fig.~\ref{fig:ConvergenceTest}.
In that figure, we show how the convergence proceeds for the low
refractive index structure described in Sec.~\ref{sec:Materials},
depending on the bandwidth of the excitation used.  For a given
bandwidth $\Delta f$ (given in the figure in units of $c/a$), we let
\[ N = \ceil{\frac{3}{2\pi\Delta f\Delta t}}, \]
and then define the excitation
\[ a_n = \frac{1}{A}\exp\brckt{2\pi if_0n\Delta t
- \frac{(2\pi\Delta fn\Delta t)^2}{2}},\quad n = -N,\ldots,N, \]
where $f_0$ is the center frequency of the excitation.  The
normalization factor $A$ is included to ensure that $\sum_{n=-N}^n
|a_n| = 1$, so that the desired mode amplitude remains roughly the
same order of magnitude through several iterations to avoid reaching
numeric limits.  Thus the excitation is a Gaussian which extends to
$\pm 3\sigma$ from the peak.  In the frequency domain, the excitation
has the amplitude spectrum $\tilde{a}(f) = e^{-(f - f_0)^2/2(\Delta
f)^2}$ in the continuous limit.  There is a trade-off in choosing the
excitation bandwidth: A narrower bandwidth results in faster
convergence per iteration, while each iteration requires more timesteps
because the excitation must be wider in the time domain.  However,
these effects are not equal.  We can expect that with the residual
determined by the relative amplitudes of the undesired modes, we would
have from the frequency spectrum that the decrease in $\log r$ per
iteration would scale as $1/(\Delta f)^2$.  On the other hand, the
number of timesteps per iteration goes as $1/\Delta f$.  Thus the
decrease in $\log r$ per timestep still increases with narrower
bandwidth, as we see in Fig.~\ref{fig:ConvergenceTest}.

To speed convergence further, we use a low-dimension Krylov subspace
computation before each iteration.  If the residual is dominated by a
few slowly-decaying modes, this procedure will serve to remove them.
For a $k$-dimensional computation, we form the subspace
\[ \mathcal{K}_k = \Span \{\psi, S\psi,\ldots, S^{k-1}\psi\}. \]
We find an orthonormal basis for $\mathcal{K}_k$ and form a matrix $V$
with the basis vectors as its columns.  Then we examine the $k\times
k$ matrix $A = V^\dagger SV$.  If the eigenvalues of $A$ are all large
(i.e. close to 1), it suggests that the mode content of
$\mathcal{K}_k$ is dominated by $k$ distinct modes.  We therefore
continue increase $k$ until one of the eigenvalues of $A$ is below
0.99, and take $k$ to be the largest one with all eigenvalues of $A$
greater than 0.99.  We typically have $k = 2$ or $k = 3$, so the
computational burden of the Krylov subspace computation is marginal.
Finally, we replace $\psi$ with the vector in $\mathcal{K}_k$ which
minimizes the residual.  This eliminates the pollution from other
modes in $\mathcal{K}_k$.

\section{Mode convergence technique}
\label{sec:ModeConvergence}

The previous section described how we compute guided modes of photonic
crystal waveguides.  While that procedure computes modes for a
particular Bloch wavenumber, we wish to find the mode that lies at a
particular point on the dispersion curve, namely the speed-of-light
mode.  Once we have a waveguide mode, we can quickly converge to the
speed-of-light mode by first computing the mode's group velocity.
This can be done by first computing the power flow
\[ P = \int_A \re (\vect{E}^*\cross\vect{H})\,d^2\vect{x}, \]
where the integral is over a cross-section of the guide.  We then
compute the linear energy density in the guide
\[ U = \frac{1}{a}\int_V \mu_0|\vect{H}|^2\,d^3\vect{x}, \]
where $V$ is the volume and $a$ is the length of one period of the
waveguide (note that the electric and magnetic energies in one period
are equal for an eigenmode).  The group velocity is then simply $P/U$.

This gives us not only a point on the dispersion curve, but the slope
of the dispersion curve at that point as well.  With this information,
we can estimate the longitudinal wavenumber of the speed-of-light
mode.  In our \textsc{mpb} simulations, the wavenumber and frequency
are both real, and we proceed as follows.  Suppose we have found
frequency $\omega^{(n)}$ at wavenumber $k_z^{(n)}$.  We also know that
the slope of the dispersion curve is $\evalu{d\omega/dk_z}_{k_z^{(n)}}
= v_g$.  Approximating the dispersion curve by the linear relationship
given by these parameters, we can estimate the frequency and
wavenumber of the speed-of-light mode for the next step using the
relations
\[ ck_z^{(n+1)} = \omega^{(n+1)}
	= \omega^{(n)} + v_g\paren{k_z^{(n+1)} - k_z^{(n)}}. \]
This yields
\[ k_z^{(n+1)} = k_z^{(n)} + \frac{\omega^{(n)}/c
	-  k_z^{(n)}}{1 - \beta_g}, \]
where $\beta_g = v_g/c$.

With our FDTD-based solver, both the wavenumber and frequency can be
complex.  The frequency can be complex because the absorbing boundary
introduces loss into the simulation.  The wavenumber can be set
arbitrarily by imposing a phase factor of $e^{-ik_zz}$ between one
waveguide period and the next in the boundary conditions; here $k_z$
need not be real.  In that case, we wish to converge on a dispersion
point for which the frequency is real, since the source does not decay
in time, and is equal to $(c\re k_z)$, for phase matching.  This
requires a complex wavenumber, which corresponds to a spatially
decaying mode.  Also, in this case, the group velocity is the complex
quantity
\[ v_g = \frac{1}{U}\int_A \vect{E}^*\cross\vect{H}\,d^2\vect{x}; \]
the imaginary part of $v_g$ indicates loss.  We can estimate the
complex wavenumber for the next iteration by considering the
wavenumber and frequency as vectors in $\mathbb{R}^2$ rather than
complex quantities.  The group velocity is then represented by the
matrix
\[ V_g = \begin{pmatrix} \re v_g & -\im v_g \\ \im v_g & \re
  v_g \end{pmatrix}, \]
and the estimated dispersion relation is
\[ \omega = \omega^{(n)} + V_g(k_z - k_z^{(n)}). \]
We must solve for
\[ \omega^{(n+1)} = \begin{pmatrix} c\re k_z^{(n+1)} \\
  0 \end{pmatrix}
= \begin{pmatrix} c & 0 \\ 0 & 0 \end{pmatrix}k_z^{(n+1)}, \]
so
\[ \begin{pmatrix} c\re k_z^{(n+1)} \\ 0 \end{pmatrix}
= \omega^{(n)} + V_g(k_z^{(n+1)} - k_z^{(n)}). \]
Now define the matrix
\[ M = \begin{pmatrix} c & 0 \\ 0 & 0 \end{pmatrix} - V_g. \]
Then
\begin{align*}
M(k_z^{(n+1)} - k_z^{(n)})
&= \begin{pmatrix} c\re k_z^{(n+1)} \\ 0 \end{pmatrix}
- \begin{pmatrix} c\re k_z^{(n)} \\ 0 \end{pmatrix}
- V_g(k_z^{(n+1)} - k_z^{(n)}) \\
&= \omega^{(n)} - \begin{pmatrix} c\re k_z^{(n)} \\ 0 \end{pmatrix}.
\end{align*}
Note that the right hand side of this equation is equal to the
discrepancy between the frequency and the speed-of-light frequency.
This gives for the updated wavenumber
\[ k_z^{(n+1)} =  k_z^{(n)} + M^{-1}\brckt{\omega^{(n)}
- \begin{pmatrix} c\re k_z^{(n)} \\ 0 \end{pmatrix}}. \]

Once we have the updated speed-of-light wavenumber estimate, we rerun
the mode computation with that new wavenumber.  We continue to repeat
this process until the difference between the frequency and $ck_z$ is
less than the desired tolerance (and in the complex case, $\im\omega$
is also less than the tolerance).  Since the new fields will be close
to the fields with the old wavenumber, we start the mode computation
from the old field configuration, speeding convergence more in each
subsequent wavenumber iteration.  Thus we are able to iteratively
converge on a speed-of-light mode once we have a single mode without
repeating the full computational task of finding eigenmodes.

In the case of the FDTD-based solver, once we have converged on a
speed-of-light mode, the wavenumber will be complex.  We can compute
the loss from the imaginary part of $k_z$.  A field component has a
spatial dependence $\psi\sim e^{-ik_zz}$, so the field magnitude goes
as
\[ \abs{\psi}\sim e^{(\im k_z)z}. \]
We can then define the decay constant of the fields as
\[ \alpha = -\frac{1}{|\psi|}\frac{d|\psi|}{dz}
= -\frac{d}{dz}(\log |\psi|) = -\im k_z. \]
Note that we have $\im k_z < 0$.  In our simulations, the numerical
tolerance on the wavenumber is $10^{-6}\cdot 2\pi/a$, where $a$ is the
lattice constant.  Then a mode is lossless to within the computational
tolerance if
\[ \alpha = \abs{\im k_z} < 10^{-6}\frac{2\pi}{a}. \]
As reported in Sec.~\ref{sec:SymmMode}, this corresponds to a loss of
\unit[0.48]{dB/cm}.

\section{Beam dynamics simulations}
\label{sec:BeamDynamicsComputations}

For our computations of particle trajectories, we use the
\textsc{matlab} \texttt{ode45} solver to directly integrate the
equation of motion.  While we use the Lorentz force equation
\[ \frac{d\vect{p}}{dt} = e(\vect{E} + \vect{v}\cross\vect{B}) \]
without approximation, we make several transformations in order to use
a more appropriate coordinate system for beam dynamics.  First, we
take derivatives with respect to the $z$ coordinate rather than with
respect to $t$.  Second, we track the following phase space
coordinates.  The transverse positions $x$ and $y$ are computed in
units of meters.  The transverse momenta, on the other hand, are not computed
in SI units; rather we normalize them to $mc$, computing in terms of
the variables
\[ P_x = \frac{p_x}{mc} = \beta_x\gamma,\qquad P_y = \frac{p_y}{mc} =
\beta_y\gamma, \]
where $\vect{p}$ is the linear momentum.  Our longitudinal coordinates
are computed relative to the ideal particle, that is, the particle
traveling directly on axis with the design momentum.  We track this
design momentum $P_0$, again normalized to $mc$, along with the six
phase space variables of the particle, and use it to compute the
longitudinal phase space coordinates.  Instead of tracking the
absolute time, we instead track the optical phase $\phi = \omega t -
k_zz$.  Instead of tracking the energy, we track the relative momentum
\[ \delta = \frac{P_z}{P_0} - 1, \]
where $P_z = p_z/mc = \beta_z\gamma$.

The ODE solver requires us to provide the derivatives of the six phase
space variables $(x, y,\linebreak[0] P_x, P_y,\linebreak[0] \phi,
\delta)$ and the design momentum $P_0$, given values for those
variables as well as the longitudinal position $z$.  To that end, we
first define the normalized charge $\bar{q} = e/mc^2$.  We then
compute the useful variables
\begin{align*}
P_z &= P_0(1 + \delta), \\
\gamma &= \sqrt{1 + (\beta\gamma)^2} \\
&= \sqrt{1 + P_x^2 + P_y^2 + P_z^2}, \\
\beta_z^{-1} &= \frac{\gamma}{P_z}.
\end{align*}
The derivatives of the position variables are purely kinematic:
\[ \frac{dP_x}{dz} = \frac{P_x}{P_z},\quad
\frac{dP_y}{dz} = \frac{P_y}{P_z}. \]
For the optical phase, we assume that the phase velocity of the mode
is adjusted to match the ideal particle velocity $\beta_0c$.  We then have
\begin{align*}
\frac{d\phi}{dz} &= \frac{d}{dz}(\omega t - k_zz)
= \frac{2\pi}{\lambda}\paren{\frac{c}{dz/dt} - \frac{ck_z}{\omega}} \\
&= \frac{2\pi}{\lambda}\paren{\frac{1}{\beta_z} - \frac{1}{\beta_0}},
\end{align*}
with
\[ \frac{1}{\beta_0} = \sqrt{1 + \frac{1}{P_0^2}}. \]
For the momentum variables, we require the fields.  We use the
polynomial fits to the fields described in sec.~\ref{sec:BeamDynamics}
together with the position coordinates $(x, y, \phi)$ to obtain the
real-valued fields $\vect{E}$ and $\vect{H}$.  Note that here we have
multiplied $\vect{H}$ by $Z_0$, i.e. $\vect{H} = Z_0\mathcal{H}$ where
$\mathcal{H}$ is the physical magnetic field.  This is done for
convenience so that both $\vect{H}$ and $\vect{E}$ have the same units
of \unit{V/m}.  The Lorentz force equation then becomes
\[ \frac{d\vect{p}}{dt}
= e(\vect{E} + \boldsymbol{\beta}\cross\vect{H}). \]
For the normalized momentum $P = \vect{p}/mc$, we then have
\begin{align*}
\frac{d\vect{P}}{dz} &= \frac{1}{mc}\frac{d\vect{p}/dt}{dz/dt}
= \frac{1}{mc^2\beta_z}e(\vect{E} + \boldsymbol{\beta}\cross\vect{H}) \\
&= \frac{\bar{q}}{\beta_z}(\vect{E} + \boldsymbol{\beta}\cross\vect{H}).
\end{align*}
This yields
\begin{align*}
\frac{dP_x}{dz}
&= \frac{\bar{q}}{\beta_z}(E_x + \beta_yH_z - \beta_zH_y) \\
&= \bar{q}\paren{\frac{E_x}{\beta_z} + \frac{dy}{dz}H_z - H_y},
\end{align*}
and similarly
\[ \frac{dP_y}{dz} = \bar{q}\paren{\frac{E_y}{\beta_z} -
  \frac{dx}{dz}H_z + H_x}. \]
For the ideal particle momentum, we retrieve the fields on axis; we
know $\vect{H} = 0$ there so we simply have
\[ \frac{dP_0}{dz} = \frac{\bar{q}E_{z0}}{\beta_0}, \]
where $E_{z0}$ is the accelerating field on axis.  We then compute
\[ \frac{dP_z}{dz} = \bar{q}\paren{\frac{E_z}{\beta_z} +
  \frac{dx}{dz}H_y - \frac{dy}{dz}H_x}. \]
From this we can finally compute
\[ \frac{d\delta}{dz} = \frac{1}{P_0}\frac{dP_z}{dz} -
\frac{P_z}{P_0^2}\frac{dP_0}{dz}. \]
This allows us to compute derivatives of the phase space coordinates
as required by the \textsc{matlab} solver.  Before we do so, we check
that the transverse position of the particle is still within the
waveguide aperture.  If it isn't, we generate a \textsc{matlab} error
so that the particle trajectory can be recorded as having exited the
waveguide.

To carry out this computation, \textsc{matlab} requires approximately
12.5 CPU seconds per particle propagated for \unit[3]{m} on a Sun Fire
V20z computer with a 2.0 GHz AMD Opteron processor and 4 GB RAM.


\bibliographystyle{apsrev}
\bibliography{allrefs}

\end{document}